\documentclass[longauth]{aa}  

\usepackage{graphicx}
\usepackage{txfonts}
\usepackage{sidecap}
\def\deg{\hbox{$^\circ$}}
\begin{document}

\title{Physical parameters of selected Gaia mass asteroids}

\authorrunning{Podlewska-Gaca et al.}
\titlerunning{Physical parameters of Gaia mass asteroids}

\author{E. Podlewska-Gaca
  \inst{1},
  A. Marciniak\inst{1}, V. Al{\'i}-Lagoa\inst{2}, P. Bartczak \inst{1},
  T. G. M{\"u}ller\inst{2}, R. Szak{\'a}ts\inst{3}, R. Duffard\inst{4},
 L. Moln{\'a}r \inst{3,5}, A. P{\'a}l\inst{3,6}, M. Butkiewicz-Bąk\inst{1},
 G. Dudziński\inst{1}, K. Dziadura\inst{1},  P. Antonini\inst{7} 
  V. Asenjo\inst{8}, M. Audejean\inst{9}, Z. Benkhaldoun\inst{10}, R.
Behrend\inst{11}, L. Bernasconi\inst{12},
 J. M. Bosch\inst{13}, A. Chapman\inst{14}, B. Dintinjana\inst{25}  A. Farkas\inst{3}, 
 M. Ferrais\inst{15}, S. Geier\inst{16,17}, J. Grice\inst{18},  R. Hirsh\inst{1}, H.
 Jacquinot\inst{19} E. Jehin\inst{15},
 A. Jones\inst{20}, D. Molina\inst{21}, N. Morales\inst{4},
 N.~Parley\inst{22}, R. Poncy\inst{23},
R.Roy\inst{24}, T. Santana-Ros\inst{26,27},  B. Seli\inst{3},  K. Sobkowiak\inst{1},
 E. Vereb{\'e}lyi\inst{3},  K. Żukowski\inst{1}  
}

\institute{Astronomical Observatory Institute, Faculty of Physics, Adam Mickiewicz University,
  Słoneczna 36, Poznań, Poland\\
  \email{edypod@amu.edu.pl}
  \and
  Max-Planck-Institut für extraterrestrische Physik (MPE), Giessenbachstrasse 1, 85748
  Garching, Germany
  \and
  Konkoly Observatory, Research Centre for Astronomy and Earth Sciences, Hungarian
  Academy of Sciences, H-1121 Budapest, Konkoly Thege Miklós út 15-17, Hungary 
  \and
  Instituto de Astrofísica de Andalucía (CSIC), Glorieta de la Astronomía s/n, 18008
  Granada, Spain.
 \and
  MTA CSFK Lend{\"u}let Near-Field Cosmology Research Group, Budapest,
  Hungary
 \and
  Astronomy Department, E\"otv\"os Lor\'and University, P\'azm\'any P. s. 1/A, H-1171 Budapest, Hungary
  \and
Observatoire des Hauts Patys, F-84410 Bedoin, France
\and 
 Asociación Astronómica Astro Henares, Centro de Recursos Asociativos El Cerro
C/ Manuel Azaña, 28823 Coslada, Spain 
\and
B92 Observatoire de Chinon, Chinon, France
  \and
Oukaimeden Observatory, High Energy Physics and Astrophysics Laboratory, Cadi Ayyad University, Marrakech,
Morocco
\and 
Geneva Observatory, CH-1290 Sauverny, Switzerland
\and
Observatoire des Engarouines, 1606 chemin de Rigoy, F-84570 Malemort-du-Comtat, France
\and
B74, Avinguda de Catalunya 34, 25354 Santa Maria de Montmagastrell
(Tàrrega), Spain 
\and
I39, Cruz del Sur Observatory, San Justo city, Buenos Aires,  Argentina. 
\and
 Space Sciences, Technologies and Astrophysics Research Institute,
Universit{\'e} de Li{\`e}ge,
 All{\'e}e du 6 Août 17, 4000, Li{\`e}ge, Belgium
\and
 Instituto de Astrof{\'i}sica de Canarias, C/ V{\'i}a Lactea, s/n, 38205 La Laguna, Tenerife, Spain
\and
Gran Telescopio Canarias (GRANTECAN), Cuesta de San Jos{\'e} s/n, E-38712, Bre{\~n}a Baja, La Palma, Spain
\and
  School of Physical Sciences, The Open University, MK7 6AA, UK
\and
Observatoire des Terres Blanches, F-04110 Reillanne, France
\and
I64, SL6 1XE Maidenhead, UK
\and
Anunaki Observatory, Calle de los Llanos, 28410 Manzanares el Real, Spain
\and
  The IEA, University of Reading, Philip Lyle Building, Whiteknights Campus, Reading, RG6
  6BX, UK 
\and
 Rue des Ecoles 2, F-34920 Le Cres, France
\and
Observatoire de Blauvac, 293 chemin de St Guillaume, F-84570 Blauvac, France
\and
University of Ljubljana, Faculty of Mathematics and Physics Astronomical
Observatory, Jadranska 19 1000 Ljubljana, Slovenia
\and
Departamento de F\'isica, Ingenier\'ia de Sistemas y Teor\'ia de la Se\~nal, Universidad de Alicante,
E-03080 Alicante, Spain
\and
Institut de Ci\'encies del Cosmos, Universitat de Barcelona (IEEC-UB),
Martı\'i i
Franqu\'es 1, E-08028 Barcelona, Spain
}

\date{Received: 25 July 2019; accepted: 20 December 2019}

\abstract{
Thanks to the Gaia mission, it will be possible to determine the
  masses of approximately hundreds of large main belt asteroids with very good
precision. 
  We currently have diameter estimates for all of them that can be used to compute their
  volume and hence their density.
  However, some of those diameters are still based on simple thermal models, which
  can occasionally lead to volume uncertainties as high as 20-30\%. 
}
{
  The aim of this paper is to determine the 3D shape models and compute the 
  volumes for 13 main belt asteroids that were selected from those targets for which Gaia will provide the
mass with an accuracy of better than 10\%.
}
{
  We used the genetic Shaping Asteroids with Genetic
Evolution (SAGE) algorithm to fit disk-integrated, dense
  photometric lightcurves and obtain detailed asteroid shape models. 
  These models were scaled by fitting them to available stellar occultation
  and/or thermal infrared observations. 
}
{
  We determine the spin and shape models for 13 main belt asteroids
  using the SAGE algorithm. Occultation fitting enables us to confirm main
  shape features and the spin state, while thermophysical modeling leads to more
  precise diameters as well as estimates of thermal inertia values. 
}
{
  We calculated the volume of our sample of main-belt asteroids for which the Gaia
  satellite will provide precise mass determinations. 
  From our volumes, it will then be possible to more accurately compute the 
  bulk density, which is a fundamental physical property needed to understand 
  the formation and evolution processes of small solar system bodies. 
}


\maketitle
%

\section{Introduction}
Thanks to the development of asteroid modeling methods
\citep{Kaasalainen2002,Viikinkoski15,Bartczak2018},
the last two decades have allowed for a better understanding of the
nature of asteroids. 
Knowledge about their basic physical properties helps us to not only understand
particular objects, but also the asteroid population as a whole. 
Nongravitational  effects  with  a  proven  direct impact  on  asteroid
evolution,  such  as  the 
Yarkovsky-O'Keefe-Radzievskii-Paddack (YORP)  and  Yarkovsky  effects, could  not  be
understood without a precise knowledge about the spin state of asteroids. 
For instance, the sign of the orbital drift induced by the Yarkovsky effect
depends on the target's sense of rotation \citep{Rubincam}. Also, spin clusters
have been observed among members of asteroid families \citep{Slivan} that are
best explained as an outcome of the YORP effect
\citep{Vokrouchlicky,Vokrouchlicky15}.

Precise determinations of the spin and shape of asteroids will be of the utmost
significance for improving the dynamical modeling of the Solar System
and also for our knowledge of the physics of asteroids. 
From a physical point of view, the mass and size of an asteroid yield its
bulk density, which accounts for the amount of matter that makes up the body and
the space occupied by its pores and fractures. For a precise density 
determination, we need a model of the body, which refers to its 3D shape
and spin state. These models are commonly obtained
from relative photometric measurements. In consequence, an estimation of the
body size is required in order to scale the model. The main techniques   used
for   size   determination \citep[for a review, see e.g.,][]{Durech2015}
are stellar occultations, radiometric techniques, or 
adaptive optics (AO) imaging, as well as the in
situ  exploration of 
spacecrafts for a dozen of visited asteroids.

The disk-integrated
lightcurves obtained from different geometries 
(phase and aspect angles) can give us a lot of information about the fundamental
parameters,
such as rotation period, spin axis orientation, and shape.
However, the shape obtained from lightcurve inversion methods is usually
scale-free. Thus, we need to use other methods to express them in kilometers
and calculate the volumes. The determination of asteroid masses is also
not straightforward, but it is expected that Gaia, thanks to its precise
astrometric measurements, will  be able to provide masses for more than a
hundred asteroids. This is possible for objects that undergo gravitational
perturbations during close approaches with other minor bodies \citep{Mouret}.

There are already a few precise sizes that are available based on quality spin and shape models
of Gaia mass targets, 
including convex inversion and 
All-Data Asteroid Modeling (ADAM) shapes \citep[some  
based on Adaptive Optics,][]{vernazza19}. However, there are still many with only
Near Earth Asteroid Thermal Model (NEATM) 
diameters. In this paper, we use 
the SAGE (Shaping Asteroids with Genetic Evolution) algorithm \citep{Bartczak2018} and combine it with 
thermo-physical models (TPM) and/or occultations to determine 
the shape, spin, and absolute scale of a list of Gaia targets in order to
calculate their densities. 
As a result, here, we present the spin solutions and 3D shape models of 13 large main belts
asteroids for which they are
expected to have mass measurements from the
Gaia mission with a precision of better than 10\%.
For some objects, we compare our results with already existing models to
test the reliability of our methods.
Thanks to 
the increased photometric datasets produced by our project,
 previously existing solutions
have been improved for the asteroids that were selected, and for two targets for which we determine the
physical properties for the first time.
We provide the scale and volume for all the bodies that are studied with realistic error
bars. These volumes combined with the masses from Gaia astrometry 
will enable precise bulk density determinations and further mineralogical
studies. 
The selected targets are mostly asteroids with diameters larger
than 100 km, which are considered to be remnants of planetesimals
\citep{morby}. 
These large asteroids are assumed to only have small macroporosity, thus
their bulk densities can be used for comparison purposes with spectra.

The paper is organized as follows. In Section 2 we present our observing
campaign,  give a brief description of the spin and shape modeling technique,
including the quality assessment of the solution, and describe the fitting to
the occultation chords and the thermophysical modeling. In Section 3
we show the results of our study of 13 main belt asteroids, and in Section 4
we summarize our findings.
Appendix A presents the results of TPM modeling, while Appendix B contains
fitting the SAGE shape models to stellar occultations.

\section{Methodology}
\subsection{Observing campaign}

\begin{figure}
   \centering
   \includegraphics[width=8cm]{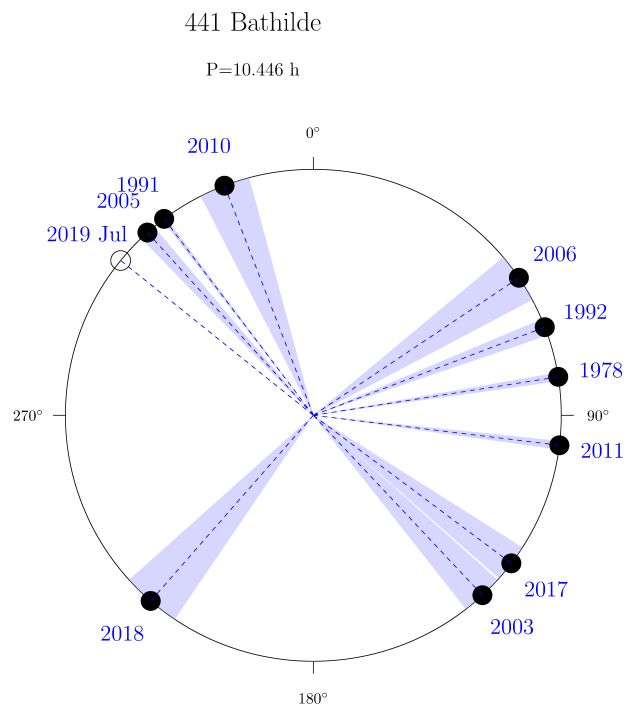}
   \caption{
     Observer-centered ecliptic longitude of asteroid (441) Bathilde 
     at apparitions with well covered lightcurves.
   }
   \label{bathilde}
  \end{figure}

In order to construct precise spin and shape models for asteroids, we
used dense photometric disk-integrated observations.
Reliable asteroid models require lightcurves from a few apparitions, that are
well distributed along the ecliptic longitude.
The available photometric datasets for selected Gaia mass targets are
 complemented by an observing campaign that provided data from
unique geometries, which improved the existing models by probing previously
unseen
parts of the surface. Using the Super-WASP (Wide Angle Search for
Planets) asteroid archive \citep{griece}
was also very helpful, as it provided data from unique observing geometries. 
Moreover, in many cases new data led to updates of sidereal period values. The coordination of observations
was also very useful for long period objects, for which the whole rotation
could not be covered from one place during one night. 
We gathered our new data during the observing campaign in the framework of the
H2020 project called Small Bodies Near And Far (SBNAF, \citealt{muller18}). The main
observing stations were located 
in La Sagra (IAA CSIC, Spain), Piszk\'estet\H{o} (Hungary), and Borowiec
(Poland), and the observing campaign was additionally supported by the GaiaGOSA web service dedicated to
amateur observers \citep{Santanaros}. For some objects, our data were
complemented
by data from the K2 mission of the Kepler space telescope \citep{szabo}
and the TRAPPIST North and
South telescopes \citep{jehin}. Gathered
photometric data went through careful analysis in order to remove any problematic
issues, such as star passages, color extinction, bad pixels,
or other instrumental effects. 
In order to exclude any unrealistic artefacts,  we decided  not to 
take into account data that were too noisy or
suspect data. The most realistic spin and shape models can be
reconstructed when the observations are spread evenly along the orbit; 
this allows one to observe all illuminated parts of the asteroid's surface.
Therefore, in this study, we particularly concentrated on the observations of objects
 for which we could cover our targets in previously unseen geometries,
which is similar to what was done for 441 Bathilde, for which data from 2018 provided
a lot of valuable information.
Fig. \ref{bathilde} shows an example of the ecliptic longitude coverage for
the asteroid 441 Bathilde. 

\subsection{Spin and shape modeling}

We used the genetic algorithm, SAGE
to calculate asteroid models \citep{Bartczak2018}. SAGE allowed us to reproduce spin and nonconvex asteroid 
shapes based exclusively on photometric lightcurves. Here, we additionally introduce the recently
 developed quality assessment system \citep{Bartczak2019}, which gives information about the reliability of the 
obtained models.  
 The uncertainty of the SAGE spin and shape solutions was calculated by the multiple cloning of the final
 models and by randomly modifying the size and radial extent of
their shape features. These clones were checked for their ability to simultaneously reproduce all the
lightcurves within their uncertainties. By lightcurve uncertainty, we
are referring to the uncertainty of each point. For the lightcurves with no uncertainty
information, we adopted $0.01$ mag. This way, the scale-free dimensions with the most extreme,
but still possible shape feature modifications, were calculated and then translated to
diameters in kilometers by fitting occultation chords. Some of the calculated
models can be compared to the solutions obtained from other methods, which often use
adaptive optics images, such as KOALA (Knitted Occultation, Adaptive-optics, and Lightcurve
Analysis, \citealt{koala})
 and ADAM \citep{Viikinkoski15}.
Such models are stored in the DAMIT Database of Asteroid Models from Inversion
Techniques (DAMIT) database
(http://astro.troja.mff.cuni.cz/projects/asteroids3D,
  \citealt{Durech2010}). 
Here, we show the nonconvex shapes that were determined with the SAGE method. We
have only used the photometric data since they
are the easiest to use
and widely available data for asteroids.
It should be noted, however, that some shape features,
 such as the depth of large craters or the height of hills, are prone to the largest uncertainty, as was shown by
\citet{Bartczak2019}.
 It is also worth mentioning here that such a comparison of two methods is
valuable as a test for the reliability of two independent methods
 and for the correctness
of the existing solutions with the support of a wider set of photometric data.      
For a few  targets from our sample, we provide
 more realistic, smoother shape solutions, which improve on the previously existing
 angular shape representations based on limited or sparse datasets.
For two targets, (145) Adeona and (308) Polyxo, the spin and shape solutions were obtained here for
the first time.

\subsection{Scaling the models by stellar occultations}
The calculated spin and shape models are usually scale-free. By using
two independent methods, the stellar occultation fitting and thermophysical modeling, we were able to provide an absolute scale for our
shape models. 
The great advantage of the occultation technique is that the dimensions of the
asteroid shadow seen on Earth can be treated as a real dimension of the object.
Thus, if enough chords are observed, we can express the size of the object
in kilometers. Moreover, with the use of multichord events, the major shape
features can be recovered from the contours. 
To scale our shape models, we used the occultation timings stored in the Planetary Data System
(PDS)
database \citep{Dunham2016}.  Only the records with at least three internally
consistent chords were taken into account. The fitting of shape contours to
events with fewer chords is burdened with uncertainties that are too large. 

Three chords also do not guarantee precise size determinations because
of substantial uncertainties in the timing of some events or the unfortunate
spatial grouping of chords.
We used the procedure implemented in \citet{durech11} to compare our
shape models with available occultation chords. 
We fit the three parameters $\xi$, $\eta$ \citep[the fundamental plane here is defined the same
as in][]{durech11}, and c, which was scaled in order to determine
the size.
 The shape models' orientations
were overlayed on the measured occultation chords and scaled to minimize
 $\chi^2$ value.
The
difference with respect to the procedure described in \citet{durech11} is that
we fit the projection silhouette to each occultation event separately,
and we took the confidence level of the nominal
solution into account as it was described in \citet{Bartczak2019}. We also did not optimize
offsets of the occultations. 
 Shape models fitting to stellar occultations with accompanying errors are presented
in Figs. \ref{junoocc}-\ref{bathildeocc}.
The final uncertainty in the volume comes from the effects of shape and
occultation timing uncertainties 
and it is usually larger than in TPM since thermal data are very sensitive 
to the size of the body and various shape features play a lesser role there.
On the other hand, precise knowledge of the sidereal period and spin axis
position is of vital importance for the proper phasing of the shape models in both
TPM and in occultation fitting. So, if a good fit is obtained by both methods, 
we consider it to be a robust confirmation for the spin parameters.  

\begin{figure*}[h]   
\centering
\includegraphics[width=0.892\textwidth]{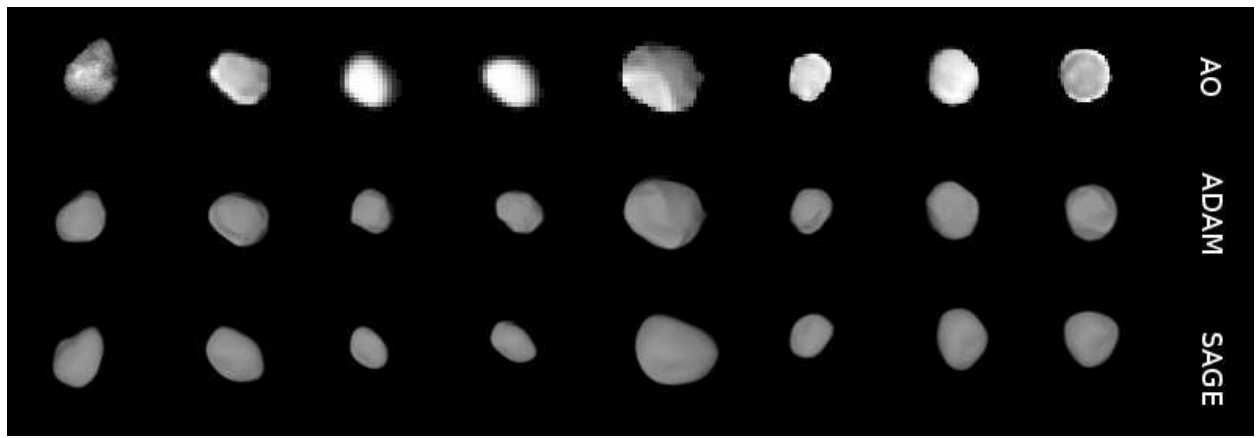}
\caption{
  Adaptive optics images of asteroid (3) Juno (top), the ADAM model sky
projection by \citet{Viikinkoski15} (middle), and the SAGE model (bottom) presented for
the same epochs.}
\label{Juno_shapes}
\end{figure*}

\subsection{Thermophysical modeling (TPM)}\label{sec:TPM}

The TPM implementation we used is based on \citet{delbo02} and \citet{ali14}.
We already described our approach in \citet{Marciniak2018} and
\citet{Marciniak2019},
which give details about the modeling of each target. So in this section, we simply
provide a brief summary of the technique and approximations we make.
In Appendix A, we include all the plots that are relevant to the modeling of each target
and we provide some additional comments. 

The TPM takes the shape model as input, and its main goal is to model the
temperature on any given surface element (facet) at each epoch at which we have
thermal 
IR (infrared) observations, so that the observed flux can be modeled. 
To account for heat conduction toward the subsurface, we solved the 1D heat
diffusion equation for each facet and we used the Lagerros approximation
for roughness
\citep[][]{Lagerros96,Lagerros98, Muller98, Muller02}.
We also consider the spectral emissivity to be 0.9 regardless of the wavelength
\citep[see, e.g.,][]{Delbo15}. 
We explored different roughness parametrizations by varying the opening 
angle of hemispherical craters covering 0.6 of the area of the facets
\citep[following][]{Lagerros96}. 
For each target, we estimated the Bond albedo that was
used in the TPM as the
average value that was obtained from the different radiometric diameters available from
AKARI and/or WISE \citep[][]{wright,Usui2011,Ali2018,Mainzer2016}, 
 and all available $H$-$G$, $H$-$G_{12}$, and $H$-$G_1$-$G_2$ values
from the Minor Planet Center (\citet{Osz11}, or \citet{Veres15}).

This approach leaves us with two free parameters, the scale of the shape
(interchangeably called the diameter, $D$), and the thermal inertia ($\Gamma$).
The diameters, which were calculated as volume-equivalent diameters, and other relevant information related to the  TPM analyses of
our targets are provided  in Table \ref{tab:tpm}.
Whenever there are not enough data to provide realistic error bar estimates, we
report the best-fitting diameter so that the models can be scaled and
compared to the scaling given by the occultations. 
On the other hand, if we have multiple good-quality thermal data, with absolute
calibration errors below 10\%, then this typically translates to a size accuracy
of around 5\% as long as the shape is not too extreme and the spin vector
is reasonably well established. 
This general rule certainly works for large main belt asteroids, that is, the Gaia mass targets. 
We do not consider the errors that are introduced by the pole orientation uncertainties
or the shapes (see \citealt{hanus16} and \citealt{Bartczak2019}); therefore, our
TPM error bars are lower estimates of the true error bars.
The previously mentioned general rule or expectation
is based on the fact 
that the flux is proportional to the square of the projected 
area, so fitting a high-quality shape and spin model to fluxes with 
10\% absolute error bars should produce a $\sim5\%$ accurate size. 
This is verified by the large asteroids that were used 
as calibrators  \citep{Muller02, harris,Muller14}.

Nonetheless, we would still argue that generally speaking, scaling 
        3D shapes, which were only determined via indirect means (such as pure LC 
        inversion) by modeling thermal IR data that were only observed close to 
        pole-on, could potentially result in a biased TPM size if the shape 
        has an over- or underestimated z-dimension \citep[e.g.,][]{Bartczak2019}. 
This also happens with at least some radar  models (e.g., \citet{rozitis}).

\section{Results}
The following subsections describe our results for each target, whereas
Tables \ref{results}, \ref{tab2}, and \ref{tab:tpm} provide the pole solutions,
the results from the occultation fitting, and the results from TPM, respectively. 
The fit of the models to the observed lightcurves can be found for each object on the ISAM\footnote{http://isam.astro.amu.edu.pl}
(Interactive Service for Asteroid Models)  web-service
\citep{marciniak2012}. On ISAM, we also show the fit of available occultation
records for all objects studied in this paper.
For comparison purposes, a few examples are given for SAGE shape models and previously
existing solutions, which are shown in Figs. \ref{Juno_shapes}-\ref{Myrrha_shapes}, as well as for previous period determinations and pole solutions, which
are given in Table \ref{tab4}.
For targets without previously available spin and shape models, we determined  the model
based on the simple lightcurve inversion method \citep[see][]{Kaasalainen2002}, such as in \citet{Marciniak2018}, and we compared the results with
those from the SAGE method.

\subsection{(3) Juno}
We used observations from 11 apparitions to model Juno's shape.
All lightcurves display amplitude variations from 0.12 to 0.22 mag., which
indicates the body has a small elongation.
Juno was already investigated with the ADAM method by
\citet{Viikinkoski15}, which was
based on 
ALMA (Atacama Large Millimeter Array) and adaptive optics data in addition to lightcurves.
The rotation period and spin axis position of both models, ADAM and SAGE, are in good agreement. However, the shapes look different from some perspectives.
 The shape contours of the SAGE model are smoother and the
main features, such as polar craters, were reproduced in both methods.
We compared our SAGE model with AO data and the results from ADAM
modeling by \citet{Viikinkoski15} in Fig. \ref{Juno_shapes}. The fit is good,
but not perfect.

A rich dataset of 112 thermal infrared measurements is available for (3) Juno, 
including unpublished Herschel PACS data \citep{Muller05}.
The complete PACS catalog of small-body data will be added to the SBNAF
infrared database once additional 
SBNAF articles are published. 
For instance, the full TPM analysis of Juno will be included in an accompanying
paper that features the rest of the PACS main-belt targets 
(Al\'{i}-Lagoa et al., in preparation). 
Here, we include Juno in order to compare the scales we obtained from TPM and
occultations. 

TPM leads to a size of $254\pm4$ km (see Tables \ref{tab2} and \ref{tab:tpm}),
which is in agreement with the ADAM solution (248 km) within the error bars.
The stellar occultations from the years 1979, 2000, and 2014 also fit well
(see Fig. \ref{junoocc} for details). The 1979 event, which had the most dense 
coverage (15 chords), leads to
 a diameter of $260^{+13}_{-12}$ km. 

\subsection{(14) Irene}
\begin{SCfigure*}[0.3][bhp]
\centering
\includegraphics[width=0.3\textwidth]{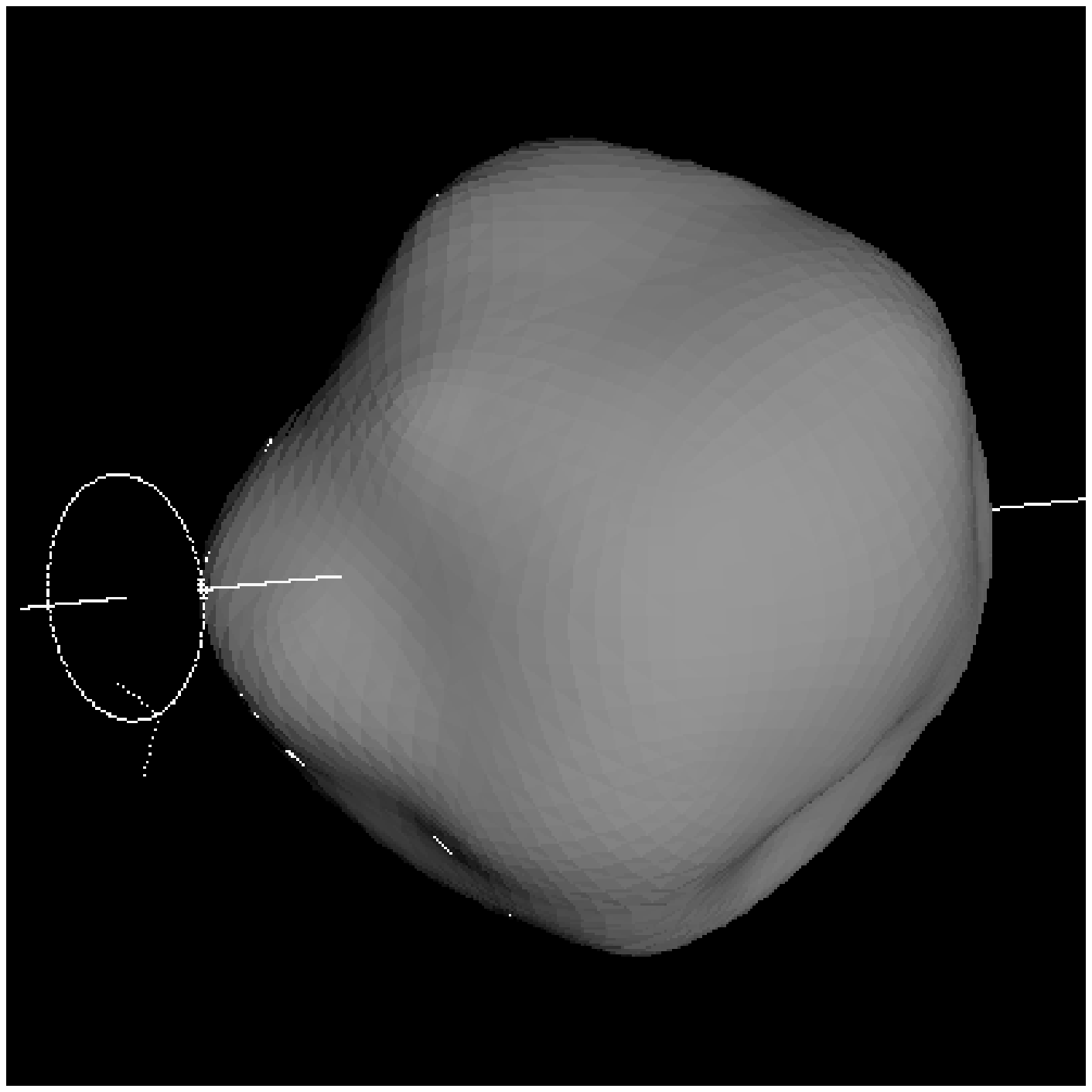}
\includegraphics[width=0.3\textwidth]{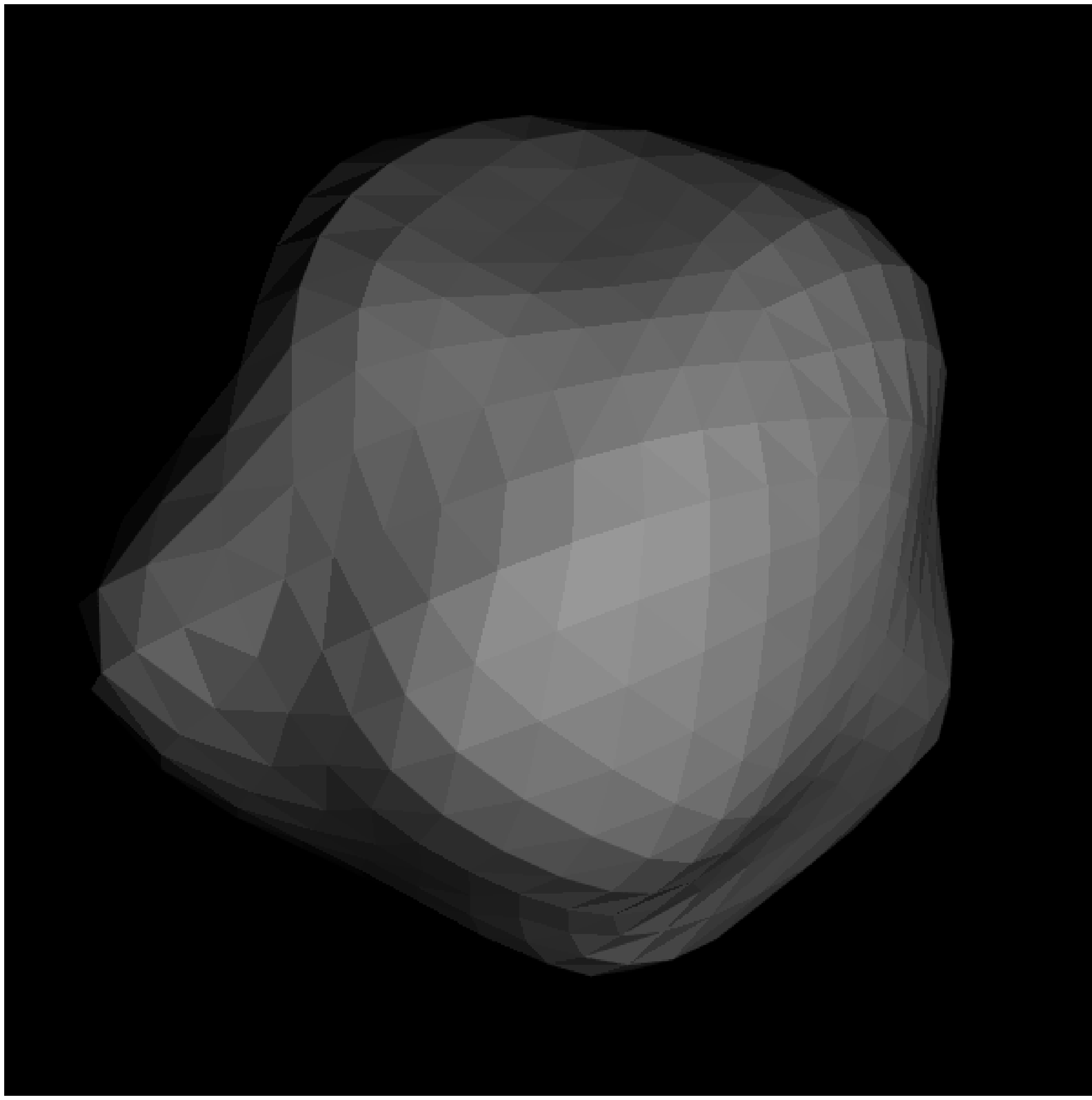}
\caption{
  Sky projections for the same epoch of SAGE (left) and ADAM (right)
  shape models of asteroid (14) Irene. Both shapes are in very good agreement.
}
\label{Irene_shapes}
\end{SCfigure*}

For (14) Irene, we gathered the lightcurves from 14
apparitions, but from very limited viewing geometries.
 The lightcurve shapes were
very asymmetric, changing character from bimodal to monomodal in some
apparitions, which indicates large aspect angle changes caused by low spin axis inclination to the orbital plane of the
body.
 The
amplitudes 
varied from 0.03 to 0.16 mag. The obtained SAGE model fits very well to the
lightcurves; the agreement is close to the noise level. The spin solution is
presented in Table \ref{results}.
The SAGE model is in very good agreement with the ADAM model, which displays the
same major shape features (see Fig. \ref{Irene_shapes}). 
This agreement can be checked for all available models by generating their sky 
projections at the same moment on the ISAM
and DAMIT\footnote{http://astro.troja.mff.cuni.cz/projects/asteroids3D}
webpages.

The only three existing occultation chords seem to point to  the slightly preferred SAGE solution from two possible mirror
solutions (Fig. \ref{ireneocc}),
 and it led to a size of
$145^{+12}_{-12}$ km for the pole 1 solution. The TPM fit resulted in a compatible size of 155 km,  which is in good
agreement within the error bars.
We note, however, that the six thermal IR data available are not substantial enough to give
realistic TPM error bars (the data are fit with an artificially low minimum
that was reduced to $\chi^2 \sim 0.1$), but nonetheless both of our size determinations
here also agree with the size of the ADAM shape model based on the following adaptive optics
imaging: 153 km $\pm$ 6km \citep{Viikinkoski2017}. 

\subsection{(20) Massalia}
Data from 13 apparitions were at our disposal to model (20) Massalia, although
some of them were grouped close together in ecliptic longitudes.
Massalia displayed regular, bimodal lightcurve shapes with amplitudes from
0.17 to 0.27 mag. New data gathered within the SBNAF and GaiaGOSA projects 
significantly improved the preliminary convex solution that exists in
DAMIT 
\citep{Kaasalainen2002}, which has a much lower pole inclination and a sidereal
period of 0.002 hours shorter. 
If we consider the long span (60 years) of available photometric data and the
shortness of the rotation period, such a mismatch causes a large shift in
rotational phase after a large number of rotations. 

The two SAGE mirror solutions have a smooth shape with a top shape appearance.
Their fit to the occultation record from 2012 led to two differing size
solutions of $106^{+6}_{-3}$ and $113^{+6}_{-10}$ km (Fig. \ref{massaliaocc}); both are smaller and outside the combined error bars
of the $145\pm2$ km solution that was obtained from the TPM. The full TPM details and the PACS data 
will be presented in Ali-Lagoa et al. (in preparation).
The SAGE shapes fit the thermal data much better than the sphere, which we
consider as an indication that the model adequately captures the relevant
shape details.
We note that (20) Massalia is one of the objects for which the stellar occultation data are rather
poor. This provides rough size determinations and underestimated uncertainties.

\subsection{(64) Angelina}
The lightcurves of (64) Angelina display asymmetric and variable
behavior, with amplitudes ranging from 0.04 mag to
0.42 mag, which indicates a spin axis obliquity around 90 degrees.
Data from ten apparitions were used to calculate the SAGE model. 
The synthetic lightcurves that were generated from the shape are in good agreement with the
observed ones. 
Although the low value of the pole's latitude of $12 \deg$ is consistent with the
previous solution by \citet{durech11} (see Table \ref{tab4} for reference), the difference of 0.0015 hours in the
period is substantial.
We favor our solution given our updated, richer dataset 
since \citet{durech11} only had dense lightcurves from three apparitions that were
complemented by sparse data with uncertainties of 0.1 - 0.2 mag 
(i.e., the level of lightcurve amplitude of this target). 
Also, the level of the occultation fit (Fig.\ref{angelinaocc}) and the TPM support our model.  
The  thermal data were well reproduced with sizes that are slightly larger but consistent
with the ones from the occultation fitting ($54$ versus $50$ km, see Tables \ref{tab2}
and \ref{tab:tpm}), and they slightly favor the same pole solution.

\subsection{(68) Leto}
\begin{SCfigure*}[0.3][bhp]
\centering
\includegraphics[width=0.29\textwidth]{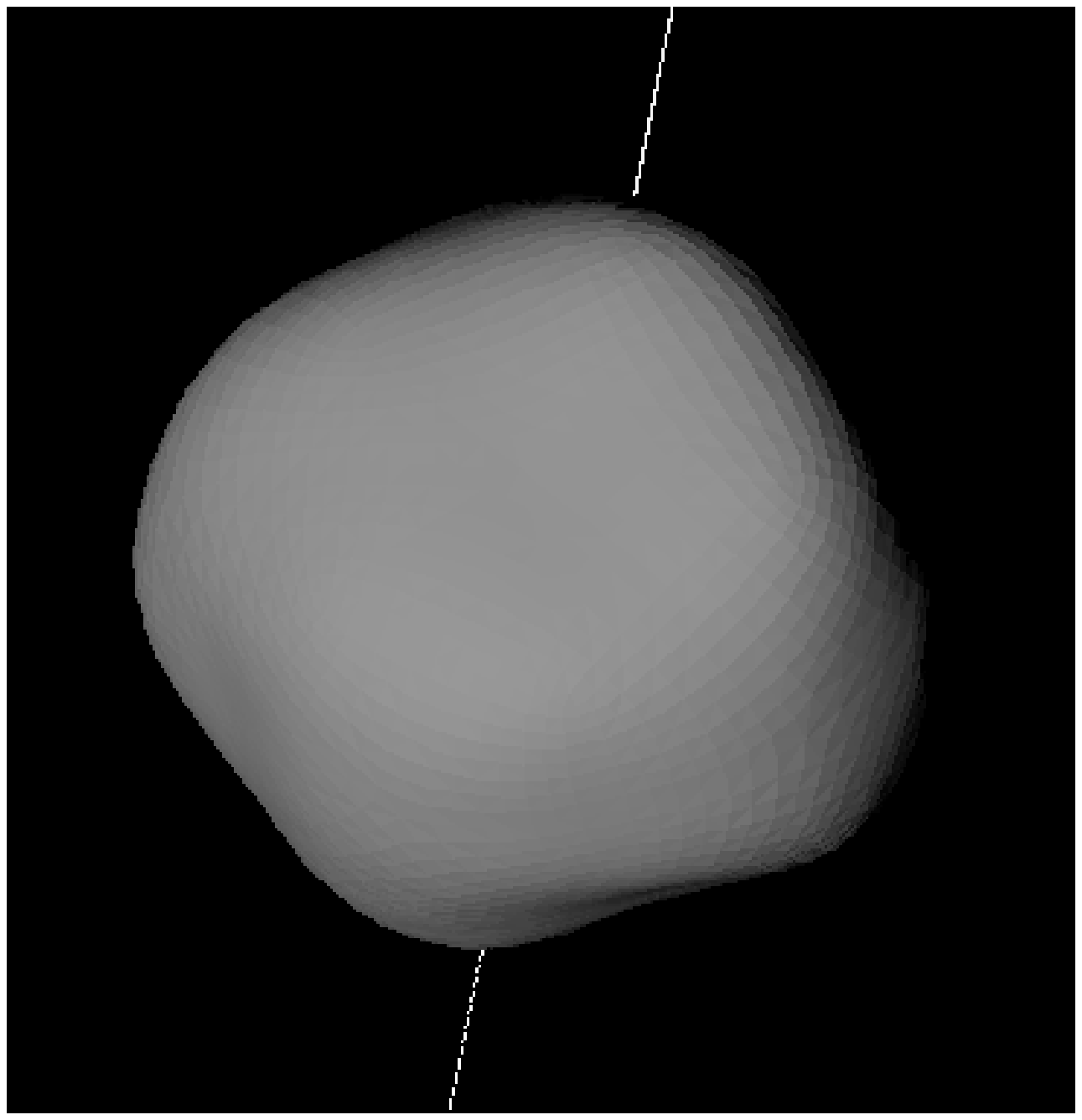}
\includegraphics[width=0.3\textwidth]{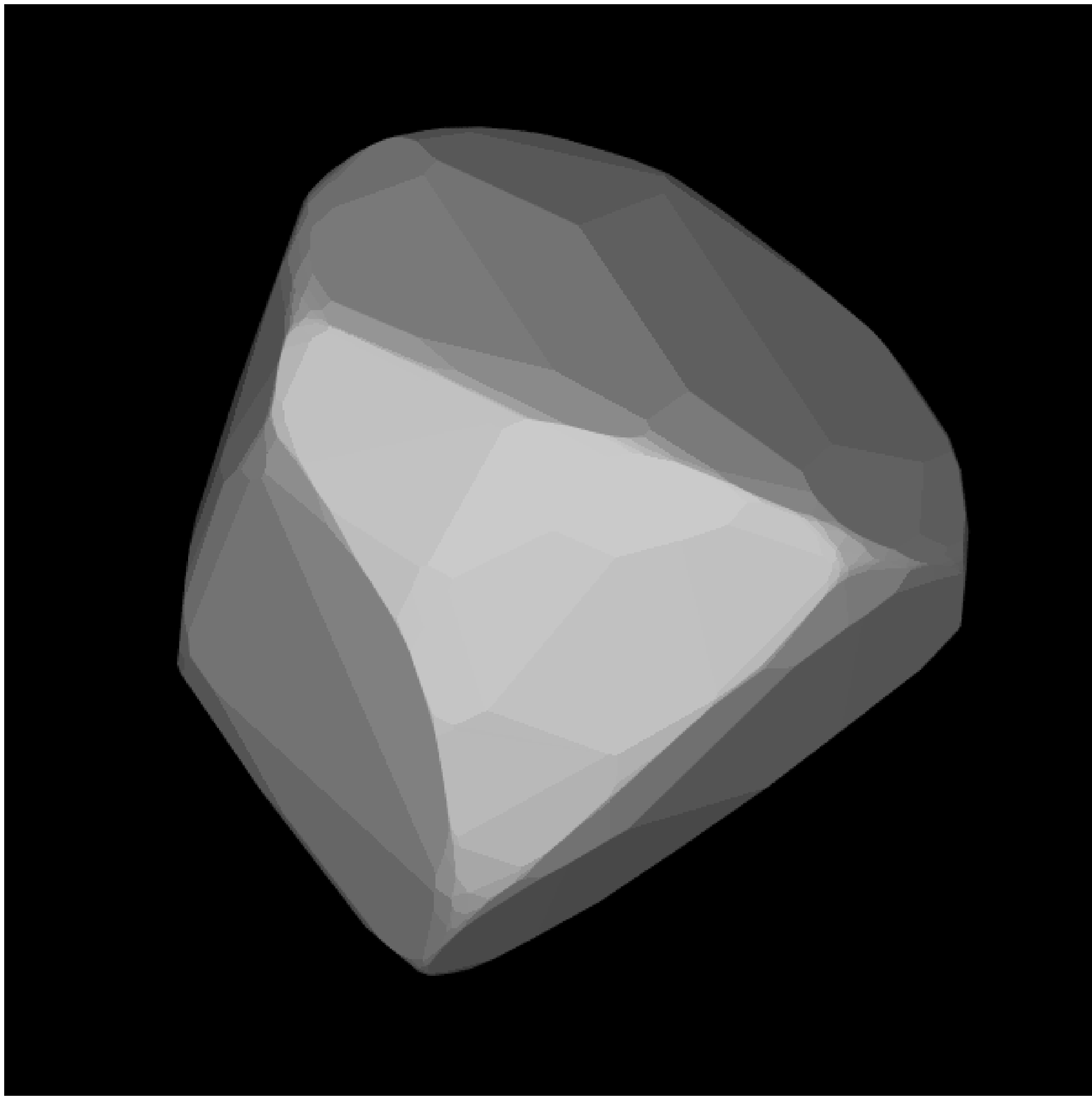}
\caption{Sky projections for the same epoch of the SAGE (left) and convex inversion (right) shape models of asteroid (68) Leto. SAGE provided a largely different and much smoother shape solution.}
\label{Leto_shapes}
\end{SCfigure*}

For Leto, data from six different apparitions consisted of somewhat asymmetric
lightcurves with unequally spaced minima. 
Amplitudes ranged from 0.10 to 0.28 mag.
The angular convex shape model published previously by \citet{hanus13},
which was mainly based on sparse data, is compared here with a much smoother SAGE
model. 
Their on-sky projections on the same epoch can be seen in
Fig. \ref{Leto_shapes}.   
The TPM analysis did not favor any of the poles.
There was only one three-chords occultation, which the models did not fit perfectly,
although pole 2 was fit better this time (see Fig. \ref{letoocc}). 
Also, the occultation size of the pole 1 solution is 30 km larger than the
radiometric one ($152^{+21}_{18}$ versus $121$ km), with similarly large error bars, whereas
the $133^{+8}_{-8}$ km size of the pole 2 solution is more consistent with the TPM and it has smaller error
bars (see Table \ref{tab2} and \ref{tab:tpm}). 

\subsection{(89) Julia}
\begin{SCfigure*}[0.3][bhp]
\centering
\includegraphics[width=0.3\textwidth]{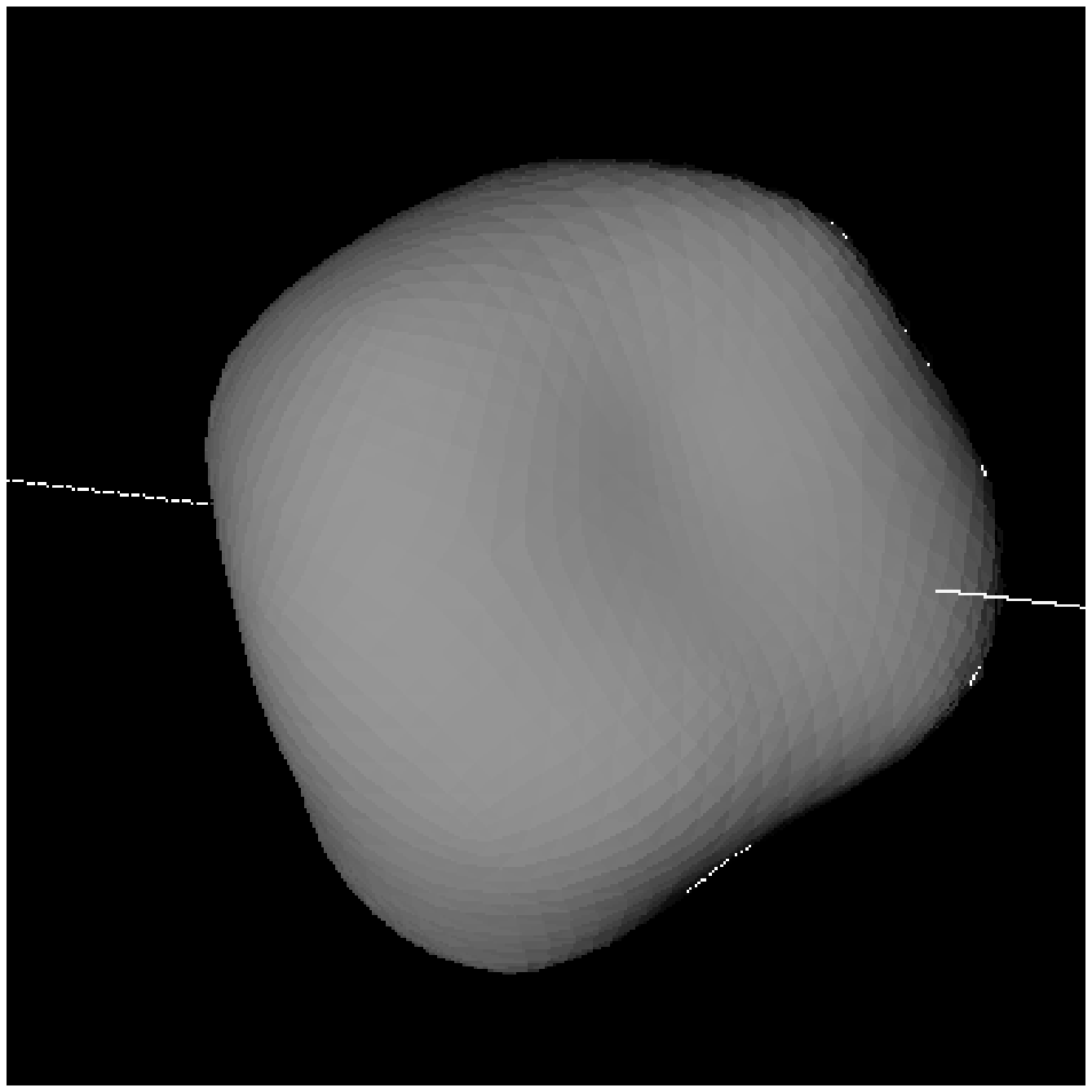}
\includegraphics[width=0.3\textwidth]{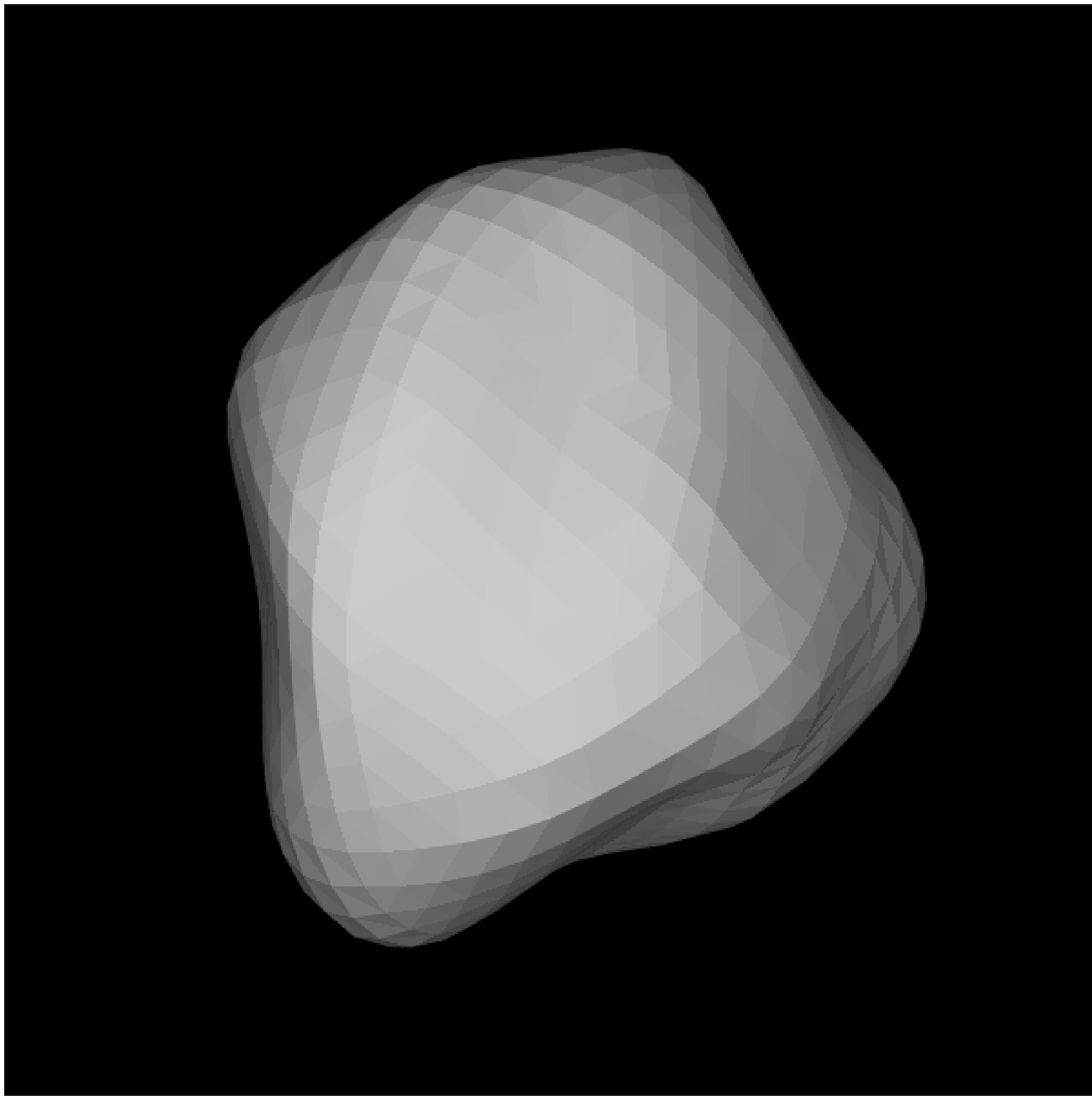}
\caption{
  Sky projections for the same epoch of SAGE (left) and ADAM (right) shape
  models of asteroid (89) Julia. A similar crater on the southern pole was reproduced by both methods.
}
\label{Julia_shapes}
\end{SCfigure*}

This target was shared with the VLT large program 199.C-0074
(PI: Pierre Vernazza), which obtained a rich set of well-resolved adaptive
optics images using VLT/SPHERE instrument.
\citet{vernazza18} produced a spin and shape model of (89) Julia using the ADAM algorithm on
lightcurves and AO images, which enabled them to reproduce major nonconvex
shape features. They identified a large impact crater that is possibly the
source region of the asteroids of the Julia collisional family. 
The SAGE model, which is based solely on disk-integrated photometry, also reproduced the
biggest crater and some of the hills present in the ADAM model
(Fig. \ref{Julia_shapes}).
Spin parameters are in very good agreement. 
Interestingly, lightcurve data from only four apparitions were used for both
models.
However, one of them spanned five months, covering a large range of
phase angles that highlighted the surface features due to various levels of
shadowing. 
Both models fit them well, but the SAGE model does slightly worse.
In the occultation fitting of two multichord events from the years 2005 and
2006, some of the SAGE shape features seem too small and others seem too large, but
overall we obtain a size (138 km) that is almost identical to the ADAM model size
($139{\pm }3$ km). 
The TPM requires a larger size ($150\pm10$ km) for this model, but it is still
consistent within the error bars.

\subsection{(114) Kassandra}
The lightcurves of Kassandra from nine apparitions (although only six have distinct
geometries) showed sharp minima of uneven depths and had amplitudes from 0.15
to 0.25 mag.
The SAGE  shape model looks quite irregular, with a deep polar crater.
It does not resemble the convex model by
\citet{durech18b}, which is provided with a warning of its wrong inertia tensor.
Nevertheless, the spin parameters of both solutions roughly agree. The SAGE model fits the
lightcurves well, except for three cases involving the same ones that the convex model
also failed to fit.
This might indicate that they are burdened with some instrumental or other
systematic errors.
Unfortunately, no well-covered stellar occultations are
available for Kassandra, so the only size determination could be done here by
TPM (see Table \ref{tab:tpm}). Despite the substantial irregularity of
the SAGE shape model, the spherical shape gives a similarly good fit to the thermal
data.

\subsection{(145) Adeona}
Despite the fact that the available set of lightcurves came
from nine apparitions, their unfortunate grouping resulted in only five distinct
viewing aspects of this body. The small amplitudes (0.04 -
0.15 mag) displayed by this target were an additional hindering
factor. Therefore, there was initially a controversy as to whether its
period is close to 8.3 or 15 hours. It was resolved by good quality data
obtained by \citet{pilcher10}, which is in favor of the latter.
SAGE model fit most of the lightcurves well, but it had problems with some
where visible deviations are apparent.  
This is the first model of this target, so there is not a previous 
model with which to compare it.
The SAGE model looks almost spherical without notable shape features, so,
as expected, the spherical shape provided a similarly good fit to the thermal
data.
The model fits the only available stellar occultation very well, which has the
volume equivalent diameter of $145^{+4.3}_{-2.7}$ km.

\subsection{(297) Caecilia}
There were data from nine apparitions available for Caecilia, which were 
well spread in ecliptic longitude. The lightcurves displayed mostly regular,
bimodal character of 0.15 - 0.28 mag amplitudes. The previous model by
\citet{hanus13} was created on a much more limited data set, 
with dense lightcurves covering only 1/3 of the orbit,  which was 
supplemented by sparse data. So, as expected, that shape model is rather crude 
compared to the SAGE model.  
Nonetheless, the period and pole orientation is in good agreement between the two models, 
and there were similar problems with both shapes when
fitting some of the lightcurves.

No stellar occultations by Caecilia are available with a sufficient
number of chords, so the SAGE model was only scaled here by TPM 
(see Table \ref{tab:tpm}). However, the diameter provided here is merely the
best-fitting value since the number of thermal IR data is too low to provide a
realistic uncertainty estimate.

\subsection{(308) Polyxo}
The available lightcurve data set has been very limited for Polyxo, 
so no model could have been previously constructed.\ However, thanks to an extensive
SBNAF observing campaign and the observations collected through GaiaGOSA, we
now have data from six apparitions, covering five different aspects. The lightcurves
were very irregular and had a small amplitude (0.08-0.22 mag), often displaying
three maxima per period.
To check the reliability of our solution, we determined the model
based on the simple lightcurve inversion method. Then, we compared the results with those from the SAGE method.
 All the parameters are in agreement within the error bars between the convex and
SAGE models. 
Still, the SAGE shape model looks rather smooth, with only small irregularities, and it fits the visible lightcurves reasonably well. 
There were three multichord occultations for Polyxo in PDS obtained in 
2000, 2004, and 2010. Both pole solutions fit them at a good level (see Fig.
\ref{polyxoocc} for details) and produced
mutually consistent diameters derived from each of the events separately
($125-133$ km, see Table \ref{tab2}). The TPM diameter ($139$ km) is slightly
larger though.\ However, in this case, there are not enough thermal data to provide a 
realistic estimate of the error bars. 

\subsection{(381) Myrrha}
\begin{SCfigure*}[0.3][bhp]
\centering
\includegraphics[width=0.3\textwidth]{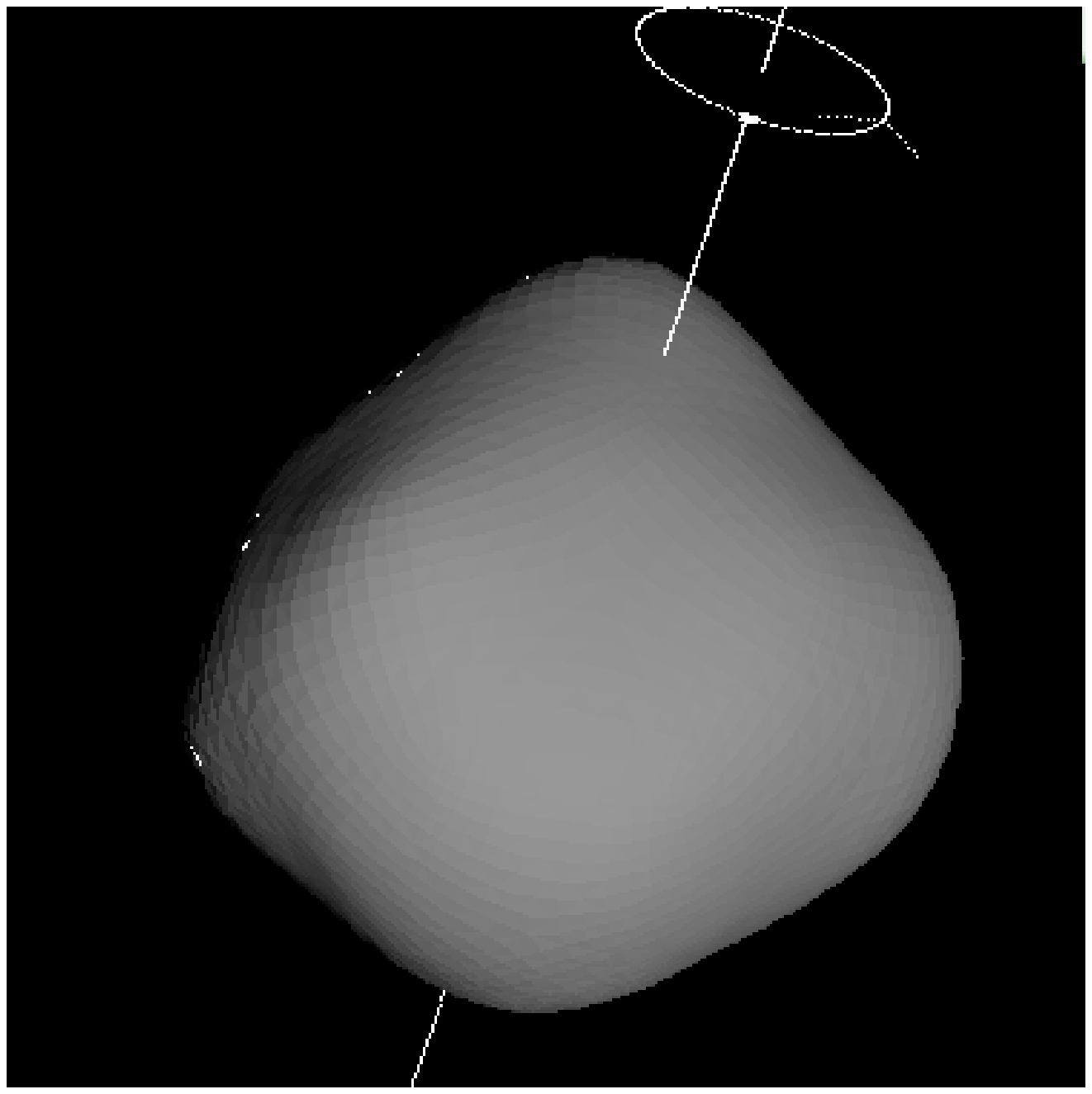}
\includegraphics[width=0.3\textwidth]{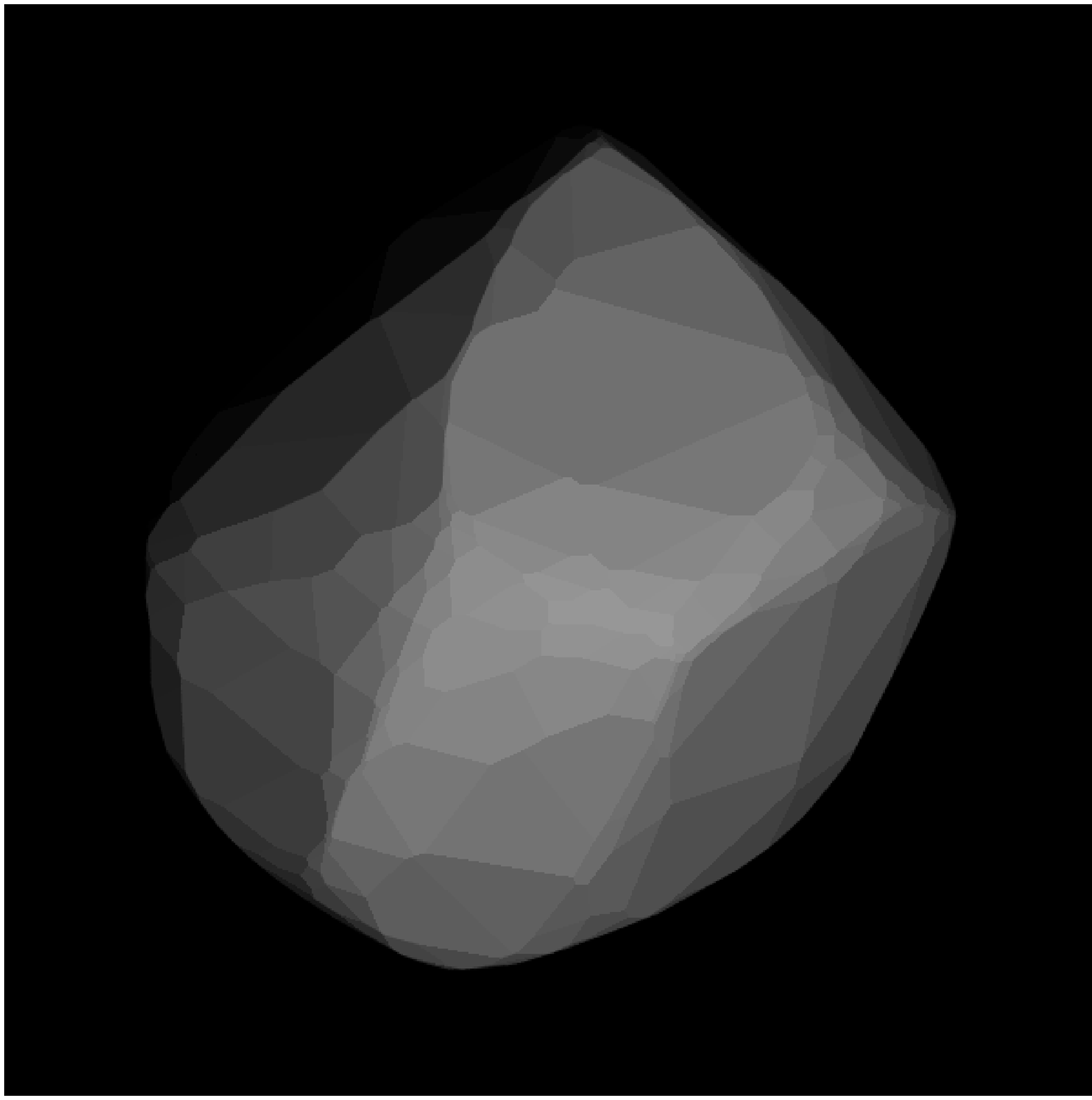}
\caption{
  Sky projections for the same epoch of SAGE (left) and convex inversion
  (right) shape models of asteroid (381) Myrrha. SAGE model is similar to the
  one from convex inversion, but it is less angular. 
}
\label{Myrrha_shapes}
\end{SCfigure*}

In the case of Myrrha, there were data from seven apparitions, but only five
different viewing aspects. The lightcurves displayed a regular shape with a large
amplitude from 0.3 to 0.36 mag. Thanks to the observing campaign that was conducted in
the framework of the SBNAF project and the GaiaGOSA observers, we were able to
determine the shape and spin state. Without the new data, the previous set of
viewing geometries would have been limited to only 1/3 of the Myrrha orbit, and
the earlier model by \citet{hanus16}
was constructed on dense lightcurves
supplemented with sparse data. As a consequence, the previous model
 looks somewhat angular
(cf. both shapes in Fig. \ref{Myrrha_shapes}). 
Due to a very high inclination of the pole to
the ecliptic plane (high value of $|\beta|$), two potential mirror pole solutions 
were very close to each other.\ As a result, an unambiguous solution for the pole position was found.
A very densely covered stellar occultation was available, although 
some of the 25 chords are mutually inconsistent and burdened with large
uncertainties (see Fig. \ref{myrrhaocc}). In the thermal IR, the SAGE model of Myrrha fits the rich
data set better than the sphere with the same pole, giving 
a larger diameter. The obtained diameter
has a small estimated error bar ($131\pm 4$ km) and it is in close agreement with
the size derived from the occultation fitting of
 timing chords ($135^{+45}_{-13}$ km).

\subsection{(441) Bathilde}
Seven different viewing geometries from ten apparitions were available for
Bathilde. The amplitude of the lightcurves varied from 0.08 to 0.22 mag. 
Similarly, as in a few previously described cases, a previous model of this target 
based on sparse and dense data was available \citep{hanus13}. The new SAGE shape
fit additional data and  it has a smoother shape.

Shapes for both pole solutions fit the only available occultation well, and the
resulting size (around $ 76$ km) is in agreement with the size from TPM ($72\pm2$ km).
Interestingly, the second solution for the pole seems to be rejected by TPM,
and the favored one fits thermal data much better than in the corresponding
sphere. The resulting diameter is larger than the one obtained from AKARI, SIMPS, and WISE (see
Tables \ref{tab2}, \ref{tab:tpm} and \ref{tab4} for comparison).

\subsection{(721) Tabora}
Together with new observations that were gathered by the SBNAF observing campaign, we have data
from five apparitions for Tabora.
Amplitudes ranged from 0.19 to 0.50 mag, and the lightcurves were sometimes
strongly asymmetric, with extrema at different levels. 
A model of Tabora has been published recently and it is based on joining sparse data 
in the visible with WISE thermal data (bands W3 and W4, \citealt{durech18a}), but
it does not have an assigned scale.
The resulting shape model is somewhat angular, but it is in agreement with
the SAGE model with respect to spin parameters.
Stellar occultations are also lacking for Tabora, and the TPM only gave
a marginally acceptable fit ($\chi^2=1.4$ for pole 1) to the thermal data, which is
nonetheless much better than the sphere. 
Thus, the diameter error bar, in this case, is not optimal ($\sim6\%$) and
additional IR data and/or occultations would be required to provide a better 
constrained volume.

\section{Conclusions}

\begin{figure}
   \centering
   \includegraphics[width=9cm]{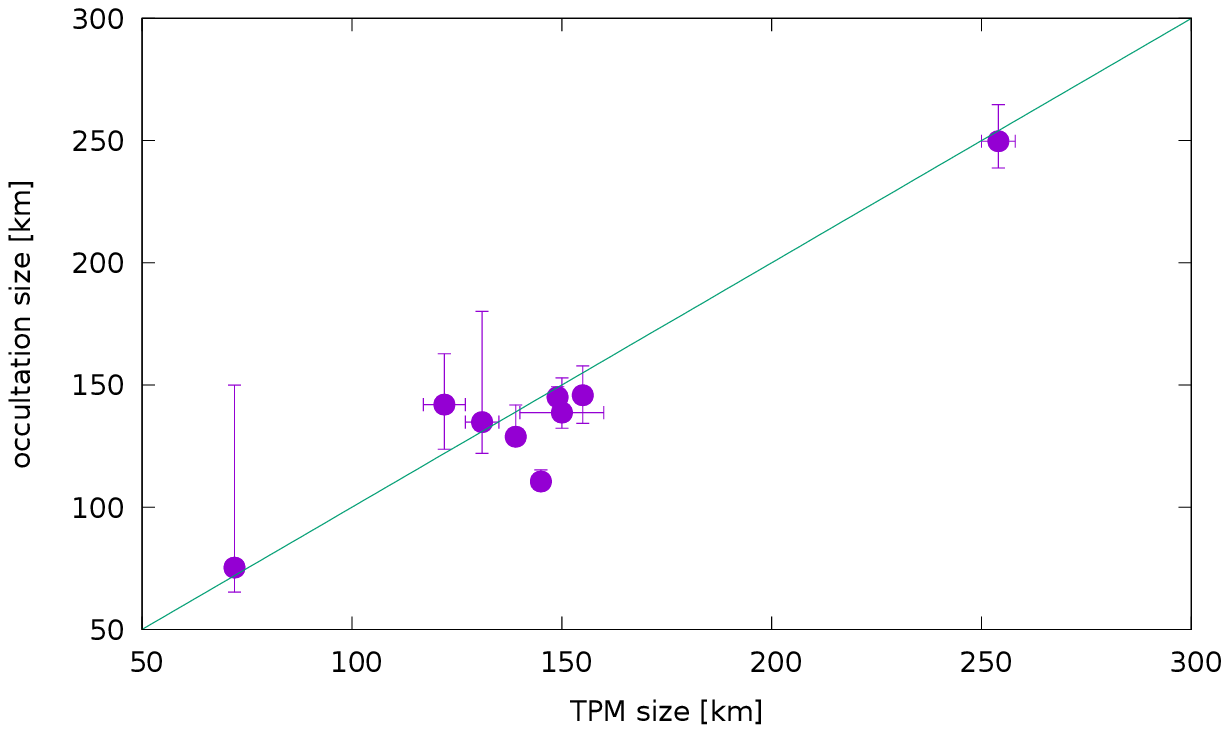}
   \caption{Set of average occultation diameters vs. diameters from TPM. The straight line is y=x.
   }
   \label{summary}
  \end{figure}

Here, we derived spin and shape models of 13 asteroids that were selected from Gaia mass
targets, using only photometric lightcurves. 
It is generally possible to recover major        
shape features of main belt asteroids, but other techniques, such as direct images or adaptive optics,
should be used to confirm the main features.
We scaled our shape models by using stellar occultation records and TPM. 
The results obtained from both techniques are usually in good agreement,
what can be seen in Fig. \ref{summary}.
In many ways, the stellar occultation fitting and thermophysical modeling are complementary to each other.
 In most cases, occultation chords match the silhouette within the error bars 
and rough diameters are provided.\ Also, thermophysical modeling resulted in more precise
size determinations, 
thus additionally constraining
the following thermal parameters: thermal inertia and surface roughness (see
Table \ref{tab:tpm}).
The diameters based on occultation fitting of complex shape models, inaccurate as they may seem here when compared to 
those from TPM, still reflect the dimensions of real bodies better than the commonly used elliptical approximation of the 
shape projection.
The biggest advantage of scaling 3D shape models by occultations is that this procedure provides volumes of these bodies,
unlike the fitting of 2D elliptical shape approximations, which only provides the lower limit for the size
of the projection ellipse.

Resulting volumes, especially those with relatively small uncertainty, are going to be a valuable input 
for the density determinations of these targets once the mass values from the Gaia
astrometry become available.
In the cases where only convex solutions were previously available,  
nonconvex solutions created here will lead to more precise volumes, and consequently better constrained densities.
In a few cases, our  solutions are the first in the literature.
The shape models, spin parameters, diameters, volumes, and 
corresponding uncertainties derived here are already available 
on the ISAM webpage.

\begin{table*}[t!]
  \centering
\begin{small}
\caption{Spin parameters of asteroid models obtained in this work, with their uncertainty values. 
The first column gives the sidereal period of rotation, next there are two sets of pole longitude and latitude. 
The sixth column gives the rms deviations of the model lightcurves from the data, and the photometric dataset 
parameters follow after (observing span, number of apparitions, and number of individual lightcurve fragments).
}
\label{results}
\begin{tabular}{rrrrccccc}
\hline
Sidereal      & \multicolumn{2}{c}{Pole 1} & \multicolumn{2}{c}{Pole 2} & rmsd  & Observing span & $N_{app}$ & $N_{lc}$ \\
period [hours]& $\lambda_p [\deg]$ & $\beta_p [\deg]$    & $\lambda_p [\deg]$ & $\beta_p
[\deg]$    & [mag] &   (years)      &           &          \\
\hline
&&&&&&&&\\
{\bf (3) Juno}  & & & & & & & & \\
${7.209533^{+0.000009}_{-0.000013}}$   & $105^{+9}_{-9}$  &$22^{+12}_{-22}$   & $-$    & $-$   &
0.015 & 1954--2015 & 11 & 28 \\  
&&&&&&&&\\
{\bf (14) Irene}  & & & & & & & & \\
${15.029892^{+0.000023}_{-0.000028}}$   & $91^{+1}_{-4}$  &$-14^{+9}_{-2}$   &
$267^{+5}_{-2}$    & $-10^{+14}_{-1}$   & 0.019 & 1953--2017 & 14 &
99 \\  
&&&&&&&&\\
{\bf (20) Massalia}  & & & & & & & & \\
${8.097587^{+0.000003}_{-0.000001}}$   & $111^{+16}_{-15}$     &
$77^{+17}_{-7}$   & $293^{+17}_{-17}$    & $76^{+20}_{-10}$   &
0.019 & 1955--2017 & 13 & 111 \\  
&&&&&&&&\\
{\bf (64) Angelina}  & & & & & & & & \\
${8.751708^{+0.000003}_{-0.000003}}$   & $135^{+4}_{-1}$ &$12^{+12}_{-14}$   & 
$313^{+3}_{-1}$    & $13^{+8}_{-11}$   & 0.020 & 1981--2017 & 10 &
81 \\  
&&&&&&&&\\
{\bf (68) Leto }  & & & & & & & & \\
${14.845449^{+0.000004}_{-0.000003}}$   & $125^{+8}_{-6}$     &$61^{+7}_{-17}$   &
 $308^{+4}_{-2}$    & $46^{+4}_{-9}$   & 0.030 & 1978--2018 & 5 & 38 \\  
&&&&&&&&\\
{\bf (89) Julia}  & & & & & & & & \\
${11.388331^{+0.000007}_{-0.000005}}$   & $125^{+8}_{-6}$     & $-23^{+8}_{-6}$   &
 $-$    & $-$   & 0.012 & 1968--2017 & 4 & 37 \\  
&&&&&&&&\\
{\bf (114) Kassandra}  & & & & & & & & \\
${10.743552^{+0.000013}_{-0.000009}}$   & $189^{+4}_{-5}$     & $-64^{+15}_{-6}$ 
  & $343^{+6}_{-3}$    & $-69^{+13}_{-11}$   & 0.019 & 1979--2018 & 8 & 43 \\  
&&&&&&&&\\
{\bf (145) Adeona}  & & & & & & & & \\
${15.070964^{+0.000038}_{-0.000044}}$   &
$95^{+2}_{-2}$
& $46^{+1}_{-4}$   & 
$-$    & $-$   & 0.12 & 1977--2018 & 9 &
78 \\  
&&&&&&&&\\
{\bf (297) Caecilia}  & & & & & & & & \\
${4.151390^{+0.000005}_{-0.000003}}$   & $53^{+6}_{-1}$     &$-36^{+11}_{-5}$   &
 $227^{+6}_{-3}$    & $-51^{+11}_{-4}$   & 0.016 & 2004--2018 & 9 &
35 \\  
&&&&&&&&\\
{\bf (308) Polyxo}  & & & & & & & & \\
${12.029587^{+0.000006}_{-0.000007}}$   & $115^{+2}_{-2}$    &$26^{+5}_{-2}$   &
 $295^{+1}_{-2}$    & $39^{+4}_{-2}$   & 0.013 & 1978--2018 & 6 & 37 \\  
&&&&&&&&\\
{\bf (381) Myrrha}  & & & & & & & & \\
${6.571953^{+0.000003}_{-0.000004}}$   & $237^{+3}_{-5}$     &
$82^{+3}_{-13}$   &
 $-$    & $-$   & 0.013 & 1987--2018 & 7 & 38 \\  
&&&&&&&&\\
{\bf (441) Bathilde}  & & & & & & & & \\
${10.443130^{+0.000009}_{-0.000005}}$   & $125^{+9}_{-7}$     &$39^{+24}_{-26}$   &
 $287^{+8}_{-15}$    & $52^{+23}_{-13}$   & 0.015 & 1978--2018 & 10 &
85 \\  
&&&&&&&&\\
{\bf (721) Tabora}  & & & & & & & & \\
${7.981234^{+0.000010}_{-0.000011}}$   & $173^{+4}_{-5}$     &$-49^{+18}_{-20}$   &
 $340^{+6}_{-9}$    & $34^{+20}_{-26}$   & 0.042 & 1984--2018 & 5 & 62 \\  
&&&&&&&&\\
\hline
\end{tabular}
\end{small}
\end{table*}

\begin{table*}

\clearpage
  \centering
\caption{ Results from the occultation fitting of SAGE models. Mirror pole solutions are 
  labeled ``pole 1'' and ``pole 2''. 
 Scaled sizes are given in kilometers as the diameters of the equivalent 
  volume spheres.} 
 \vspace{1mm}
\label{tab2}
 \begin{tabular}{cc|c  c c c c}
\hline
Number & Name & pole & 
  Year  of  occultation  & Diameter (km) & $+\sigma D$ (km) & $-\sigma D$
(km)\\
\hline\hline
3 & Juno &  &  1979-12-11 & 260.0 & 13.0 & -12.0\\
 &  &  &  2000-05-24 & 236.0 & 20.0 & -17.0\\
 &  &  &  2014-11-20 & 250.0 & 12.0 & -11.0\\
 &  &  &   &  &  &   \\
14 & Irene & 1 &  2013-08-02 & 145.8 & 12.0 & -11.5\\
 &  & 2 &  2013-08-02 & 145.2 & 91.5 & -18.1\\
 &  &  &    &  &  &  \\
20 & Massalia & 1 &  2012-10-09 & 106.5 & 4.8 & -2.8\\
 &  & 2 &  2012-10-09 & 113.5 & 6.2 & -9.9\\
 &  &  &    &  &  &  \\
64 & Angelina & 1 &  2004-07-03 & 48.9 & 3.8 & -2.3\\
 &  & 2 &  2004-07-03 & 50.7 & 2.1 & -3.0\\
 &  &  &    &  &  &  \\
68 & Leto & 1 &  1999-05-23 & 152.0 & 20.8 & -18.3 \\
 &  & 2 &  1999-05-23 & 132.8 & 8.4 & -8.0\\
 &  &  &    &  &  &  \\
89 & Julia &  &  2005-08-13 & 138.7 & 14.2 & -6.4\\
 &  &  &  2006-12-04 & 137.3 & 2.1 & -4.5\\
 &  &  &   &  &  &  \\
145 & Adeona &  &  2005-02-02 & 145 & 4.3 & -2.7\\
 &  &  &    &  &  &  \\
308 & Polyxo & 1 &  2000-01-10 & 133.5 & 5.8 & -6.3\\
 &  &  &  2004-11-16 & 125.4 & 11.1 & -8.6\\
 &  &  &  2010-06-02 & 128.8 & 3.0 & -2.8\\
 &  & 2 & 2000-01-10 & 131.2 & 5.0 & -2.9\\
 &  &  &  2004-11-16 & 125.3 & 10.7 & -8.1\\
 &  &  &  2010-06-02 & 127.8 & 3.5 & -4.3\\
 &  &  &   &  &  &  \\
381 & Myrrha &  &  1991-01-13 & 134.8 & 45.3 & -12.8\\
 &  &  &    &  &  &  \\
441 & Bathilde & 1 &  2003-01-11 & 75.3 & 74.6 & -10.0\\
 &  & 2 & 2003-01-11 & 76.8 & 15.9 & -9.1\\
    \hline
  \end{tabular}
\end{table*}

\begin{acknowledgements}
The research leading to these results has received funding from the European Union's
       Horizon 2020 Research and Innovation Programme, under Grant Agreement no 687378 (SBNAF).
\newline
Funding for the Kepler and K2 missions is provided by the
NASA Science Mission directorate. L.M. was supported by the Premium Postdoctoral Research Program
 of the Hungarian Academy of Sciences. The research leading to these results has received 
funding from the LP2012-31 and LP2018-7 Lend\"ulet grants of the Hungarian Academy of Sciences.
This project has been supported by the Lendület grant LP2012-31 of the Hungarian Academy of Sciences
and by the GINOP-2.3.2-15-2016-00003 grant of the Hungarian National Research, 
Development and Innovation Office (NKFIH).
\newline
TRAPPIST-South is a project funded by the Belgian Fonds de la Recherche Scientifique
(F.R.S.-FNRS) under grant FRFC 2.5.594.09.F. TRAPPIST-North is a
project funded by the University of Liège, and performed in collaboration with
Cadi Ayyad University of Marrakesh. EJ is a FNRS Senior Research Associate.
\newline
"The Joan Or{\'o} Telescope (TJO) of the Montsec Astronomical Observatory (OAdM)
is owned by the Catalan Government and operated by the Institute for Space Studies of Catalonia (IEEC)."
\newline
"This article is based on observations made with the SARA telescopes (Southeastern Association for Research in Astronomy),
whose node is located at the Kitt Peak National Observatory, AZ under the auspices of the National Optical Astronomy Observatory (NOAO)."
\newline
"This project uses data from the SuperWASP archive. The WASP project is currently funded and operated by Warwick University
and Keele University, and was originally set up by Queen's University Belfast, the Universities of Keele, St. Andrews,
and Leicester, the Open University, the Isaac Newton Group, the Instituto de Astrofisica de Canarias,
the South African Astronomical Observatory, and by STFC."
\newline
"This publication makes use of data products from the Wide-field Infrared Survey Explorer,
 which is a joint project of the University of California, Los Angeles, 
and the Jet Propulsion Laboratory/California Institute of Technology,
 funded by the National Aeronautics and Space Administration."
\newline
The work of TSR was carried out through grant APOSTD/2019/046 by Generalitat Valenciana (Spain)
\newline
\end{acknowledgements}

%
%

\begin{appendix}

\section{Additional tables}

\begin{table*}
\clearpage
  \centering
\caption{ Summary of TPM results, including the minimum reduced chi-squared
    ($\bar{\chi}^2_{m}$), the best-fitting diameter ($D$) and corresponding
    1$\sigma$ statistical error bars, and the number of IR data that were modeled
    ($N_{\mathrm{IR}}$).
    TLC (Yes/No) refers to the availability of at least one thermal lightcurve
    with eight or more points sampling the rotation period. The
    $\bar{\chi}^2_{m}$ obtained for a spherical model with the same spin
    properties is shown. We also provide the value of thermal inertia $\Gamma$
    and surface roughness. Whenever the two mirror solutions provided different
    optimum diameters, we show them in different lines. Acceptable solutions,    
and preferred ones whenever it applies to mirror models, are highlighted in
    bold face.
}
    \label{tab:tpm}
  \vspace{1mm}
  \begin{tabular}{|l |  c c c  c c c c l|}
    
    \hline
    Target [pole]  & $N_{\mathrm{IR}}$ & TLC & $\bar{\chi}^2_{m}$ &   $D \pm \sigma D$ (km) & $\bar{\chi}^2_{m}$ for sphere &
 $\Gamma$ [SI units] & Roughness   & Comments\\
    \hline\hline
    \textbf{(3) Juno}  &  112  & No & 1.3  & 254 $\pm$ 4 & 1.0 &$70^{+30}_{-40}$ &$\gtrsim$1.00 & Borderline acceptable\\
 & & & & & & & & fit. Sphere does \\
 & & & & & & & & better.\\
    (14) Irene 1  & 6  & No  & 0.1  & 155 & 0.4 &70 &0.80 & Very few data to \\
 & & & & & & & & provide realistic  \\
 & & & & & & & &  error bars.\\
    (14) Irene 2  & 6  & No  & 0.2  & 154 & 0.2 &70 &0.80 & \emph{Idem} \\
    \textbf{(20) Massalia  1,2}  & 72 & No  & 0.5 & 145$\pm$2 & 1.6 & $35^{+25}_{-10}$  & $\lesssim$0.20 & Mirror solutions
pro-\\
 & & & & & & & &vide virtually same fit \\
    \textbf{(64) Angelina  1} & 23 & Yes & 0.8 & 54 $\pm$2 & 1.10 &$35^{+25}_{-20}$ &0.20 & Did not model MSX data \\
    (64) Angelina  2 & 23 & Yes & 1.16 & 54 $\pm$2 & 1.24 &  $20^{+25}_{-10}$ &
    0.25& \emph{Idem} \\
    \textbf{(68) Leto  1} & 55 & Yes & 0.6    & 121 $\pm$ 5 & 0.83 &
    $40^{+25}_{-20}$&0.50 & Small offset between mir-\\
 & & & & & & & &    ror solutions (not stat.\\
 & & & & & & & &   significant)\\
    (68) Leto  2          & 55 & Yes & 0.7    & 123 $\pm$ 5 & 0.87 &$35^{+45}_{-25}$ &
0.45& \emph{Idem} \\
    \textbf{(89) Julia}  & 27  & No  & 1.0  & 150 $\pm$ 10 & 1.5 &$100^{+150}_{-50}$  & $\gtrsim$0.90 &  Only northern aspect \\
 & & & & & & & &angles covered($A<70\deg$) \\
    & & & & & & & &  in the IR. Unexpectedly  \\%
 & & & & & & & & high thermal inertia \\
 & & & & & & & & fits better probably \\
 & & & & & & & & because the phase angle\\
 & & & & & & & & coverage is not well \\
 & & & & & & & & balanced (only 3 measu-\\
 & & & & & & & & rements with $\alpha>0$). \\
    \textbf{(114) Kassandra  1,2} & 46 & Yes &  0.6    & 98  $\pm$ 3 & 0.70 &
$20^{+30}_{-20}$ &0.55 & Quite irregular but spheres\\
 & & & & & & & & provide similar fit \\
    (145) Adeona    & 17 & No & 0.47  & 149 $\pm$10 &
0.23 &
$70^{+130}_{-70}$ & 0.60 &  Phase angle coverage\\
 & & & & & & & & is not well balanced  \\
& & & & & & & &  between pre- and \\
& & & & & & & &  post-opposition \\
    (297) Caecilia  & 13  &  No & 0.9 & 41 & 0.9 &10 & 0.35& Too few data to give\\
 & & & & & & & & realistic error bars \\
    (308) Polyxo 1,2   & 13 & No & 0.4  & 139 & 0.35 &50 &0.45 &Too few data to give\\
 & & & & & & & & realistic error bars \\
    \textbf{(381) Myrrha}  & 73 & Yes & 0.40  & 131$\pm$4 & 1.6 & $80^{+40}_{-40}$ &$\gtrsim$1.00 & Good fit but some
small\\
 & & & & & & & & waviness in
    residuals\\
& & & & & & & & vs. rot. phase    plot\\
    \textbf{(441) Bathilde 1} & 26 & Yes & 0.7   & 72 $\pm$ 2 & 1.7 & $180^{+20}_{-60}$ &$\gtrsim$0.90 & Very high thermal inertia  \\
    (441) Bathilde 2 & 26 & Yes & 1.6   & -- & $>2$ & -& -& Bad fit \\
    \textbf{(721) Tabora 1}  & 40 & Yes & 1.4  & 78$\pm$5 & $>5$ & $6^{+14}_{-6}$
    &0.65 & Borderline acceptable fit,\\
 & & & & & & & & still better than sphere \\
    (721) Tabora 2  & 40 & Yes & 2.1  & -- & $>5$ &$-$ &$-$ & Bad fit \\
    \hline
  \end{tabular}
\end{table*}

\begin{table*}

  \centering
\caption{ Results from the previous solutions available in the literature. Mirror pole solutions are 
  labeled ``pole 1'' and ``pole 2''. 
 Scaled sizes are given in kilometers as the diameters of the equivalent 
  volume spheres. For objects marked with $^*$ we have taken the sizes from the
AKARI, SIMPS, and WISE \citep{Usui2011,tedesco, Mainzer2016} missions, respectively, for which the sizes were often calculated with an STM
approximation of the spherical shape, and often without a known pole solution. } 
 \vspace{1mm}
\label{tab4}
\begin{tabular}{rrrrccccc} 
\hline
Sidereal      & \multicolumn{2}{c}{Pole 1} & \multicolumn{2}{c}{Pole 2} & D &reference   & & \\
period [hours]& $\lambda_p$ & $\beta_p$    & $\lambda_p$ & $\beta_p$    & km         & & & \\
\hline
&&&&&&&&\\
{\bf (3) Juno}  & & & & & & & & \\
${7.20953}$   & $105\deg$  &$21\deg$   & $-$    & $-$   & $248\pm5$ &
\citet{Viikinkoski15} &  &    \\  
&&&&&&&&\\
{\bf (14) Irene}  & & & & & & & & \\
${15.02987}$   & $91\deg$  &$-15\deg$   &
$-$    & $-$   & $153\pm6$&\citet{Viikinkoski2017} &  & 
 \\  
&&&&&&&&\\
{\bf (20) Massalia}  & & & & & & & & \\
${8.09902}$   & $179\deg$     &
$39\deg$   & $360\deg$    & $40\deg$   &
${131.56/145.5/-}^*$ &\citet{Kaasalainen2002}   & &  \\  
&&&&&&&&\\
{\bf (64) Angelina}  & & & & & & & & \\
${8.75033}$   & $138\deg$ &$14\deg$   & 
$317\deg$    & $17\deg$   & $52\pm10$&\citet{durech11}   &  &
 \\  
&&&&&&&&\\
{\bf (68) Leto }  & & & & & & & & \\
${14.84547}$   & $103\deg$     &$43\deg$   &
 $290\deg$    & $23\deg$   & $112\pm14$ &\citet{hanus13}   &  &  \\  
&&&&&&&&\\
{\bf (89) Julia}  & & & & & & & & \\
${11.388332}$   & $14\deg$     & $-24\deg$   &
 $-$    & $-$   & $140\pm3$&\citet{vernazza18}  &  &  \\  
&&&&&&&&\\
{\bf (114) Kassandra}  & & & & & & & & \\
${10.74358}$   & $196\deg$     & $-55\deg$ 
  & $4\deg$    & $-58\deg$   & ${93.91/99.65/100}^*$&\citet{durech18b}  &  &  \\  
&&&&&&&&\\
{\bf (145) Adeona}  & & & & & & & & \\
$-$   & $-$     &$-$   &
 $-$    & $-$   & ${141.39/151.14/151}^*$&  &  &
 \\  
&&&&&&&&\\
{\bf (297) Caecilia}  & & & & & & & & \\
${4.151388}$   & $47\deg$     &$-33\deg$   &
 $223\deg$    & $-53\deg$   & ${42.28/39.48/-}^*$&\citet{hanus13}  &  &
 \\  
&&&&&&&&\\
{\bf (308) Polyxo}  & & & & & & & & \\
$-$   & $-$     &$-$   &
 $-$    & $-$   & ${135.25/140.69/144.4}^*$&  &  &
 \\  
&&&&&&&&\\
{\bf (381) Myrrha}  & & & & & & & & \\
${6.57198}$   & $3\deg$     &
$48\deg$   &
 $160\deg$    & $77\deg$   &${117.12/120.58/129}^*$& \citet{hanus16}  &  &  \\  
&&&&&&&&\\
{\bf (441) Bathilde}  & & & & & & & & \\
${10.44313}$   & $122\deg$     &$43\deg$   &
 $285\deg$    & $55\deg$   & ${59.42/70.32/70.81}^*$& \citet{hanus13}  & & \\  
&&&&&&&&\\
{\bf (721) Tabora}  & & & & & & & & \\
${7.98121}$   & $172\deg$     &$53\deg$   &
 $343\deg$    & $38\deg$   & ${81.95/76.07/86.309}^*$&\citet{durech18a}  &  &  \\  
&&&&&&&&\\
    \hline
  \end{tabular}
\end{table*}

\clearpage

\section{TPM plots and comments}

The data we used was collected in the SBNAF infrared database\footnote{https://ird.konkoly.hu/}.
In this section, we provide observation-to-model ratio (OMR) plots produced for
the TPM analysis. 
Whenever there was a thermal lightcurve available within the data
set of a target, this was also plotted (see Table~\ref{tab:tpm}). 
In general, IRAS data have larger error bars, carry lower weights,
and, therefore, their OMRs tend to present larger deviations from one. 
On a few occasions, some or all of them were even removed from the
$\chi^2$ optimization, as indicated in the corresponding figure caption. 
To save space, we only include the plots for one of the mirror solutions 
either because the TPM clearly rejected the other one or because the differences
were so small that the other set of plots are redundant. Either way,
that information is given in Table~\ref{tab:tpm}. 
Table~\ref{tab:key} links each target to its corresponding plots in
this section.
\begin{table}[!h]
  \centering
  \caption{List of targets and references to the relevant figures.}
  \label{tab:key}
  \begin{tabular}{l l l}
    \hline
    \hline
    Target  & OMR plots & Thermal lightcurve\\
    \hline
    \hline
    & & \\
    (3) Juno        & Fig.~\ref{fig:00003_OMR} & --\\
    (14) Irene      & Fig.~\ref{fig:00014_OMR} & --\\
    (20) Massalia   & Fig.~\ref{fig:00020_OMR} & --\\
    (64) Angelina   & Fig.~\ref{fig:00064_OMR} & Fig.~\ref{fig:00064_ThLC} (left)\\
    (68) Leto       & Fig.~\ref{fig:00068_OMR} & Fig.~\ref{fig:00064_ThLC} (right)\\
    (89) Julia      & Fig.~\ref{fig:00089_OMR} & --\\
    (114) Kassandra & Fig.~\ref{fig:00114_OMR} & Fig.~\ref{fig:00114_ThLC} (left)\\
    (145) Adeona    & Fig.~\ref{fig:00145_OMR} & --\\
    (308) Polyxo    & Fig.~\ref{fig:00308_OMR} & --\\
(381) Myrrha    & Fig.~\ref{fig:00381_OMR} & Fig.~\ref{fig:00114_ThLC} (right)\\
    (441) Bathilde  & Fig.~\ref{fig:00441_OMR} & Fig.~\ref{fig:00441_ThLC} (left)\\
    (721) Tabora    & Fig.~\ref{fig:00721_OMR} & Fig.~\ref{fig:00441_ThLC} (right)\\    
    & & \\
    \hline
    \hline
  \end{tabular}
\end{table}

\begin{figure*}
  \centering
  \includegraphics[width=0.42\linewidth]{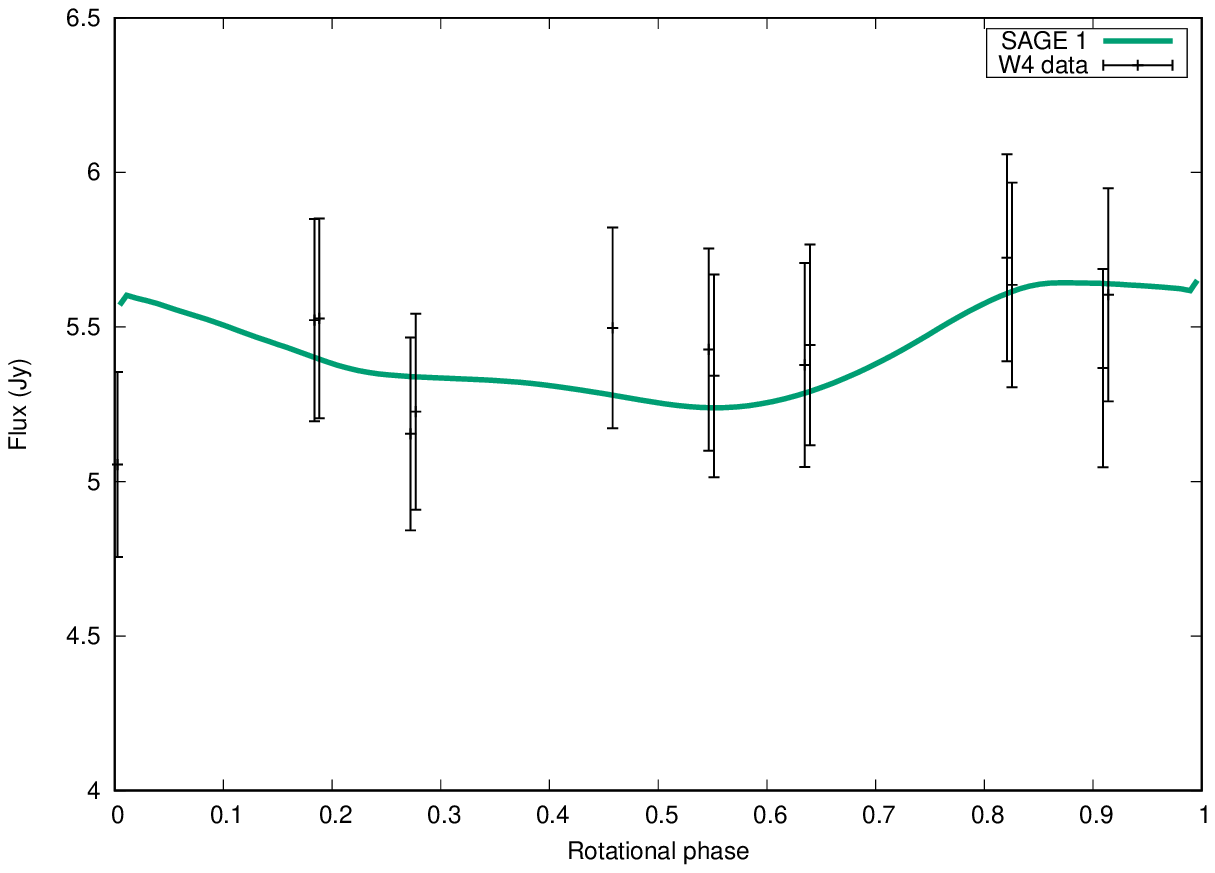}
  \includegraphics[width=0.42\linewidth]{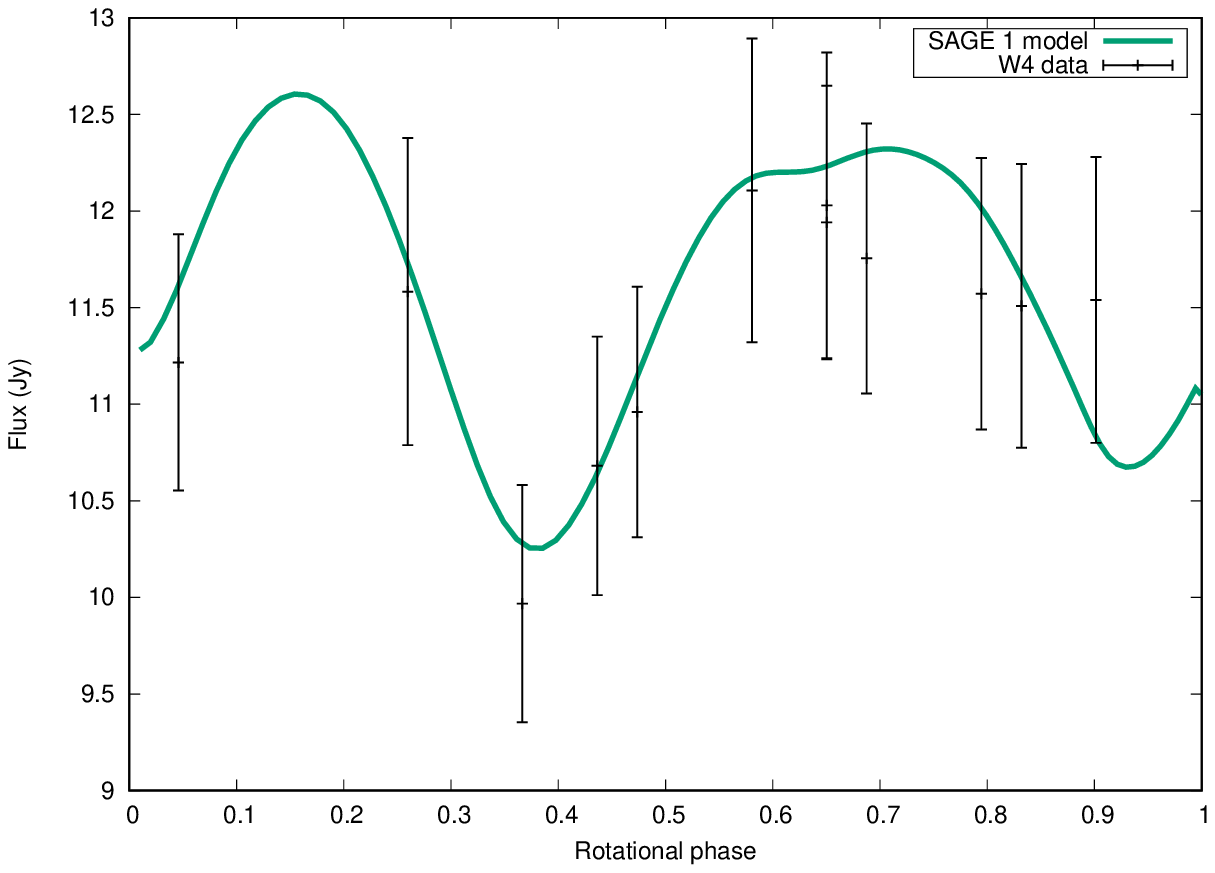}
  \caption{W4 data and model of thermal lightcurves that were generated with the
    best-fitting thermal parameters and size. Left: (64) Angelina's SAGE
    pole 1 model. Right: (68) Leto, also Pole 1. 
  }\label{fig:00064_ThLC}
  
  \centering
  \includegraphics[width=0.42\linewidth]{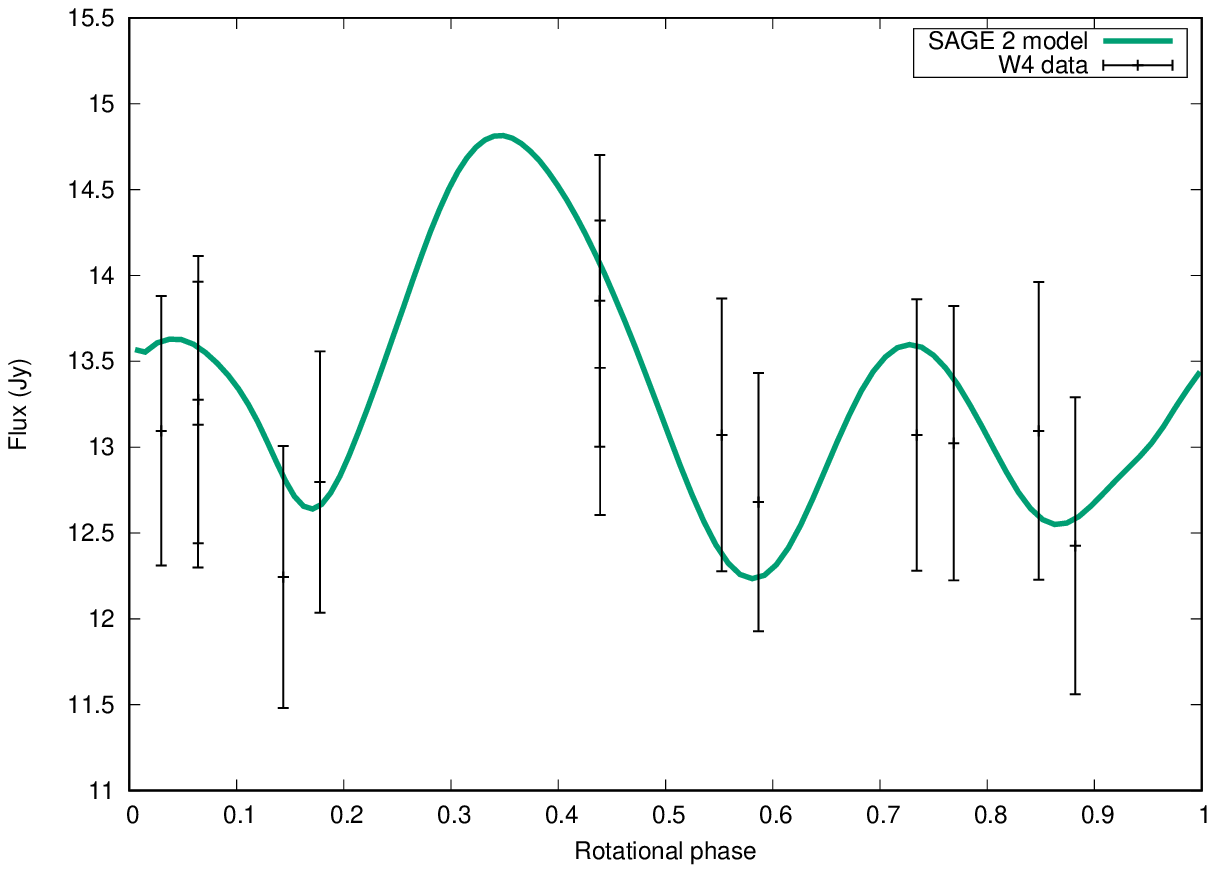}
  \includegraphics[width=0.42\linewidth]{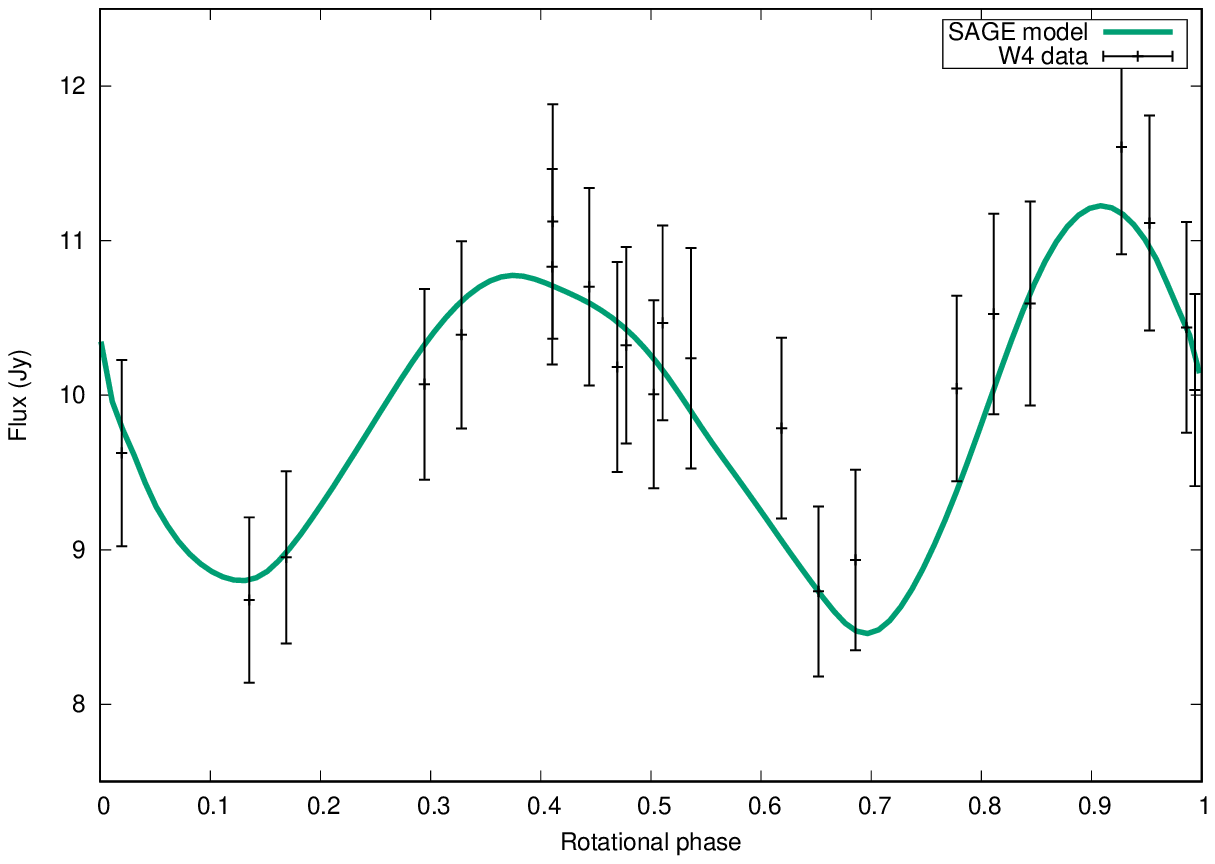}
  \caption{Left: (114) Kassandra. Right: (381) Myrrha. 
  }\label{fig:00114_ThLC}

  \includegraphics[width=0.42\linewidth]{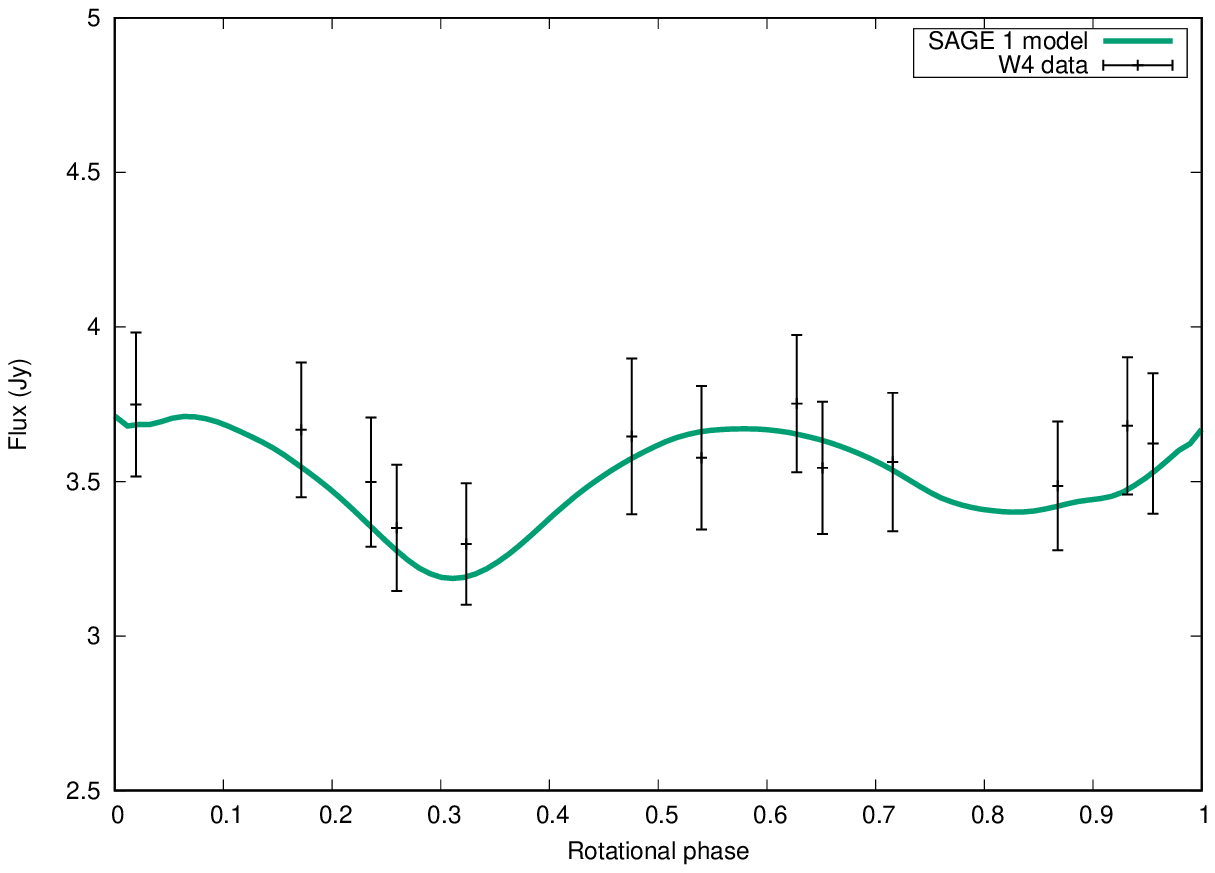}
  \includegraphics[width=0.42\linewidth]{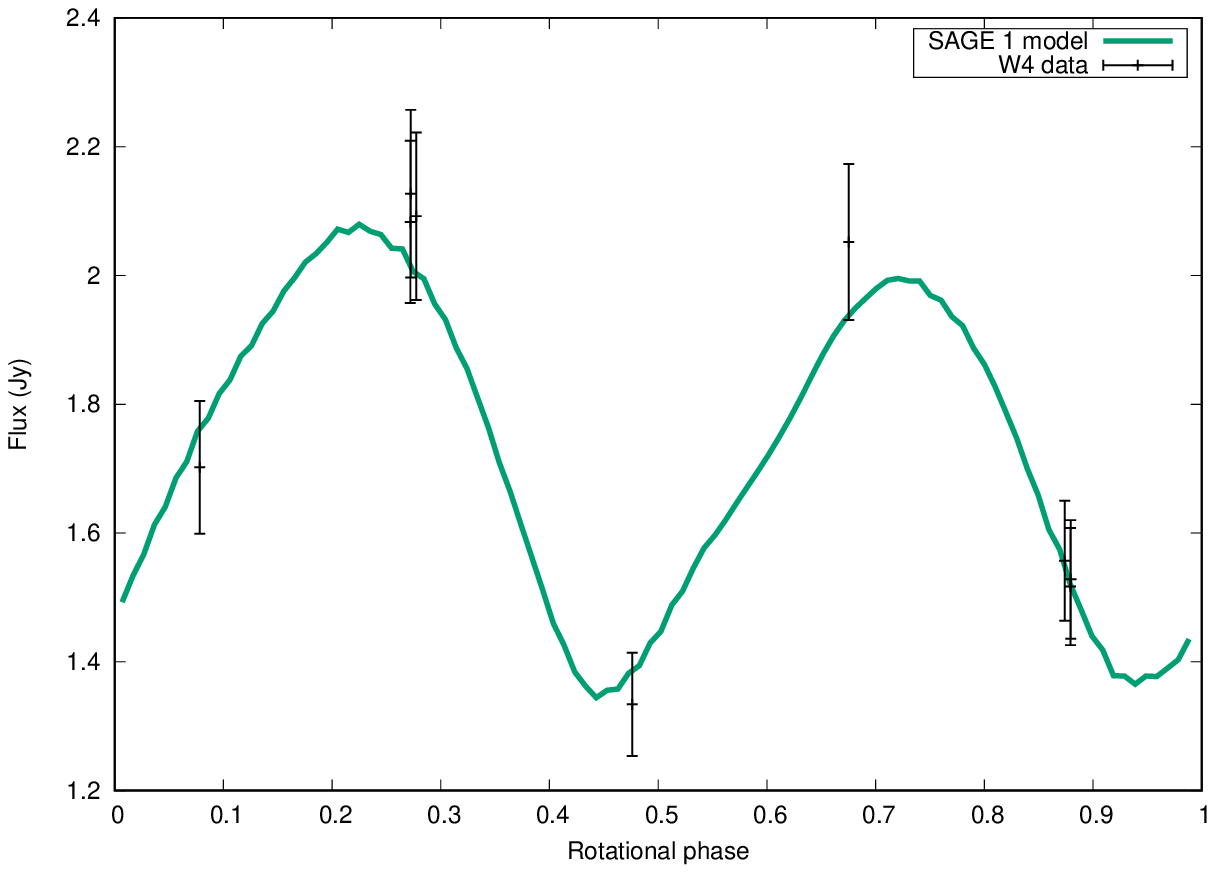}
  \caption{Left: (441) Bathilde. Right: (721) Tabora. 
  }\label{fig:00441_ThLC}
\end{figure*}

\begin{figure}
  \centering
  \includegraphics[width=0.8\linewidth]{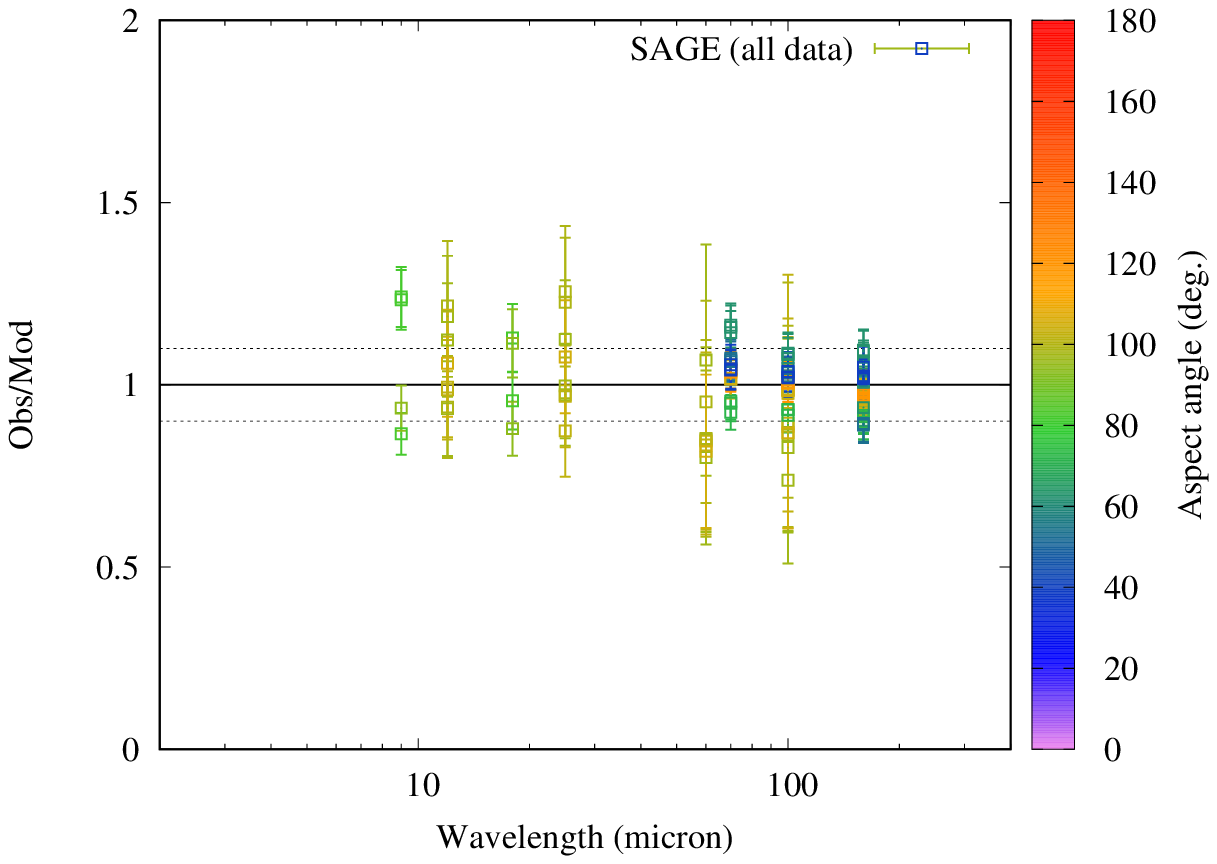}
  
  \includegraphics[width=0.8\linewidth]{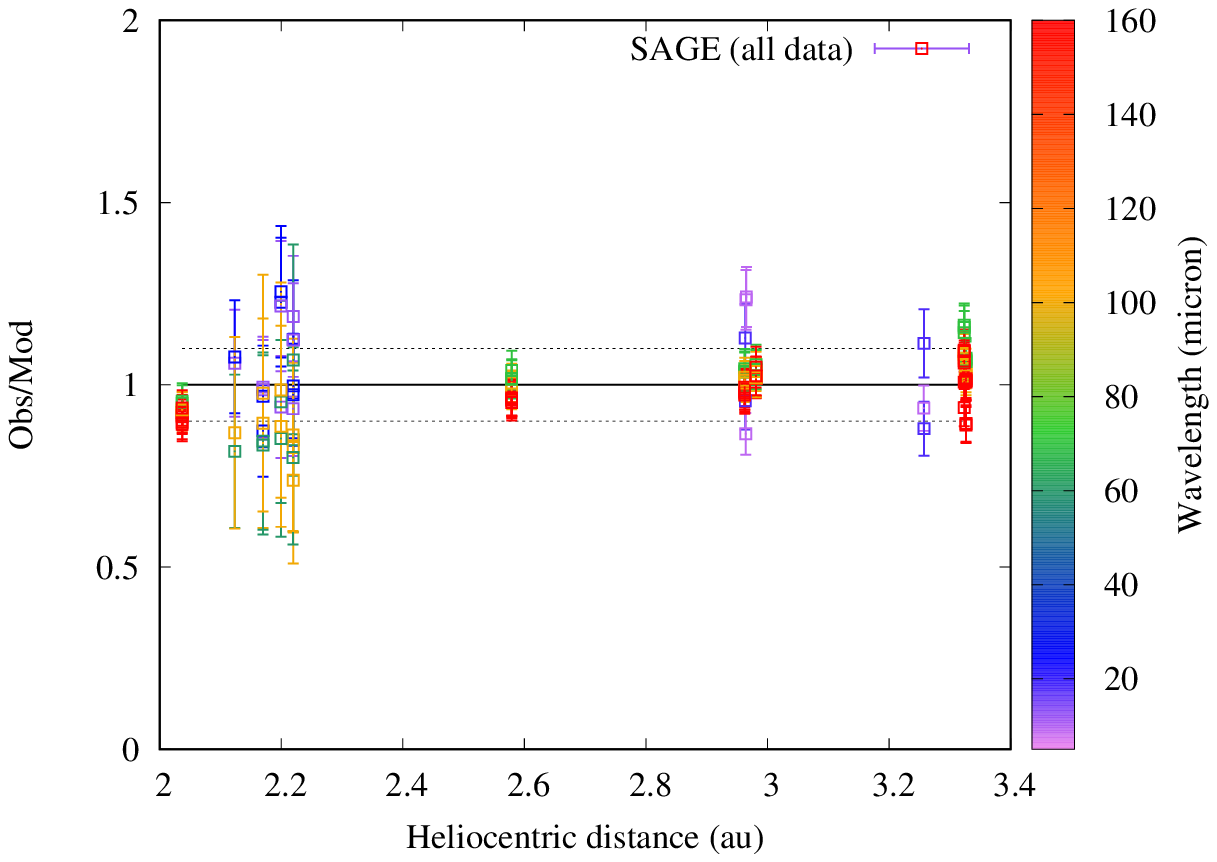}
  
  \includegraphics[width=0.8\linewidth]{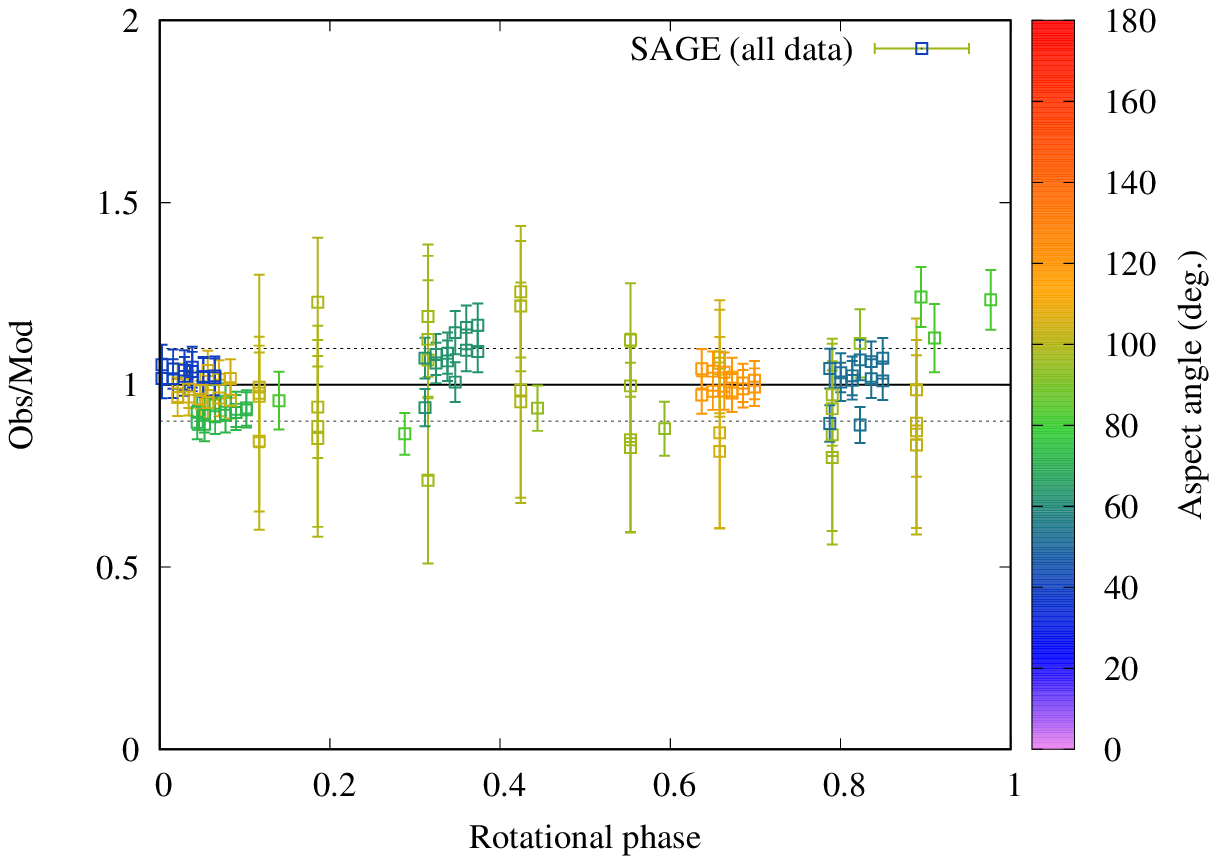}
  
  \includegraphics[width=0.8\linewidth]{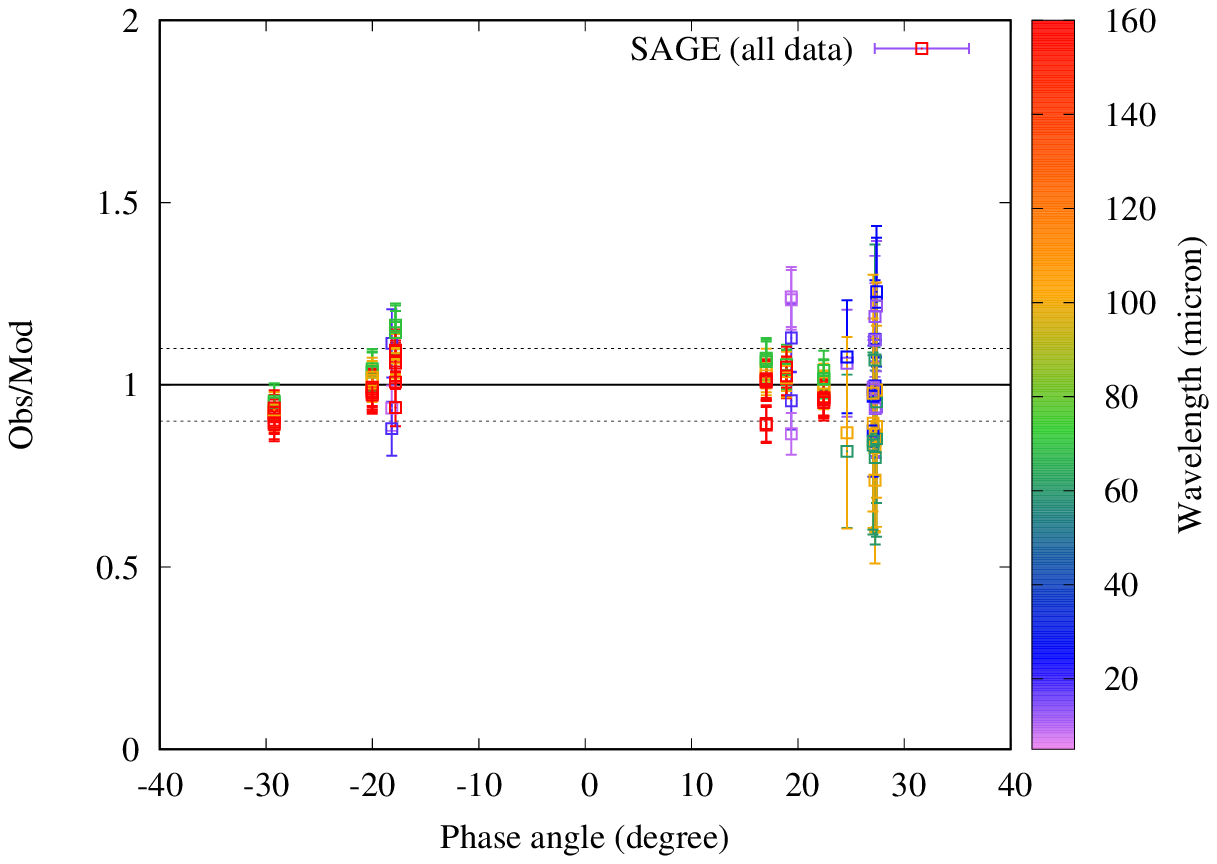}
  
  \caption{(3) Juno (from top to bottom): observation-to-model ratios versus
    wavelength, heliocentric distance, rotational phase, and phase angle. 
    The color bar either corresponds to the aspect angle or to the wavelength 
    at which each observation was taken.
    There are some systematics in the rotational phase plot, which indicate
    there could be some small artifacts in the shape. 
  }\label{fig:00003_OMR}
\end{figure}

\begin{figure}
  \centering
  
  \includegraphics[width=0.8\linewidth]{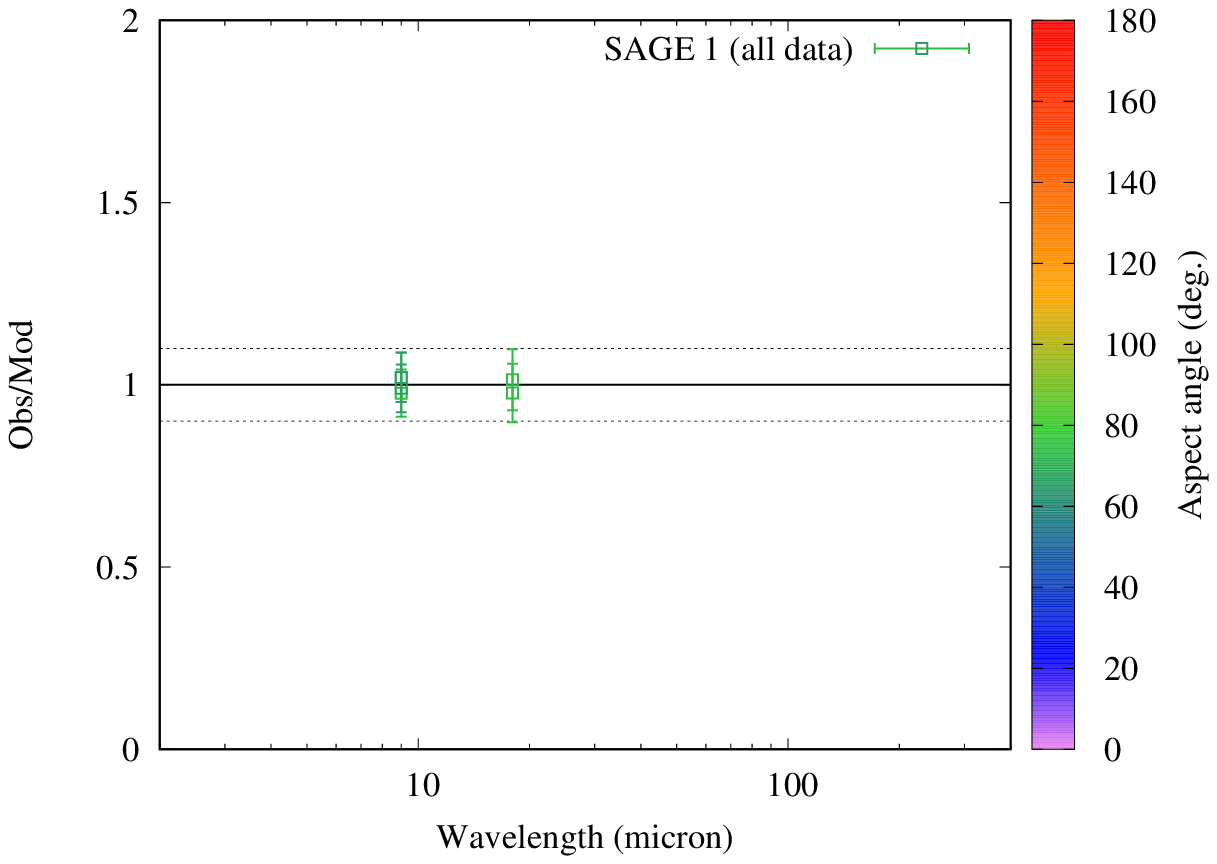}

  \includegraphics[width=0.8\linewidth]{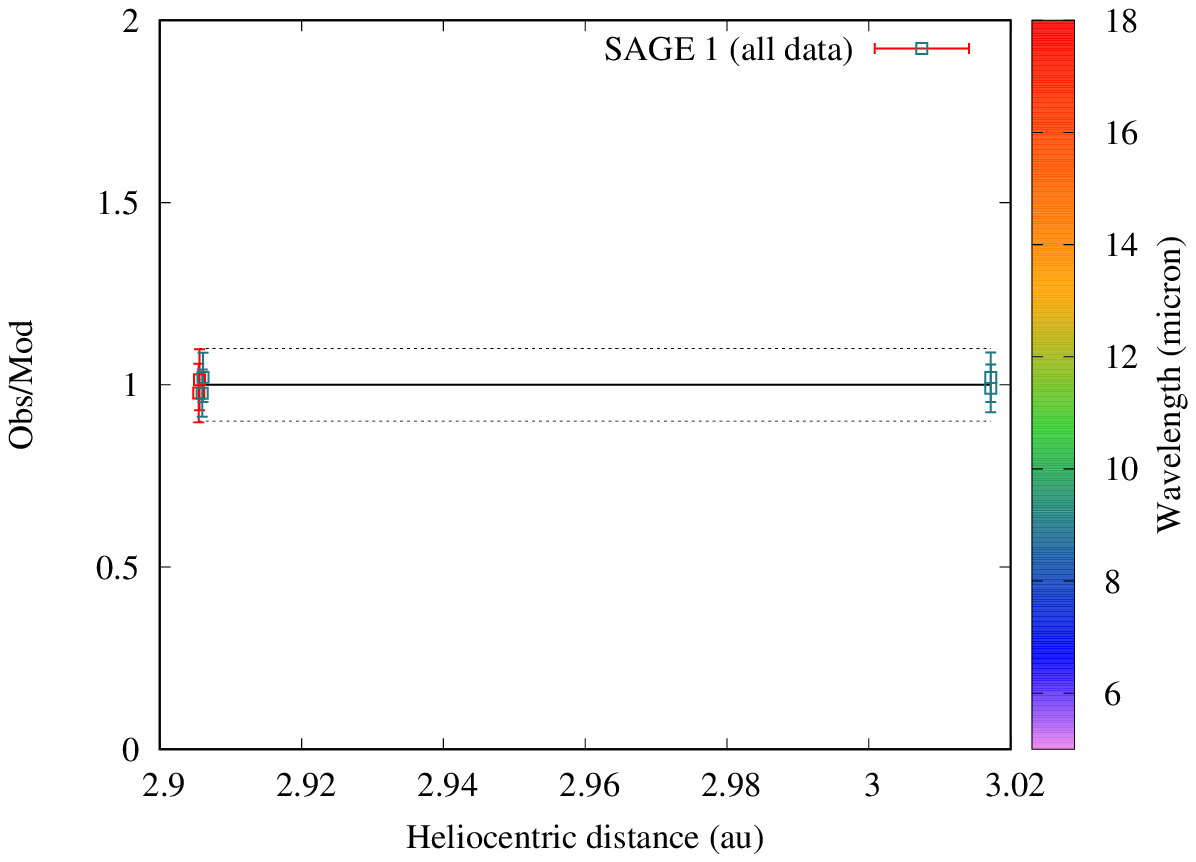}
  
  \includegraphics[width=0.8\linewidth]{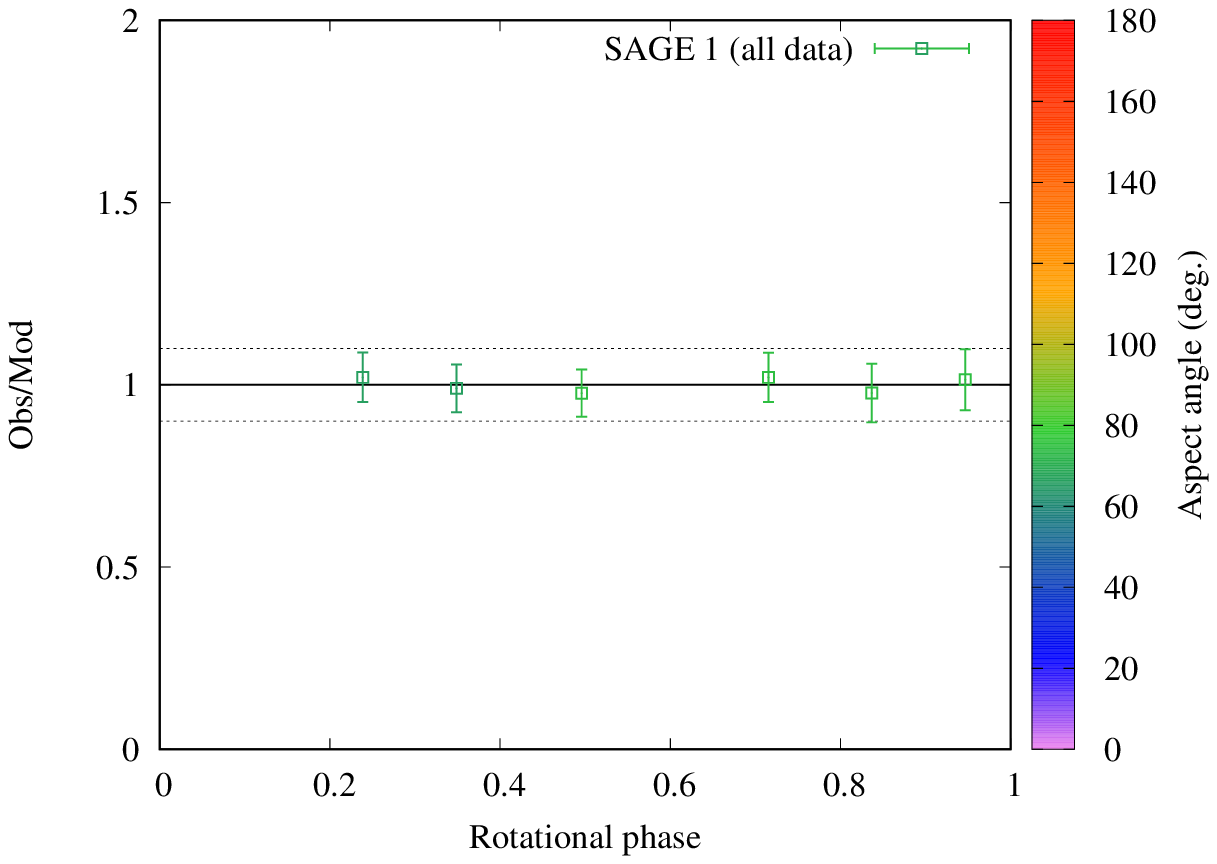}
  
  \includegraphics[width=0.8\linewidth]{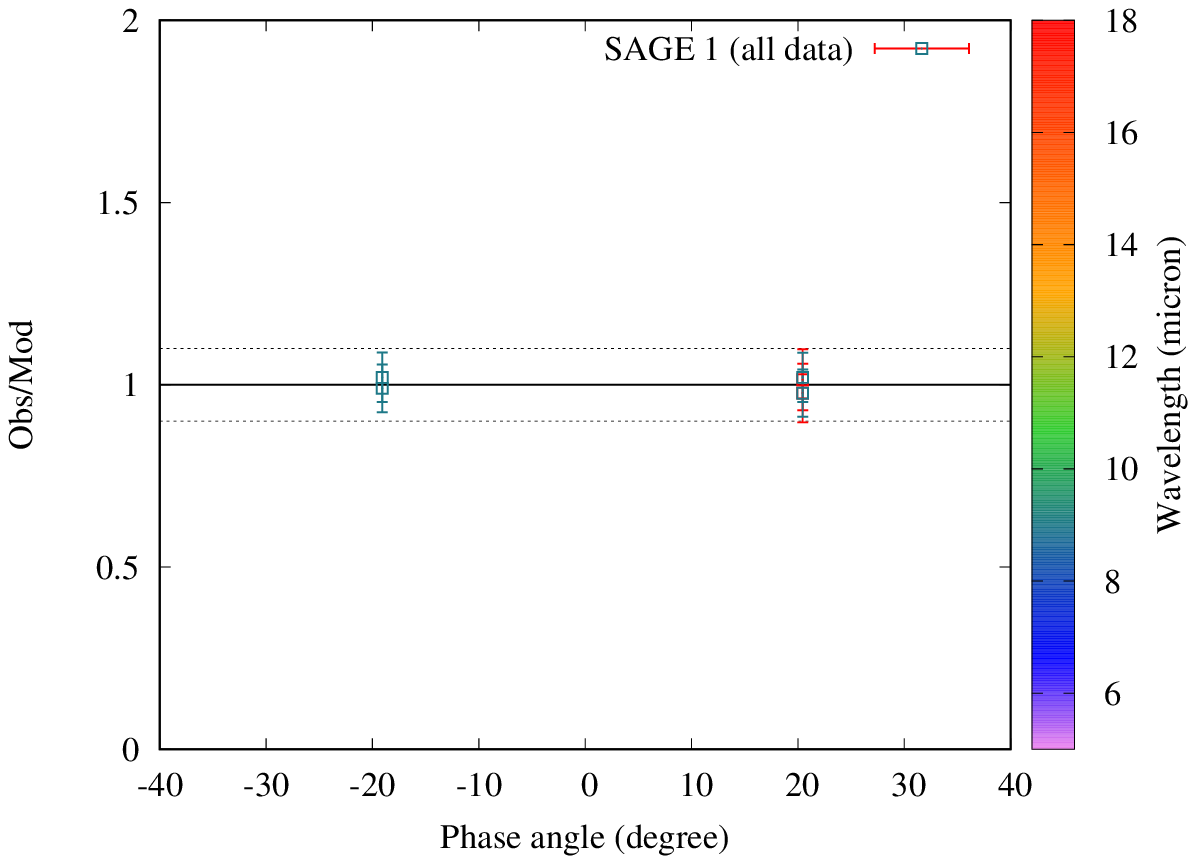}
  
  \caption{(14) Irene (from top to bottom): observation-to-model ratios versus
    wavelength, heliocentric distance, rotational phase, and phase angle. 
    The plots that correspond to the pole 2 solution are very similar. 
  }\label{fig:00014_OMR}

\end{figure}

\begin{figure}
  \centering
  
  \includegraphics[width=0.8\linewidth]{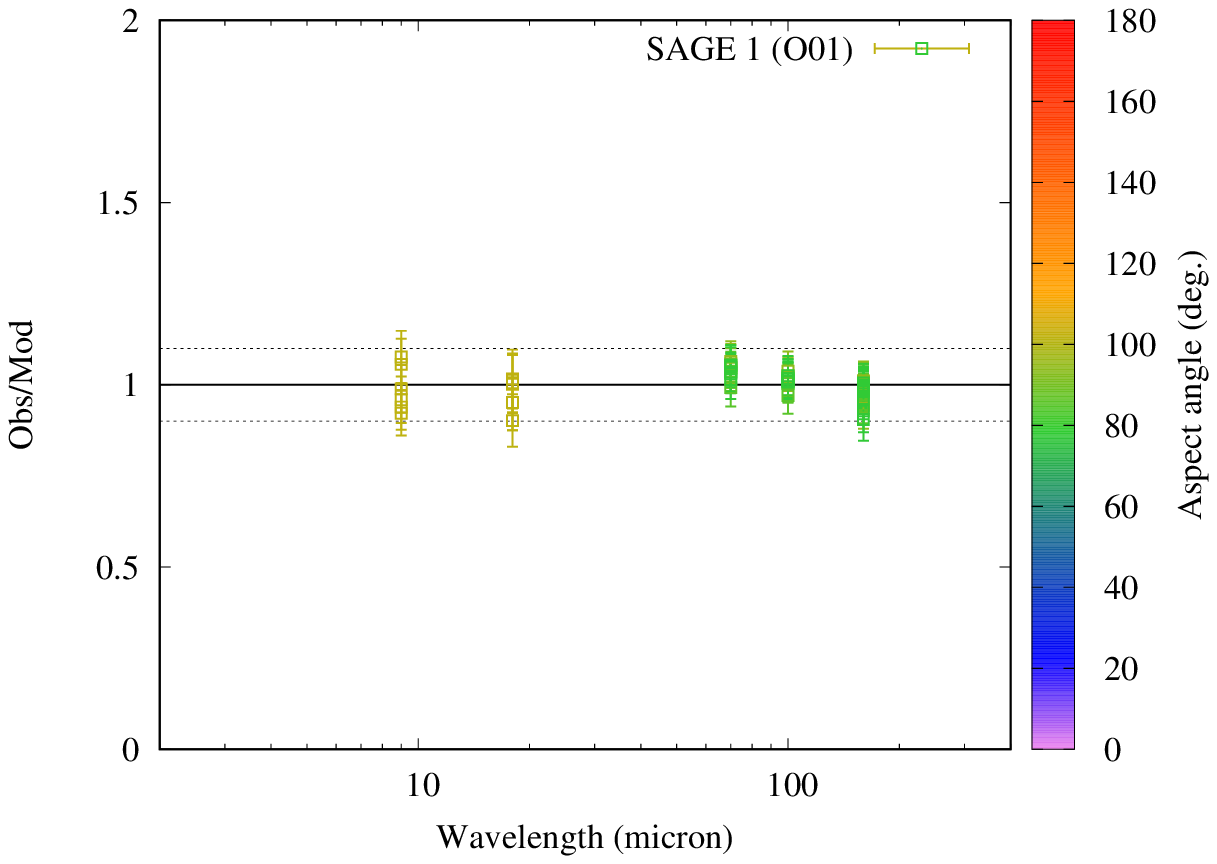}

  \includegraphics[width=0.8\linewidth]{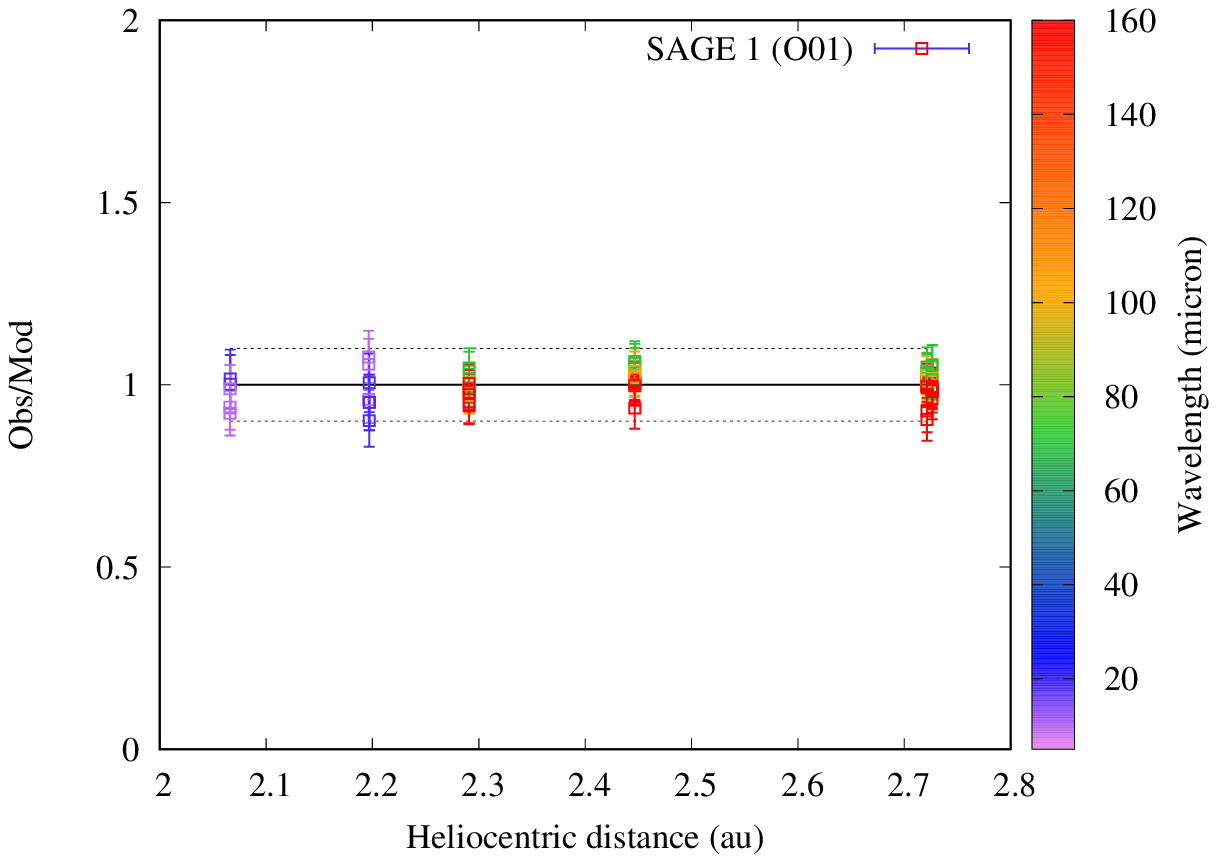}

  \includegraphics[width=0.8\linewidth]{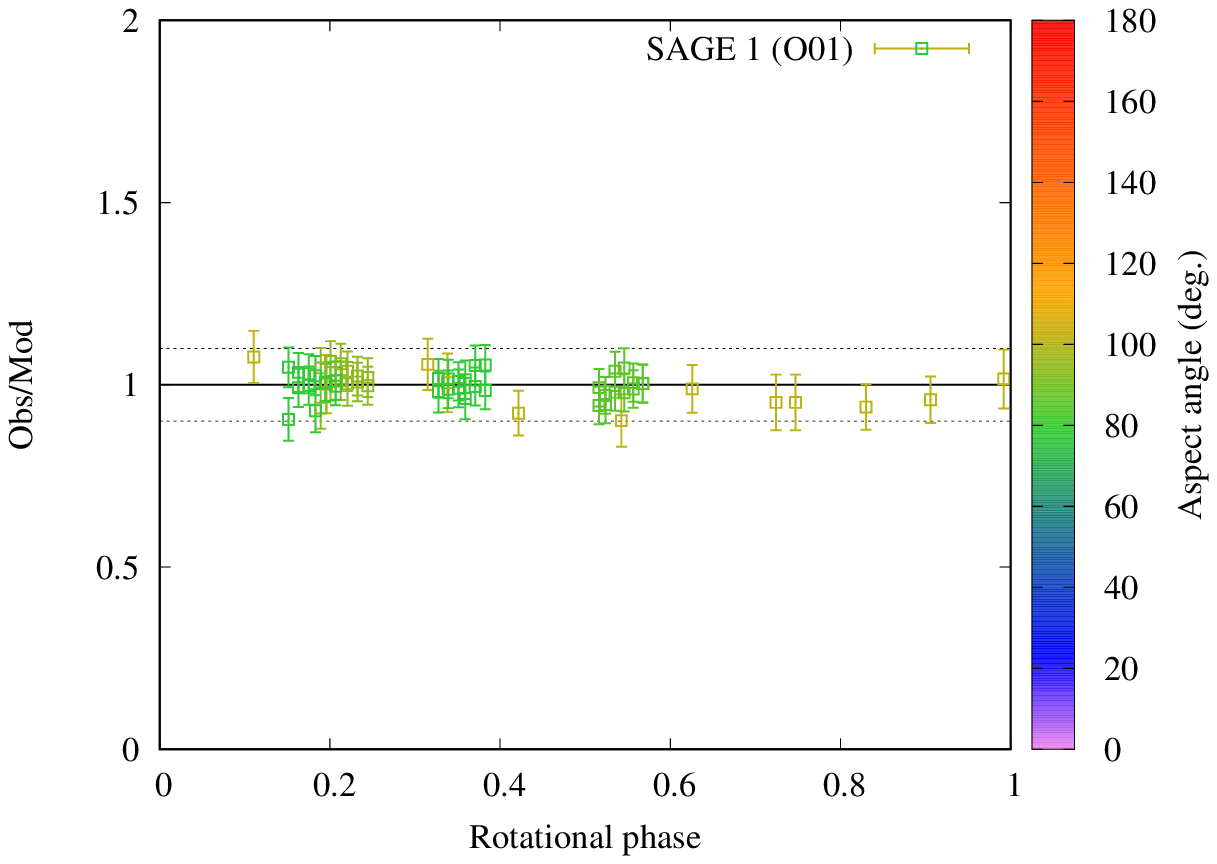}

  \includegraphics[width=0.8\linewidth]{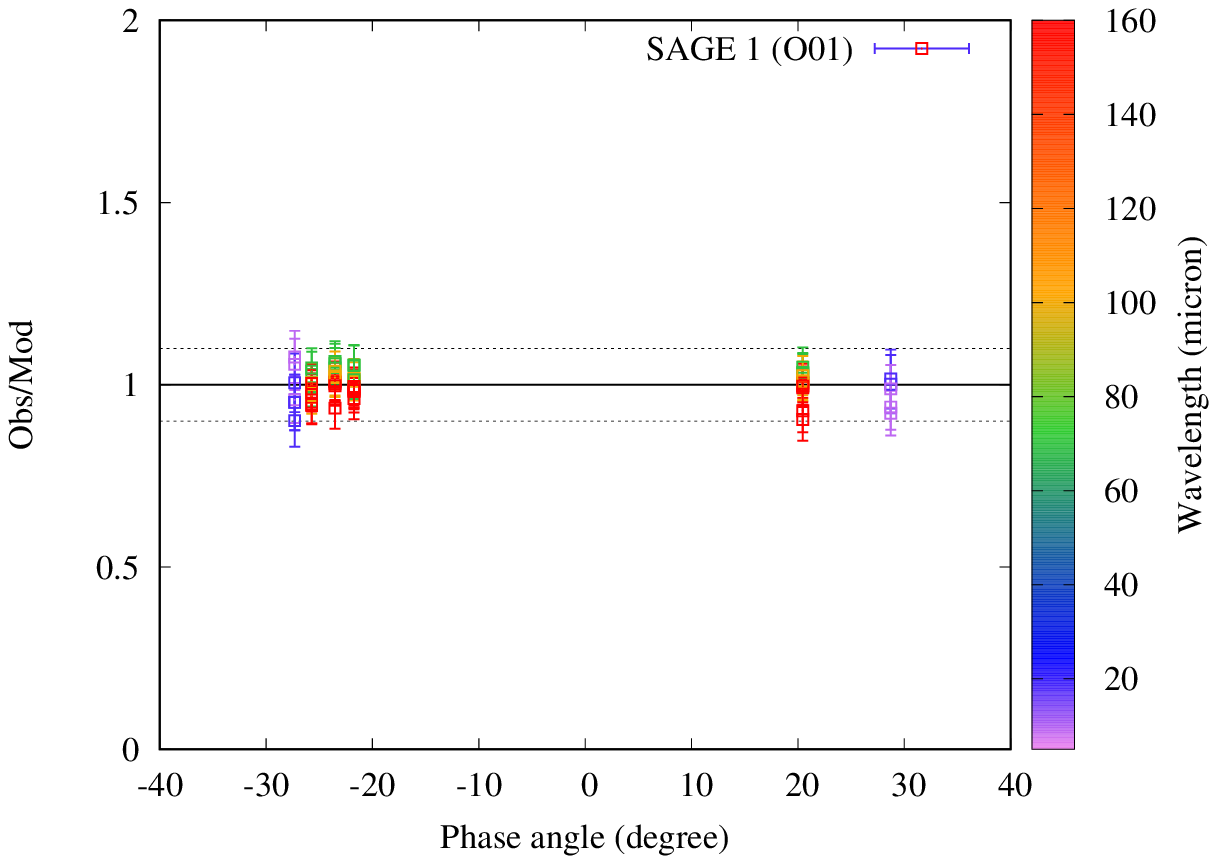}

  \caption{(20) Massalia.  
    The O01 label indicates that the IRAS data were removed from the analysis, in
    this case because their quality was too poor. 
  }\label{fig:00020_OMR}
\end{figure}

\begin{figure}
  \centering

  \includegraphics[width=0.8\linewidth]{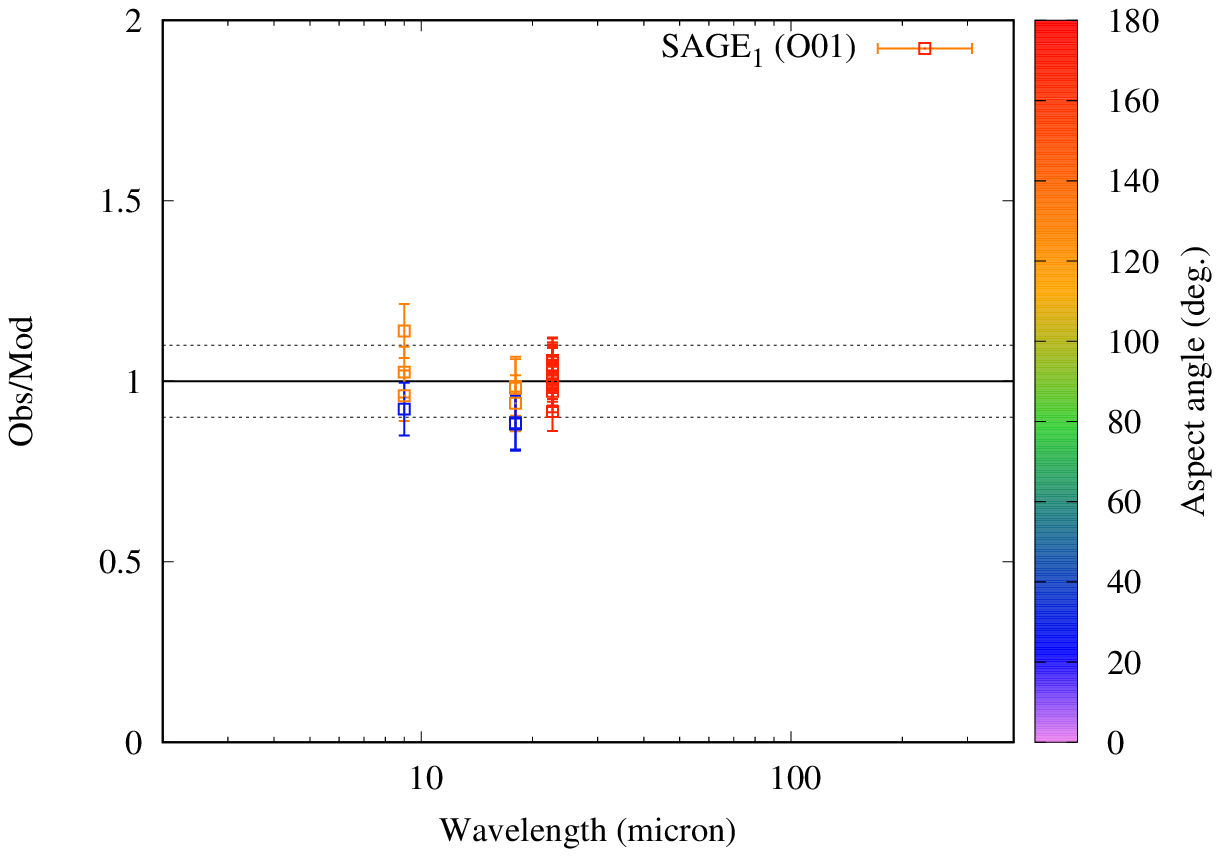}

  \includegraphics[width=0.8\linewidth]{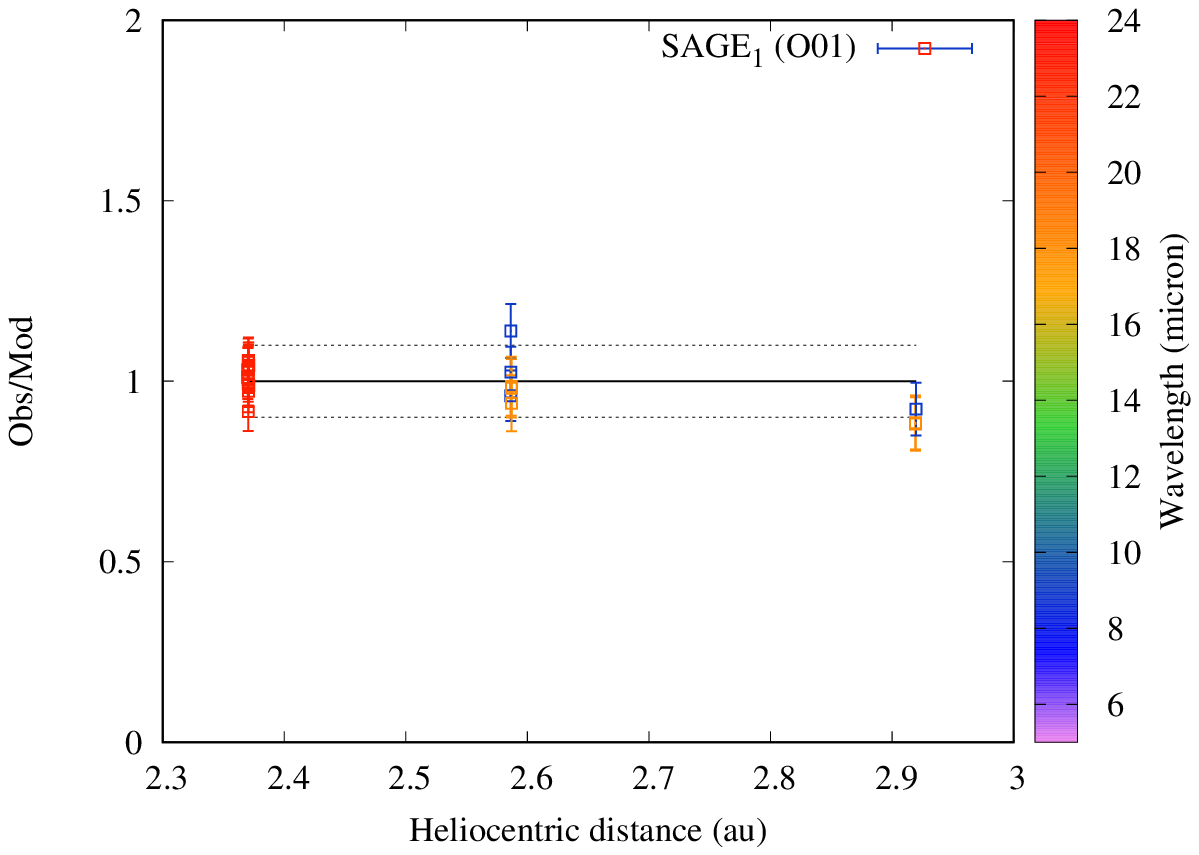}

  \includegraphics[width=0.8\linewidth]{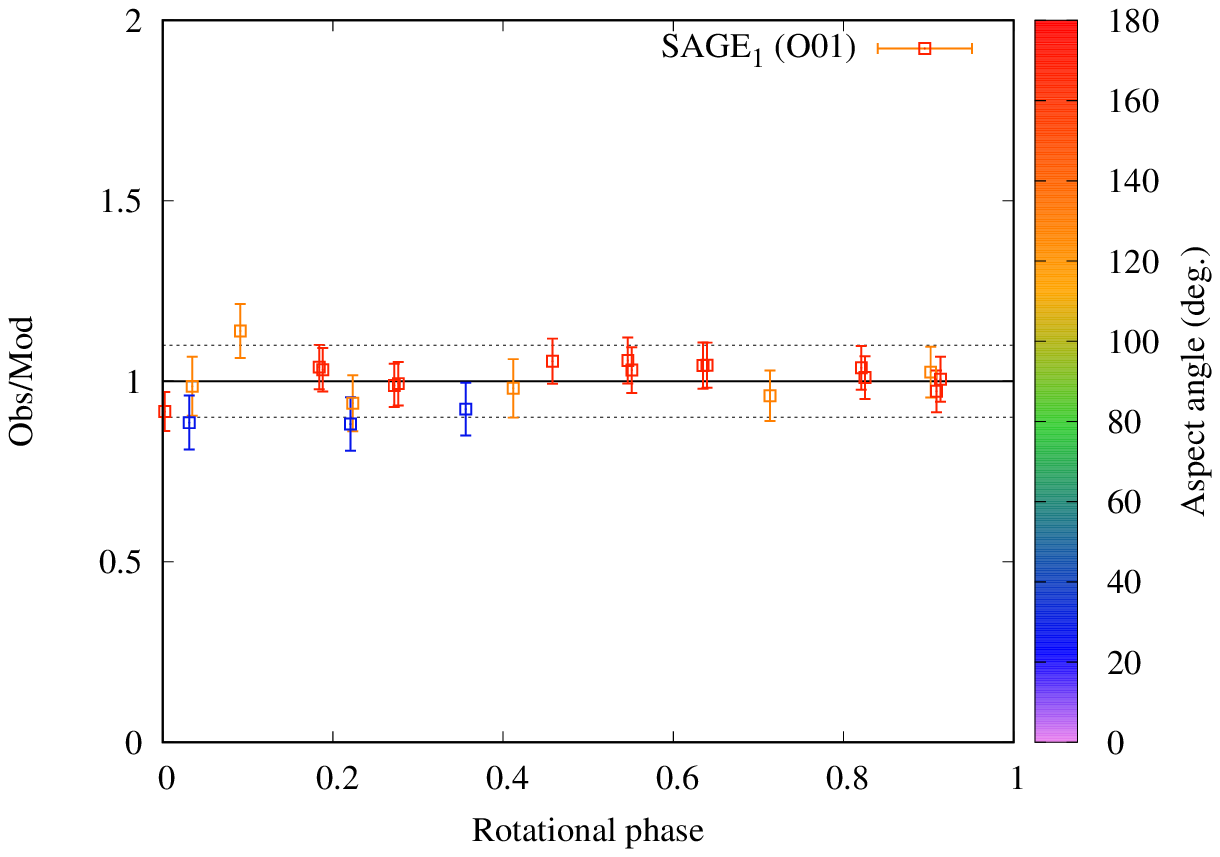}

  \includegraphics[width=0.8\linewidth]{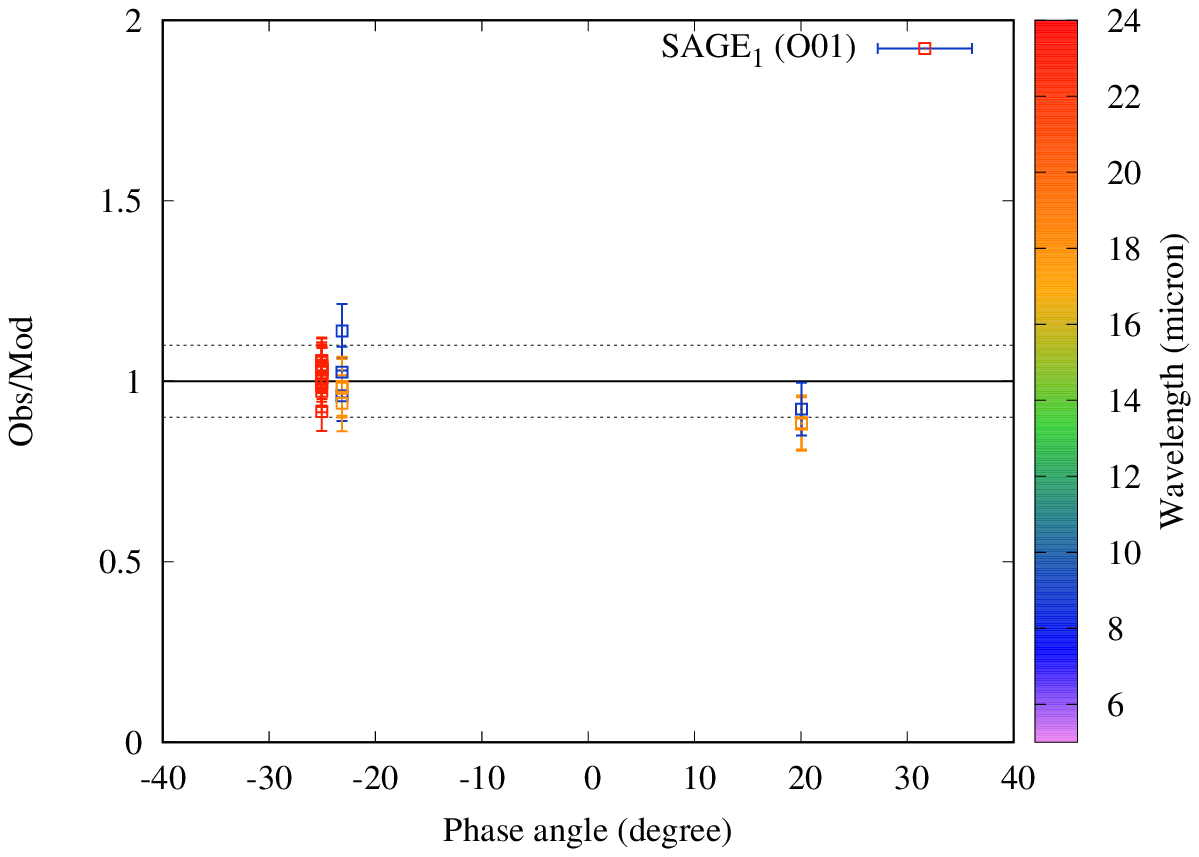}

  \caption{(64) Angelina. 
    Pole 1 was favored in this case because it provided a significantly lower
    minimum $\chi^2$. 
    The O01 label indicates that the very few MSX were clear outliers and were
    removed from the analysis. 
  }\label{fig:00064_OMR}
\end{figure}

\begin{figure}
  \centering
  
  \includegraphics[width=0.8\linewidth]{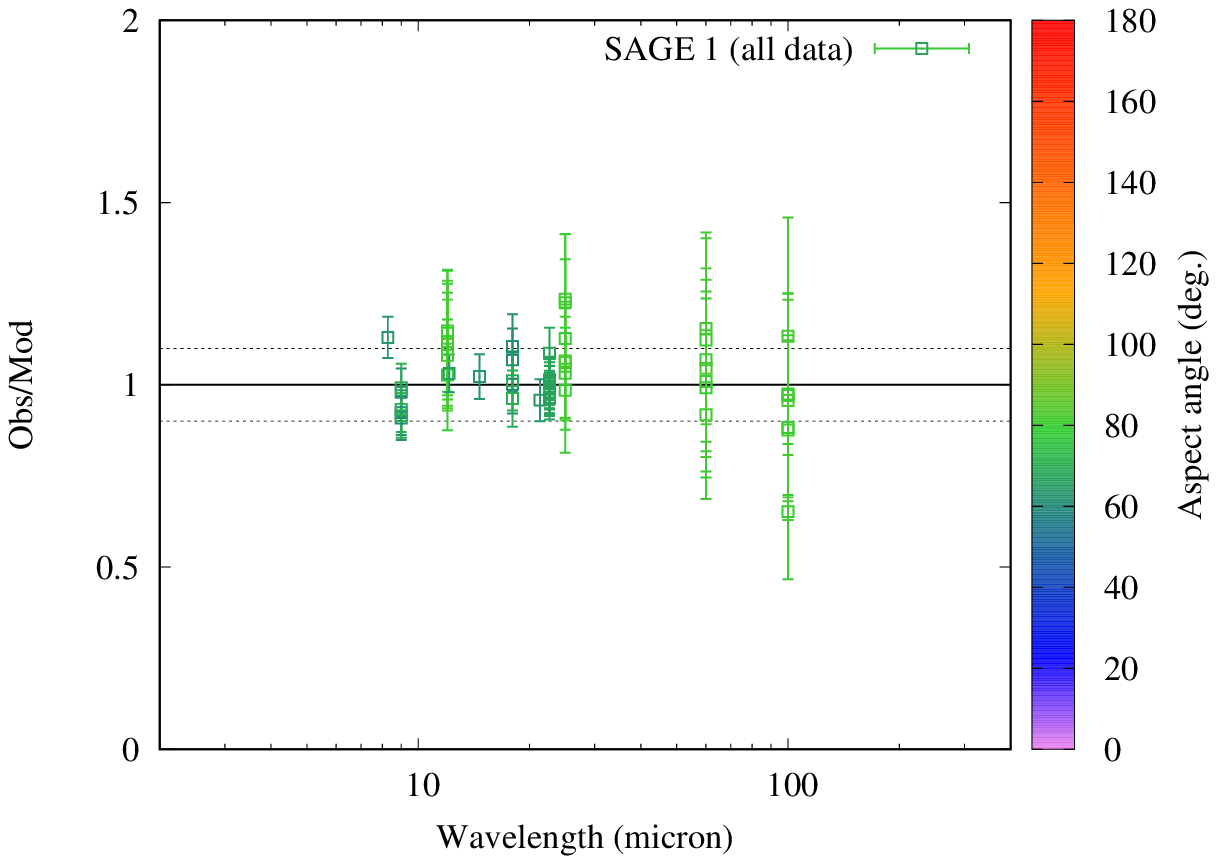}\\

  \includegraphics[width=0.8\linewidth]{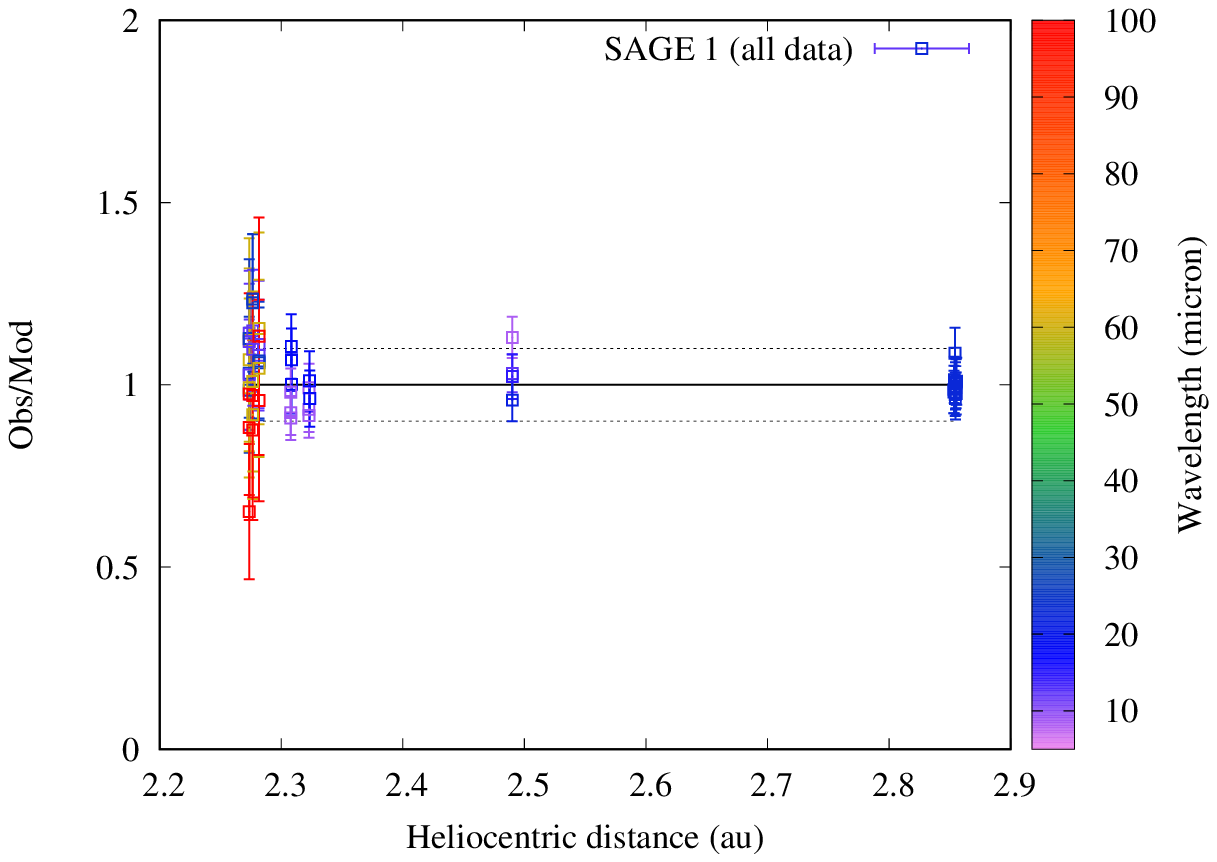}\\
  
  \includegraphics[width=0.8\linewidth]{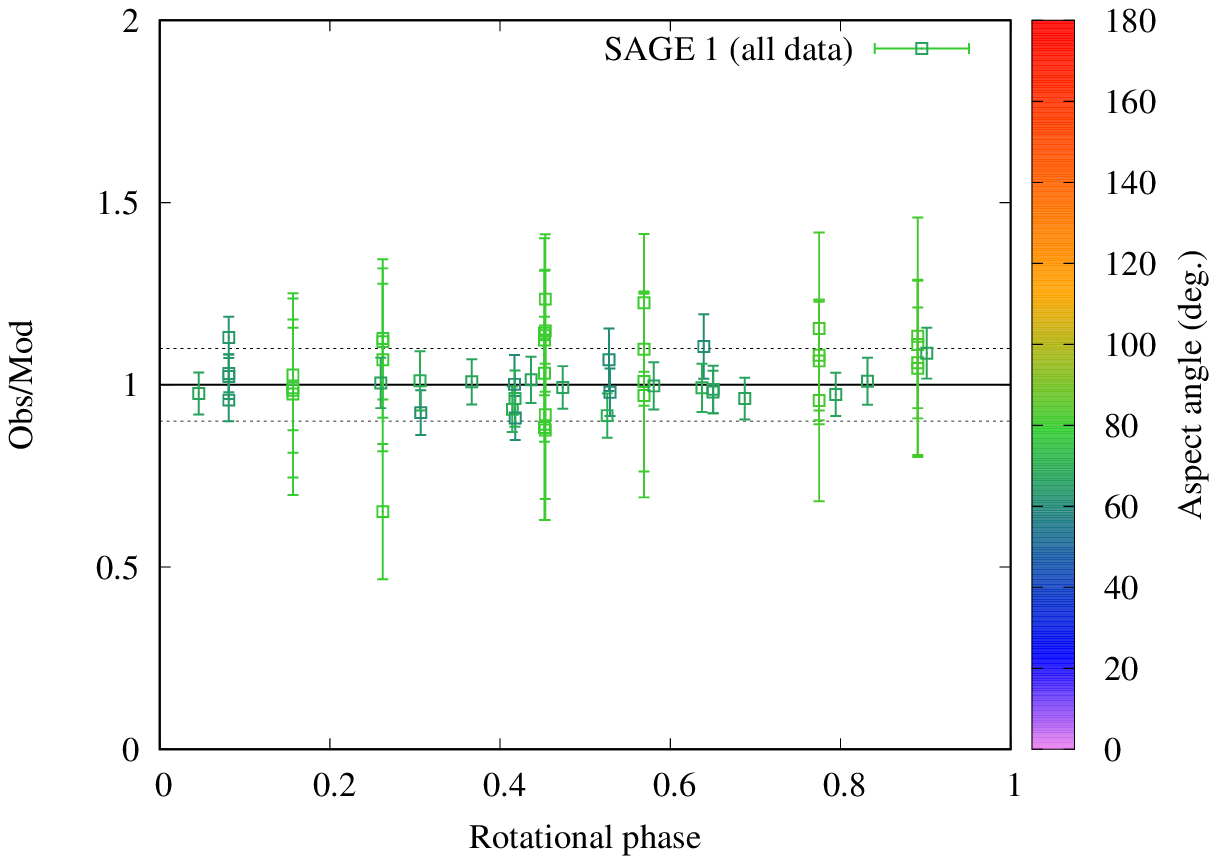}\\

  \includegraphics[width=0.8\linewidth]{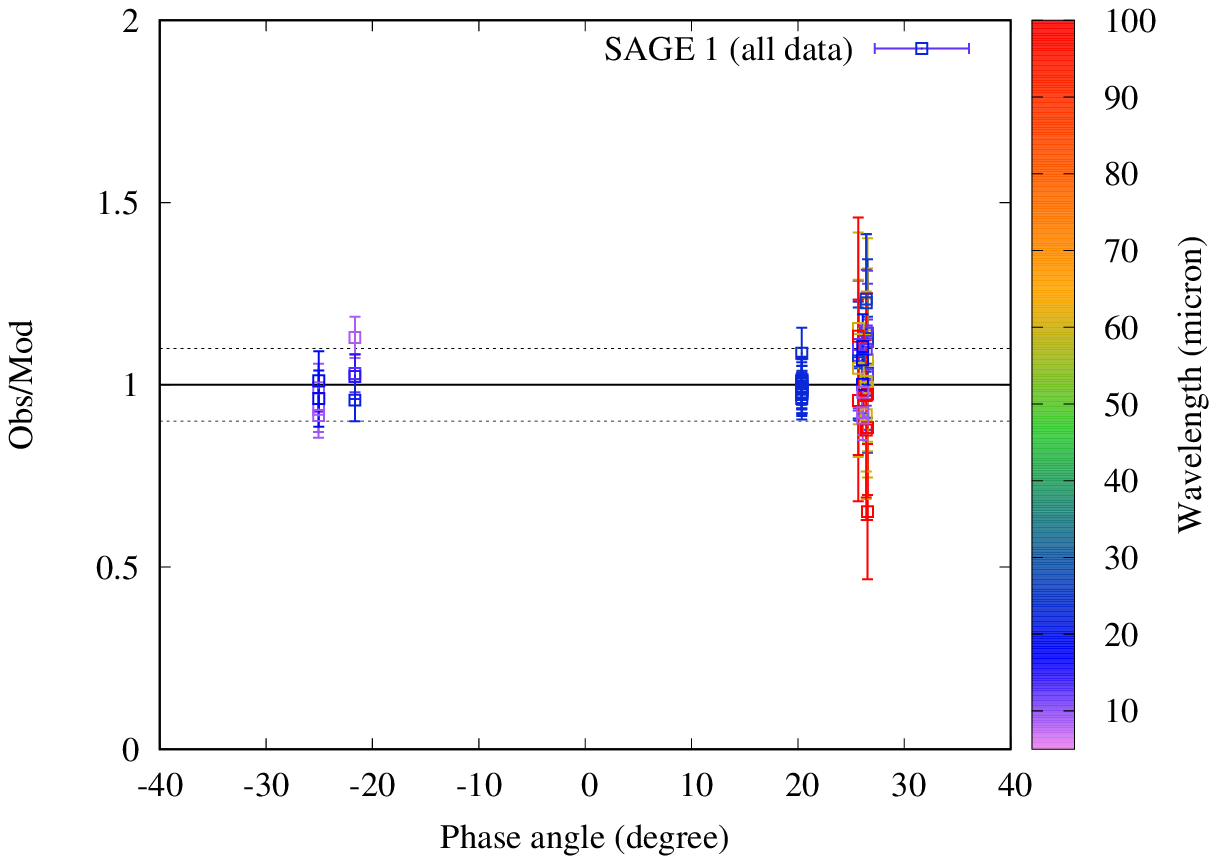}\\

  \caption{(68) Leto. The two mirror solutions fitted the data statistically
    equally well. 
  }\label{fig:00068_OMR}

\end{figure}

\begin{figure}
  \centering
  
  \includegraphics[width=0.8\linewidth]{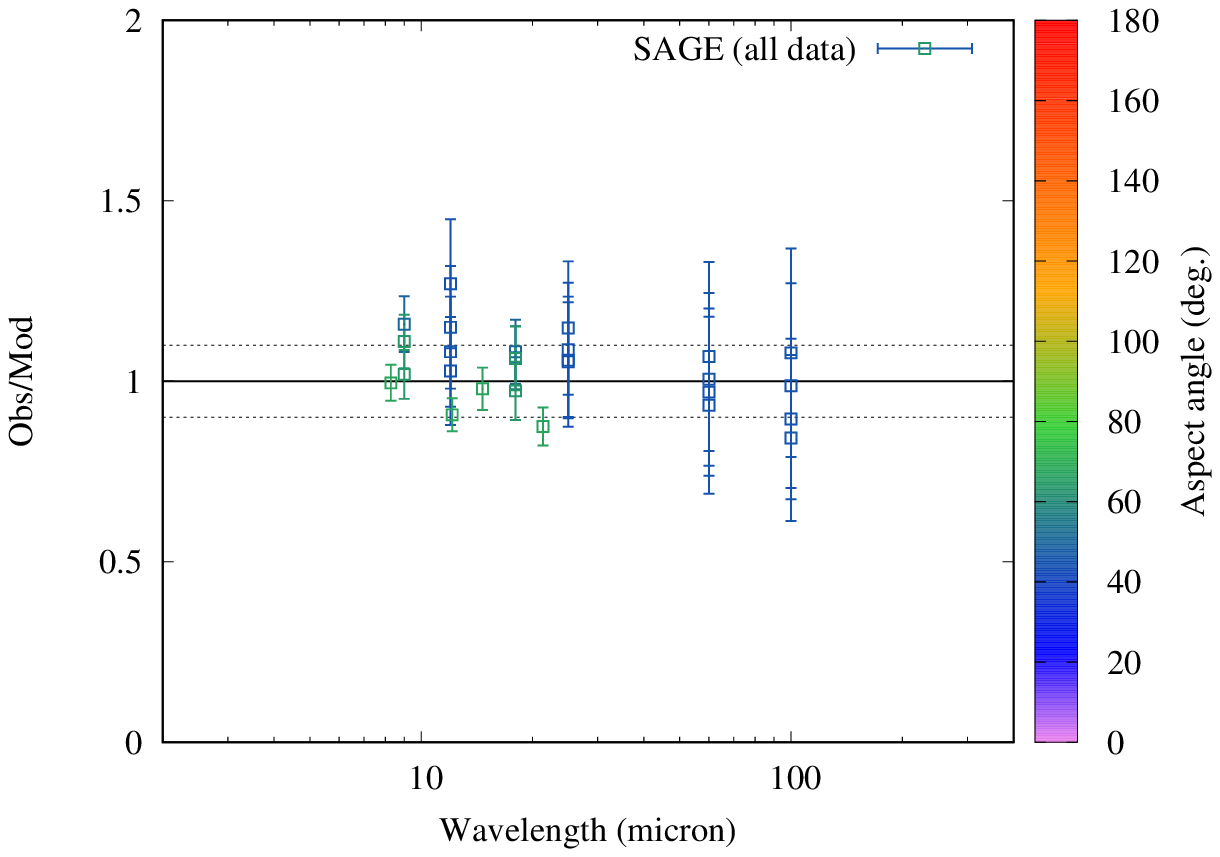}
  
  \includegraphics[width=0.8\linewidth]{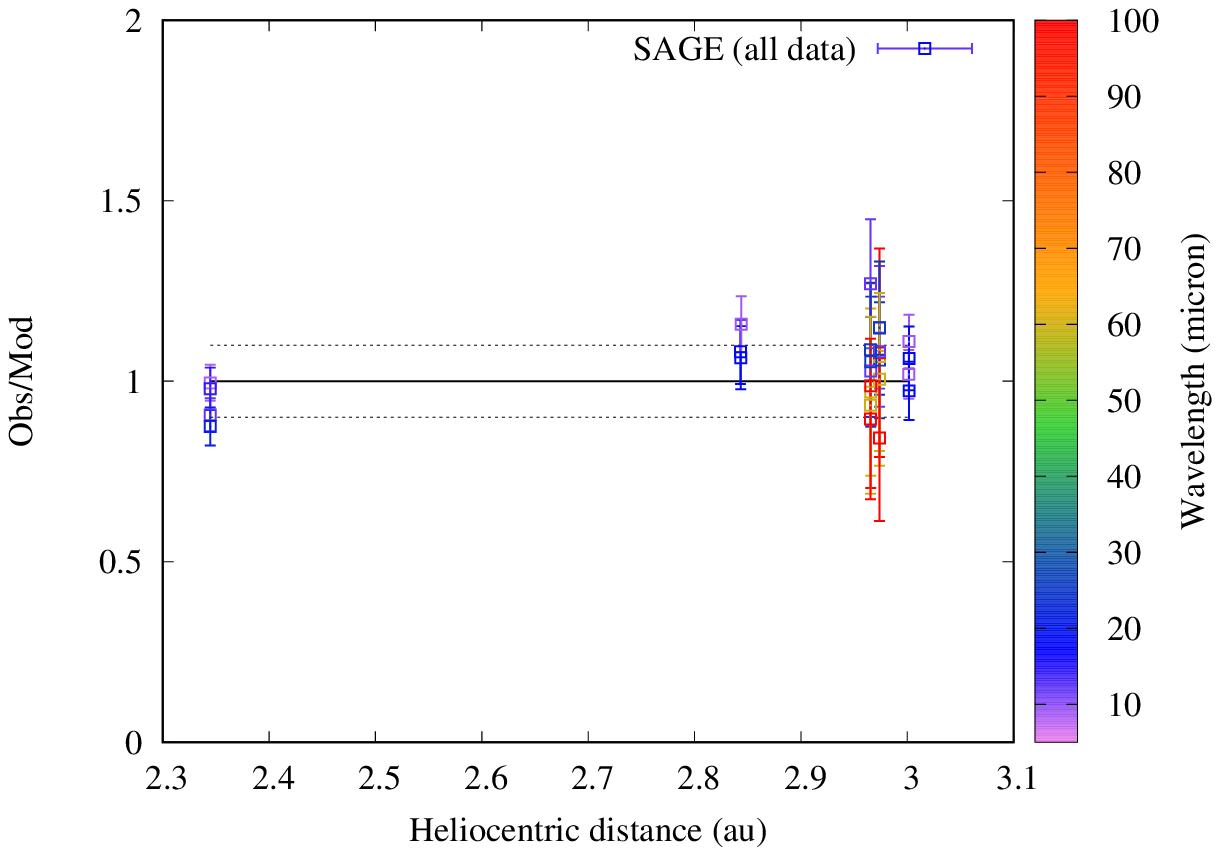}
  
  \includegraphics[width=0.8\linewidth]{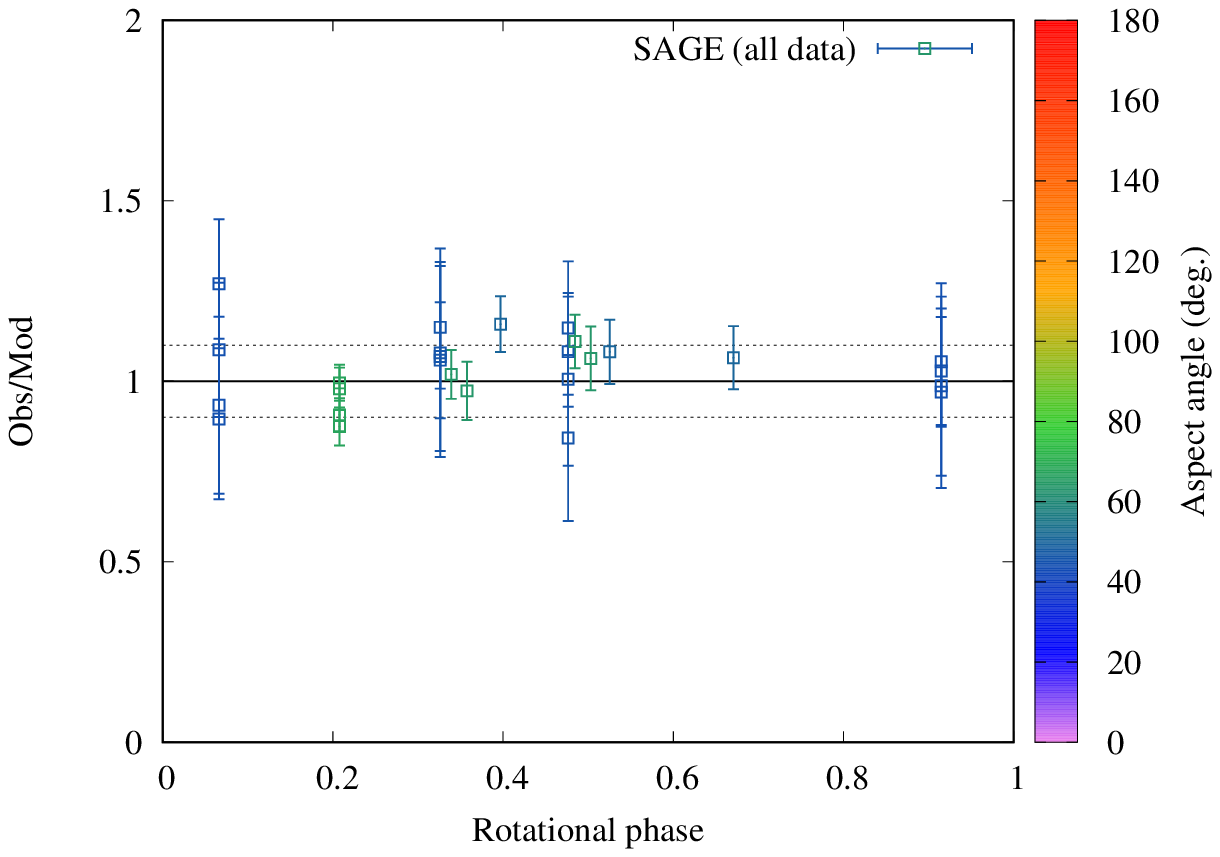}
  
  \includegraphics[width=0.8\linewidth]{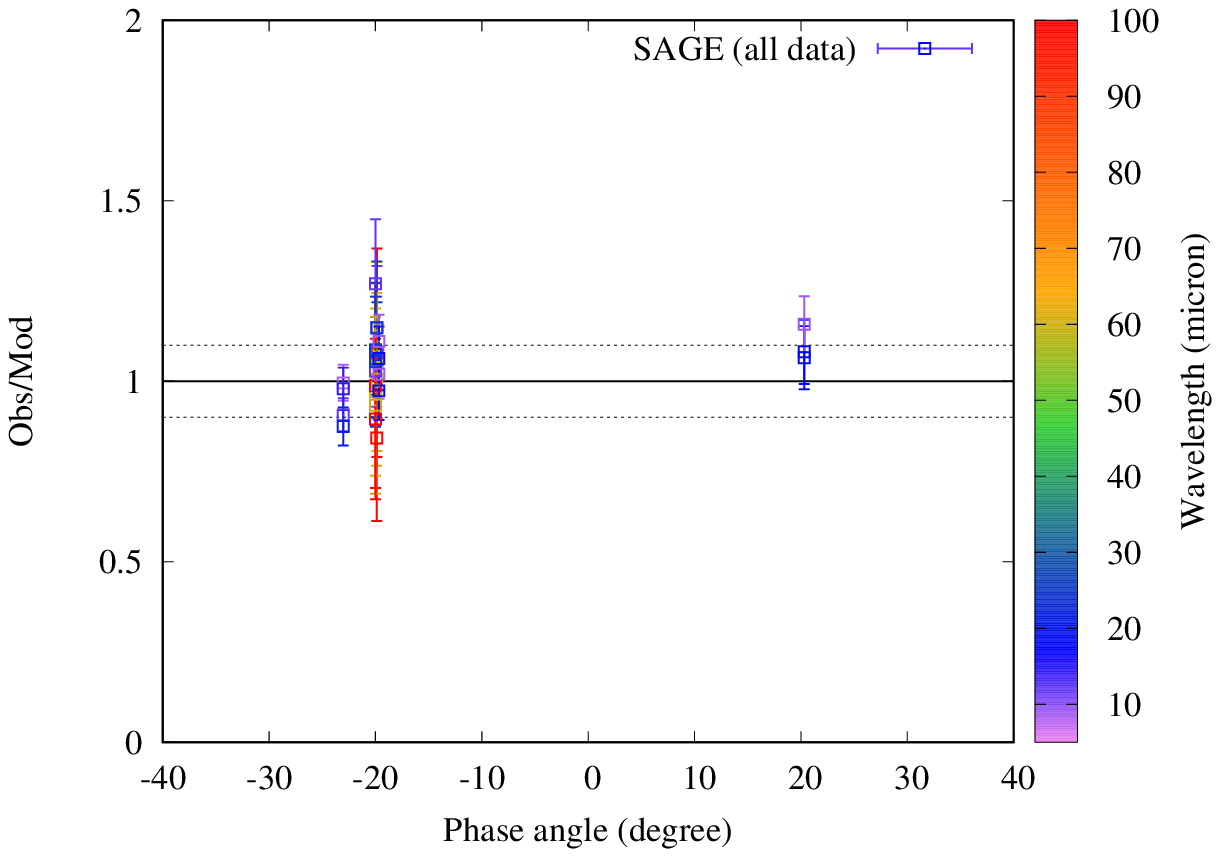}
  
  \caption{(89) Julia.
    The SAGE model provided a formally acceptable fit to the data
    (See Table~\ref{tab:tpm}) but the optimum thermal inertia (150\,SI units)
    is higher than expected for such a large main-belt asteroid. It is probably
    an artefact and manifests itself in the strong slope in the wavelength plot.
    The bias could be caused by two possible factors:
    We did not consider the dependence of thermal inertia with temperature 
    (see e.g., Marsset et al. 2017, Rozitis et al. 2018) and the data  were taken
    over a wide range of heliocentric distances; the thermal inertia
    is not well constrained because we have very few observations at
    positive phase angles. 
  }\label{fig:00089_OMR}
\end{figure}

\begin{figure}
  \centering
  
  \includegraphics[width=0.8\linewidth]{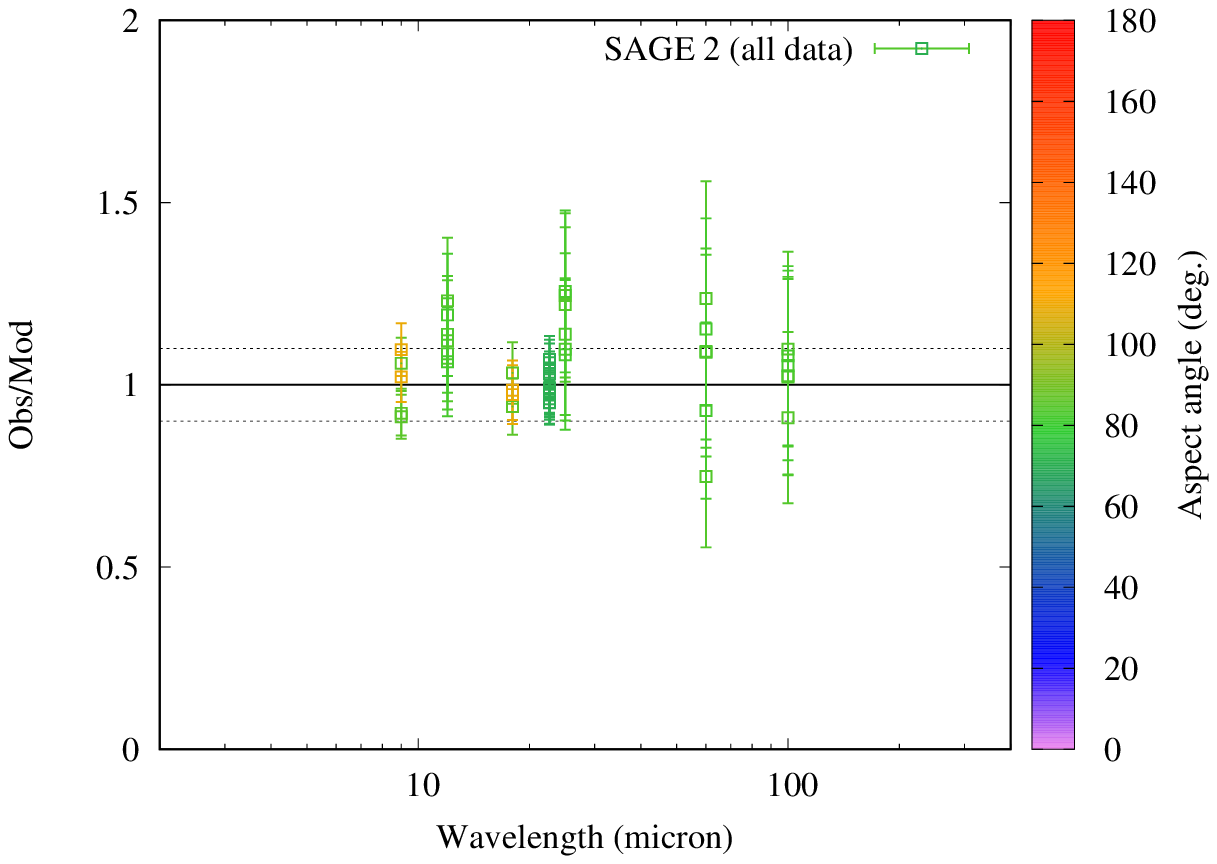}
  
  \includegraphics[width=0.8\linewidth]{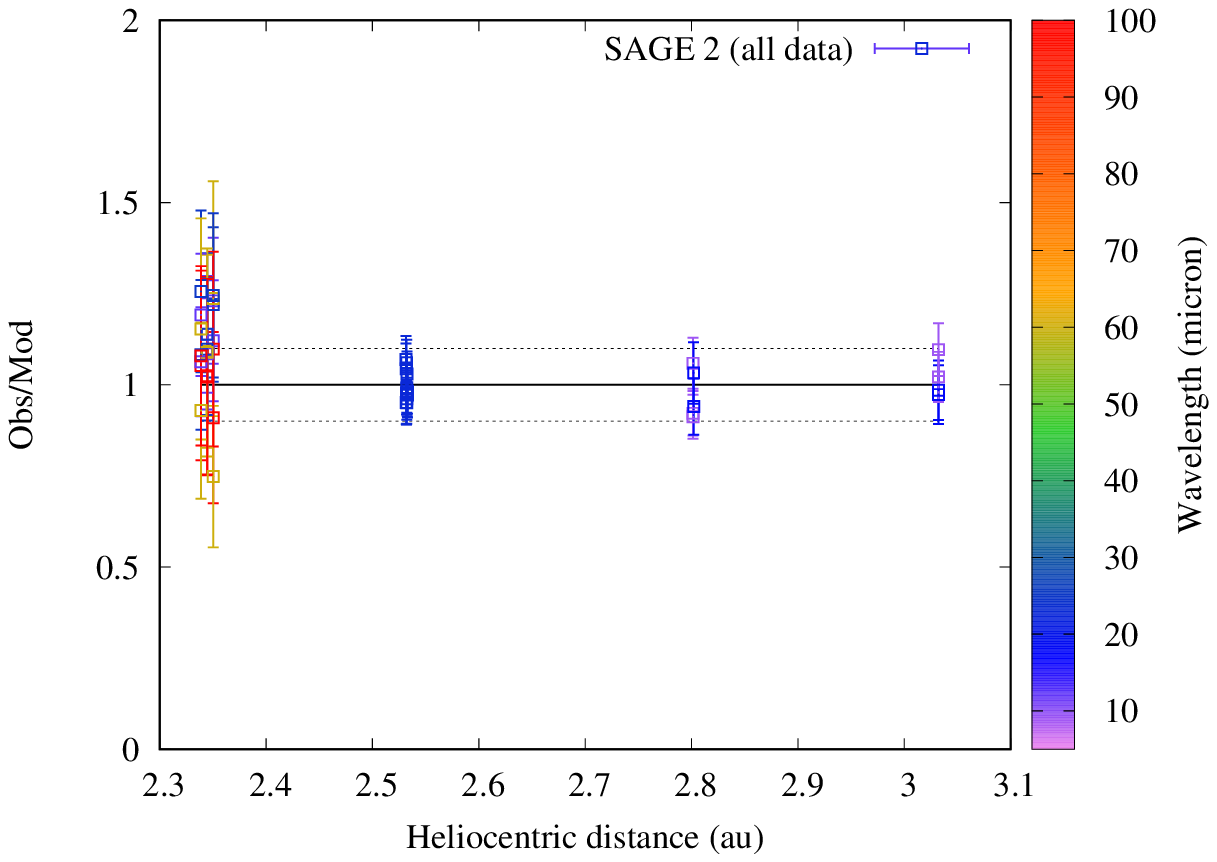}
  
  \includegraphics[width=0.8\linewidth]{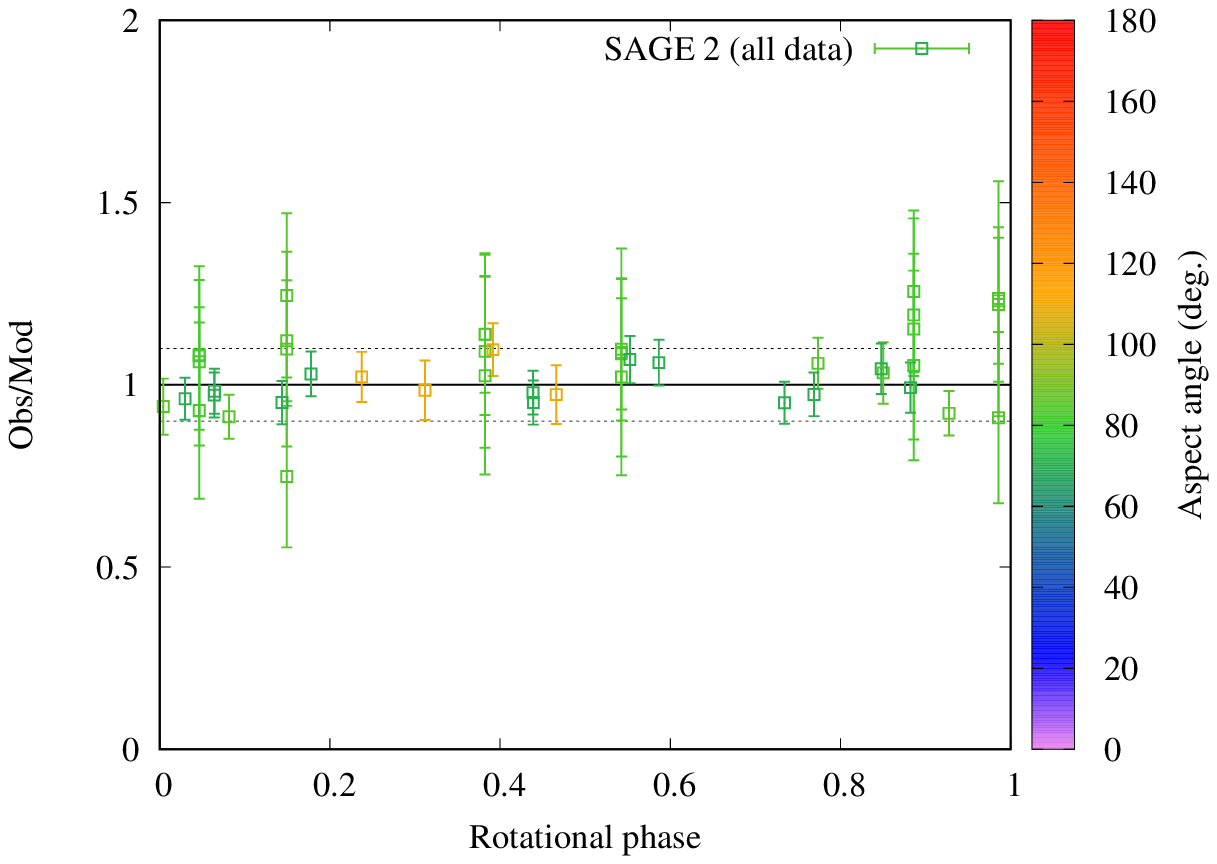}
  
  \includegraphics[width=0.8\linewidth]{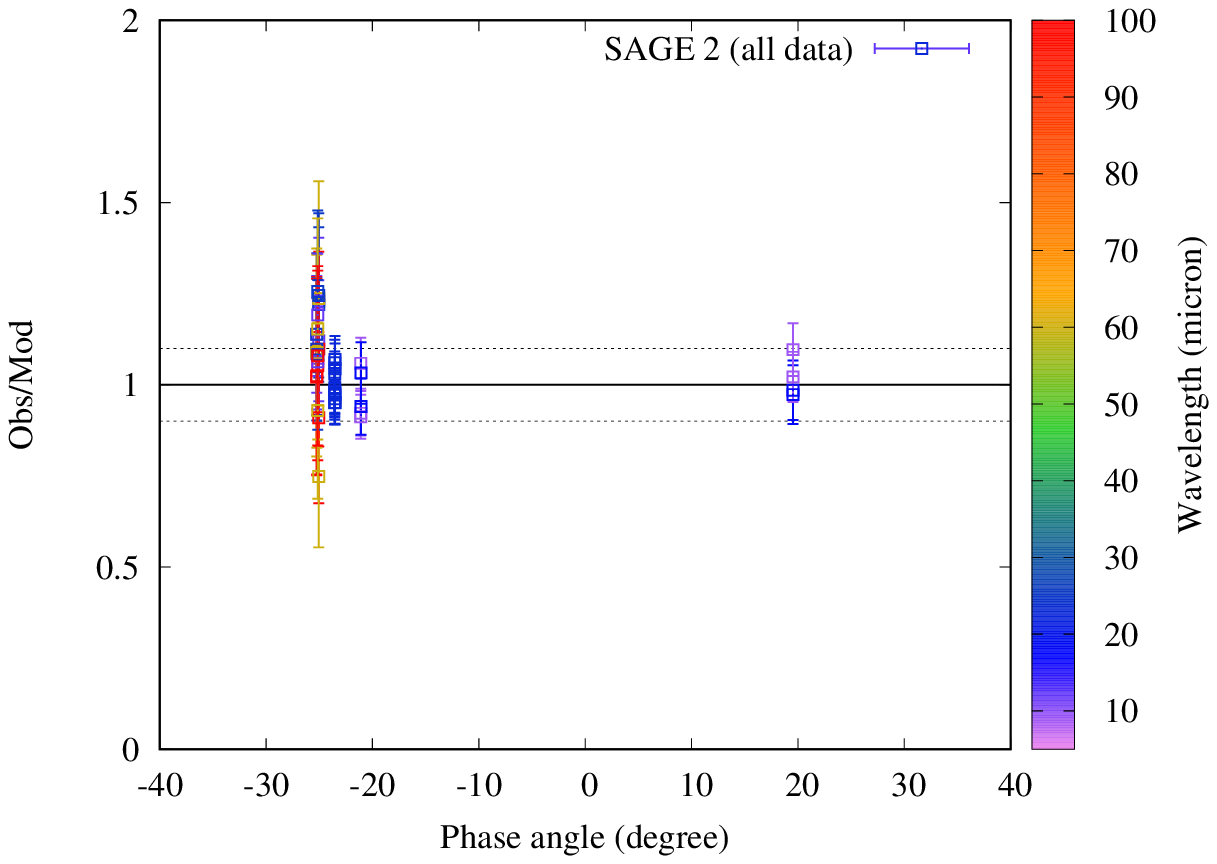}

 \caption{(114) Kassandra. 
  }\label{fig:00114_OMR}
\end{figure}

\begin{figure}
  \centering
  
  \includegraphics[width=0.8\linewidth]{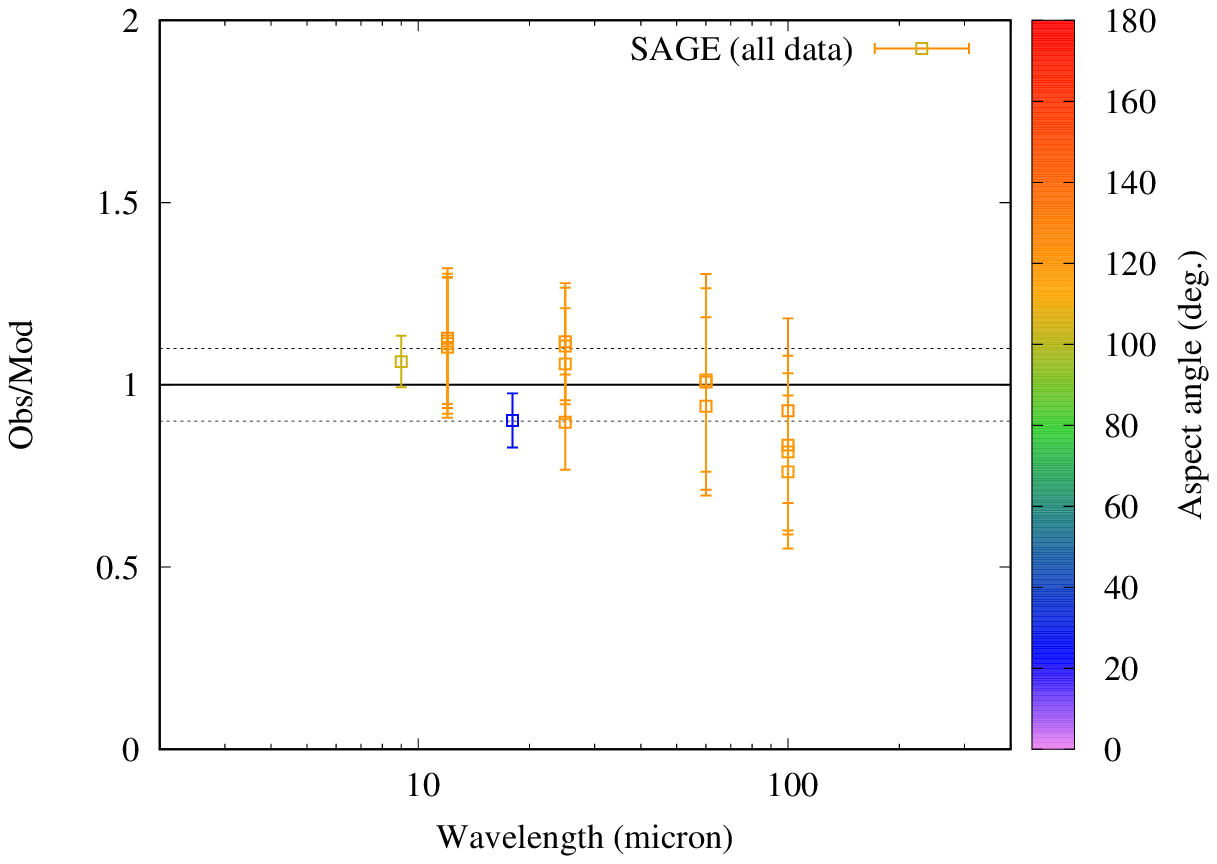}
  
  \includegraphics[width=0.8\linewidth]{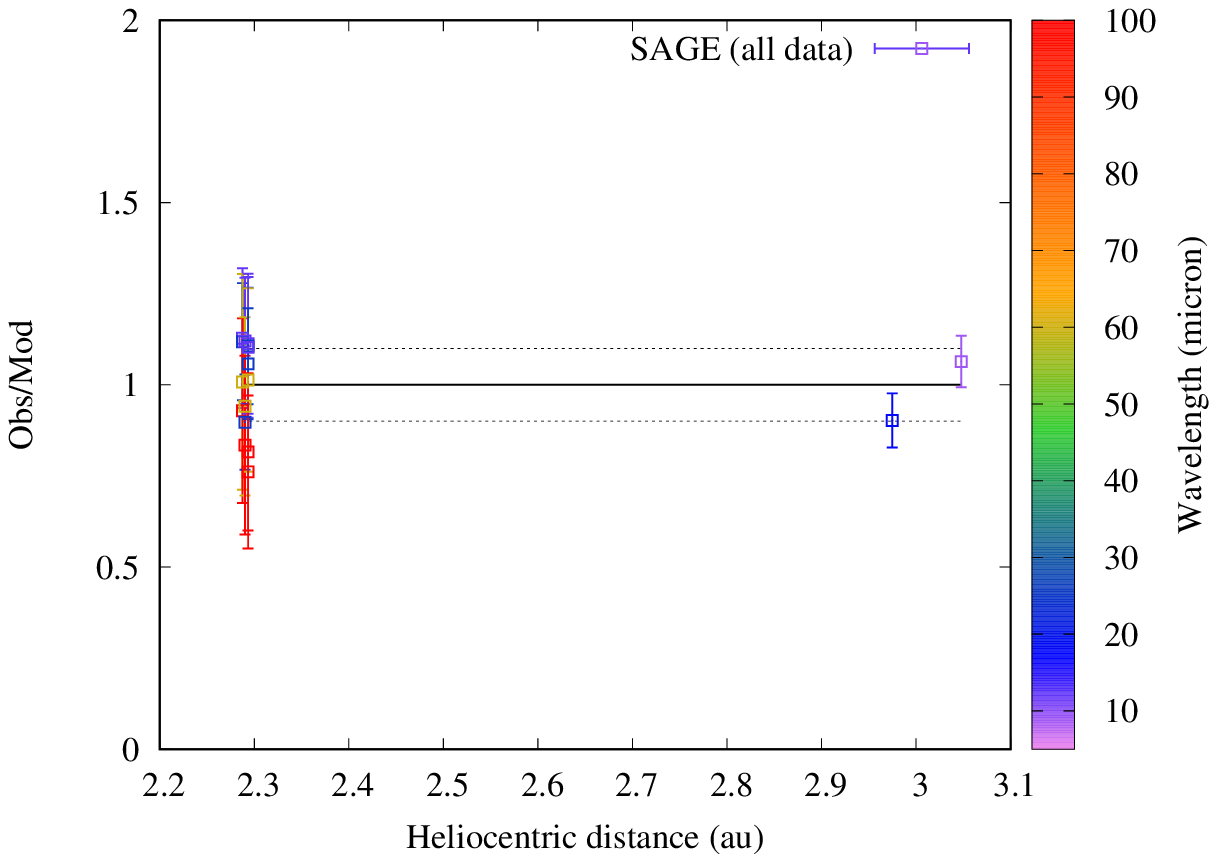}
  
  \includegraphics[width=0.8\linewidth]{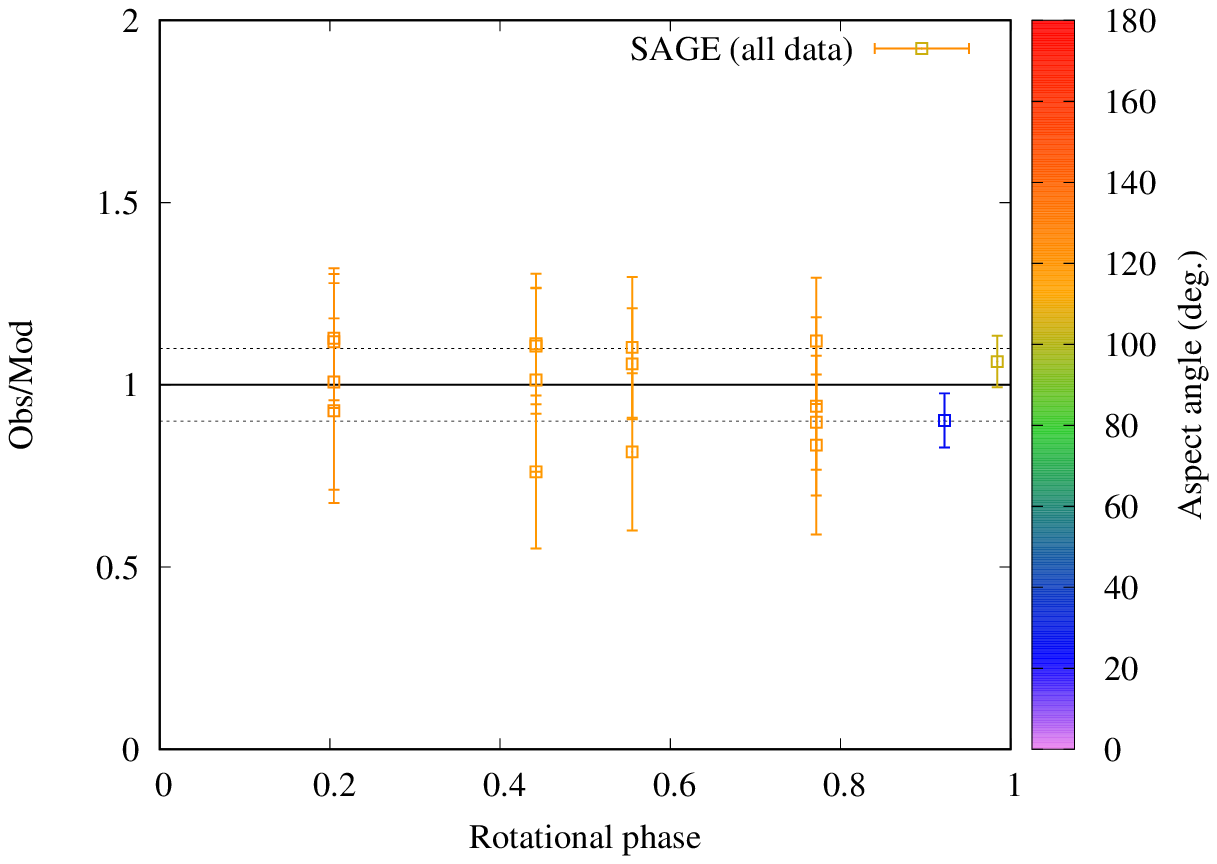}
  
  \includegraphics[width=0.8\linewidth]{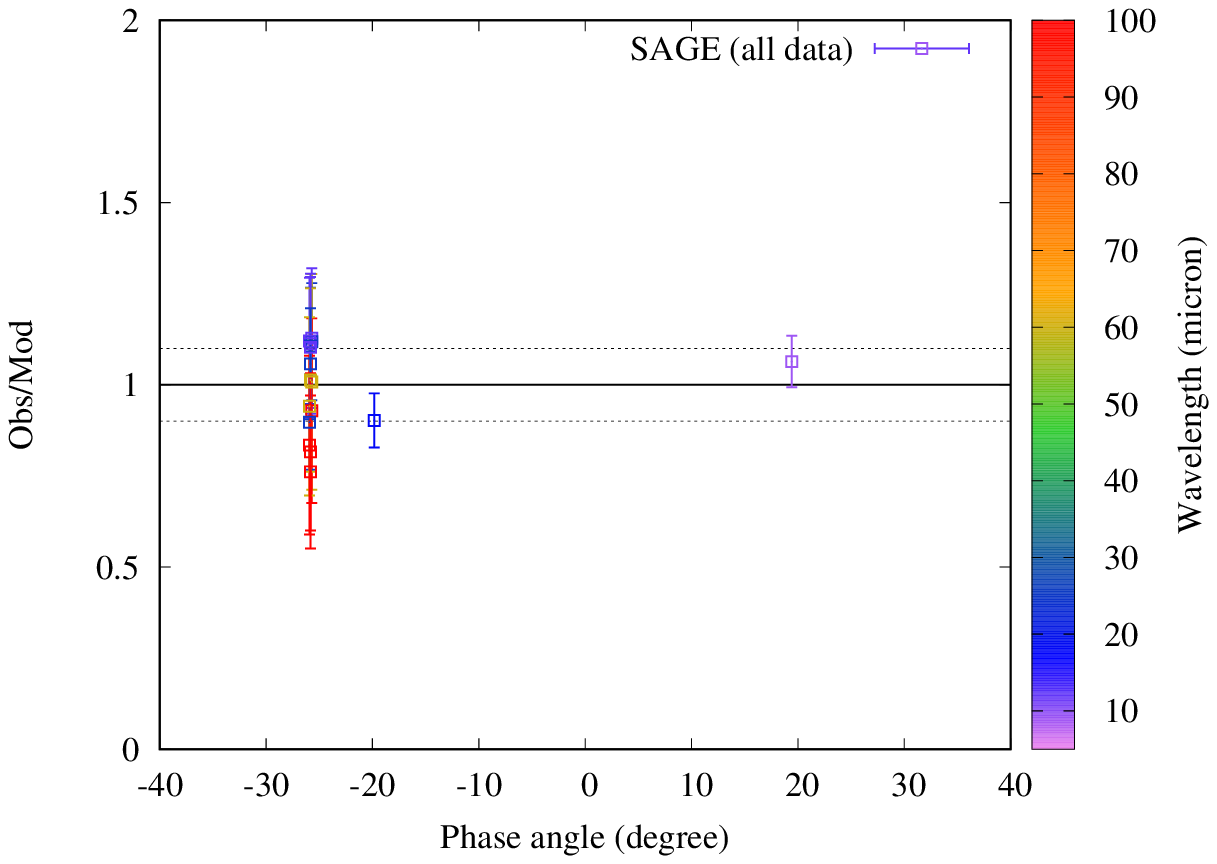}

  \caption{(145) Adeona. 
  }\label{fig:00145_OMR}
\end{figure}

\begin{figure}
  \centering
  
  \includegraphics[width=0.8\linewidth]{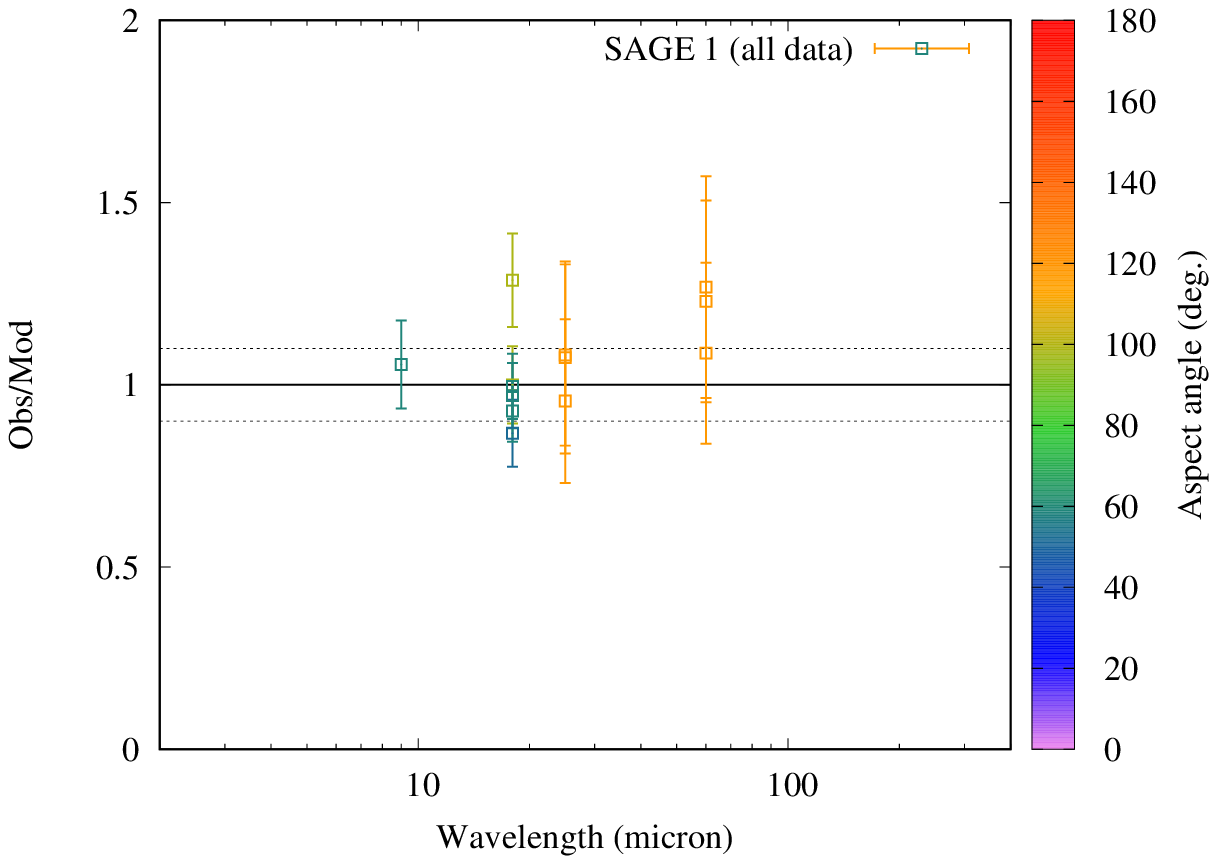}
  
  \includegraphics[width=0.8\linewidth]{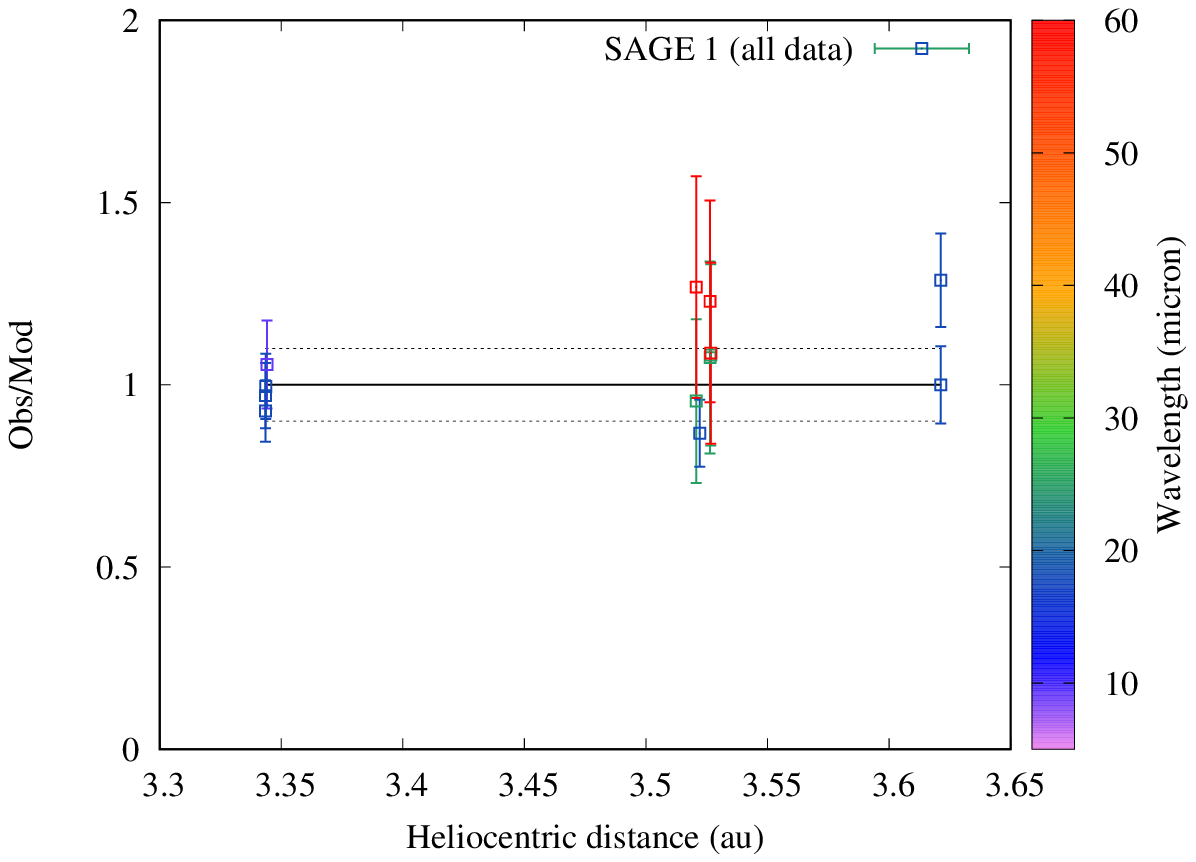}
  
  \includegraphics[width=0.8\linewidth]{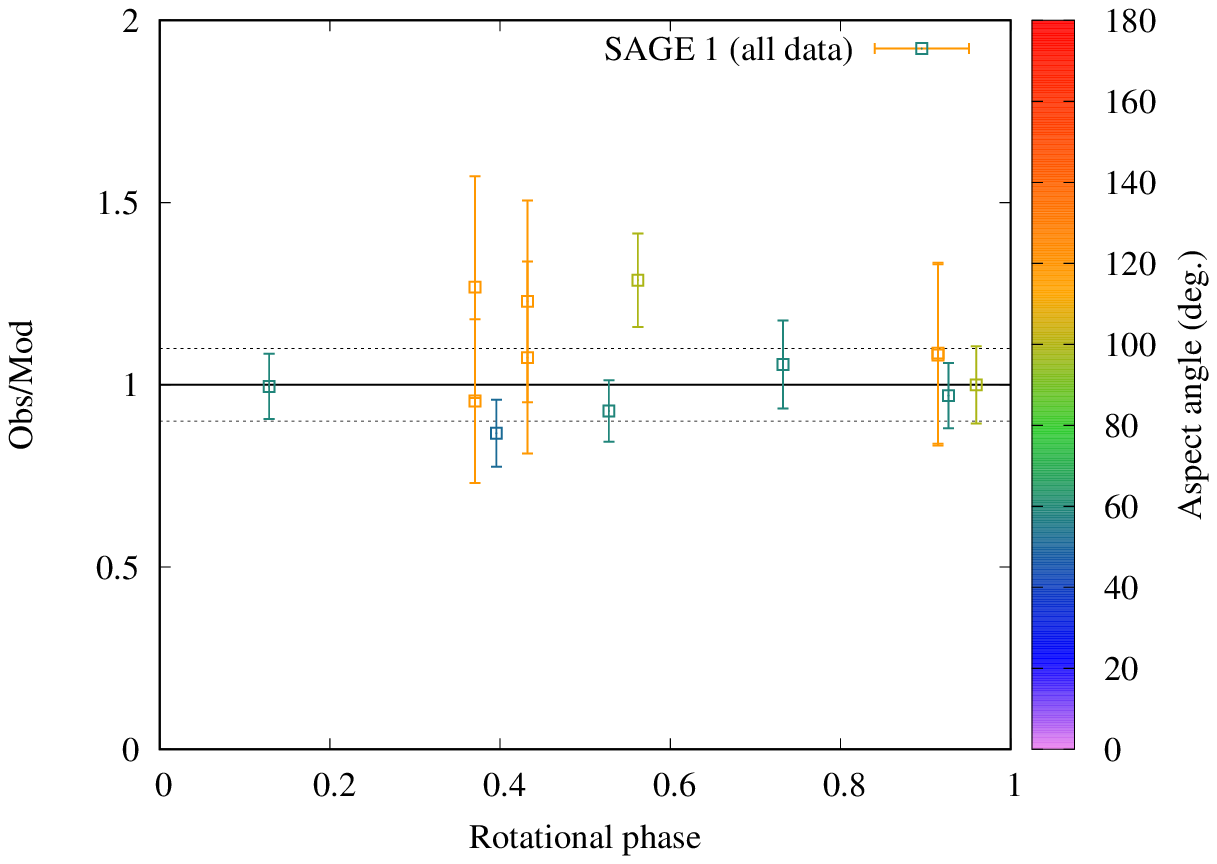}
  
  \includegraphics[width=0.8\linewidth]{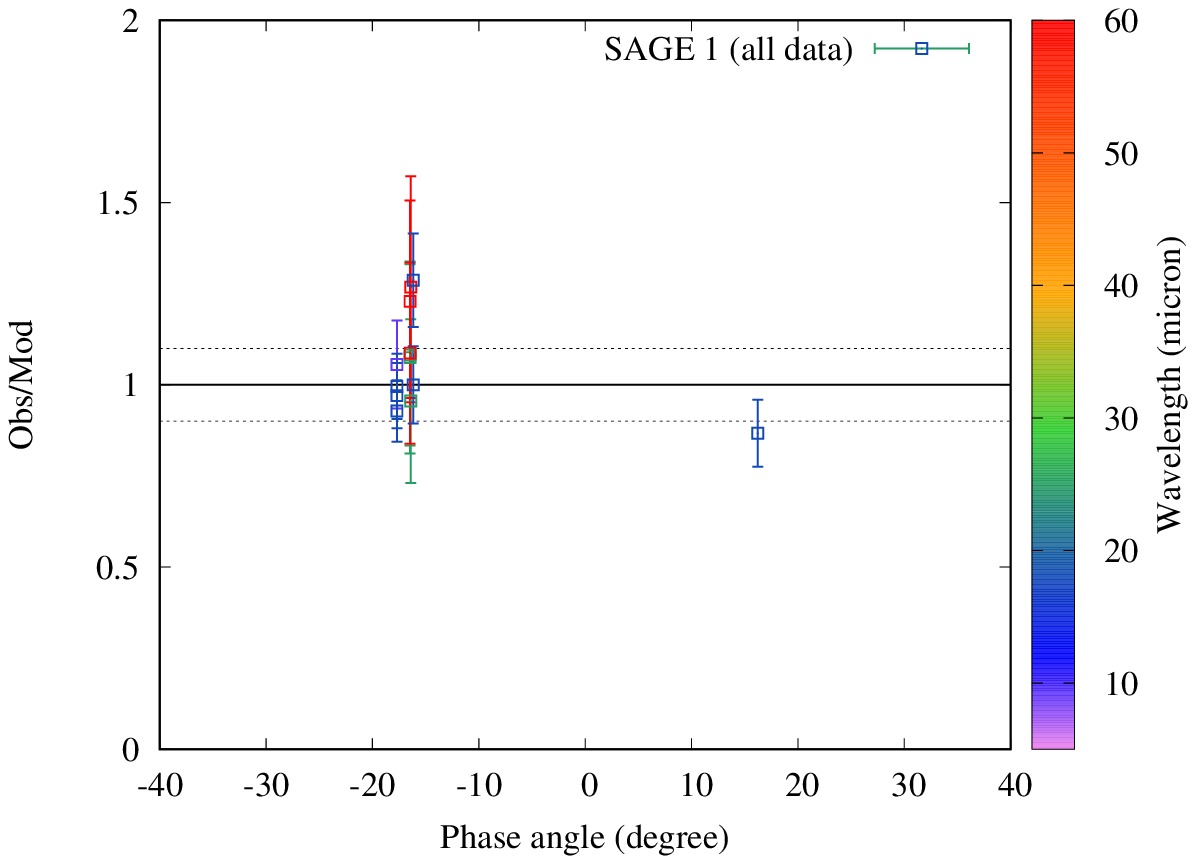}
  
  \caption{(297) Caecilia. 
    There is not good phase angle coverage. There were not enough data to
    provide realistic error bars for the size. More thermal IR data are
    clearly needed. 
  }\label{fig:00297_OMR}
\end{figure}

\begin{figure}
  \centering
  
  \includegraphics[width=0.8\linewidth]{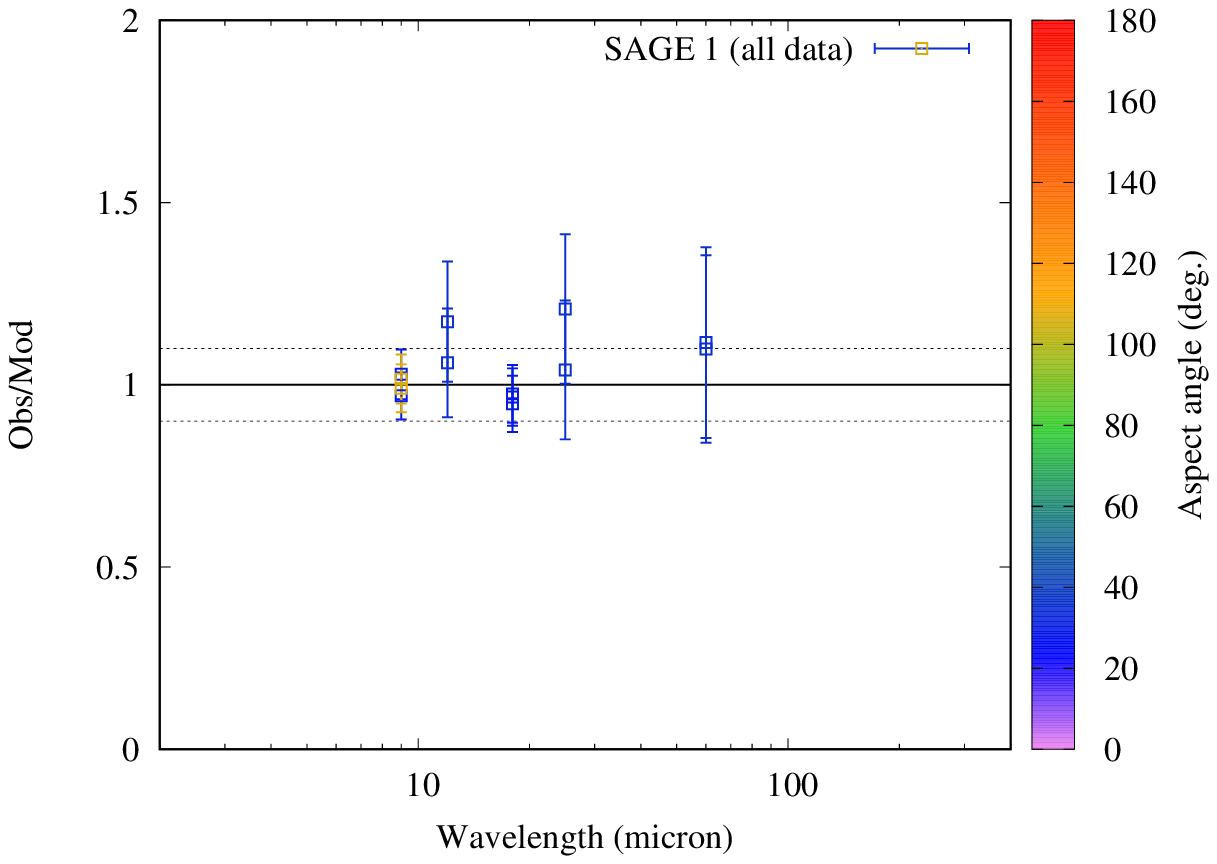}
  
  \includegraphics[width=0.8\linewidth]{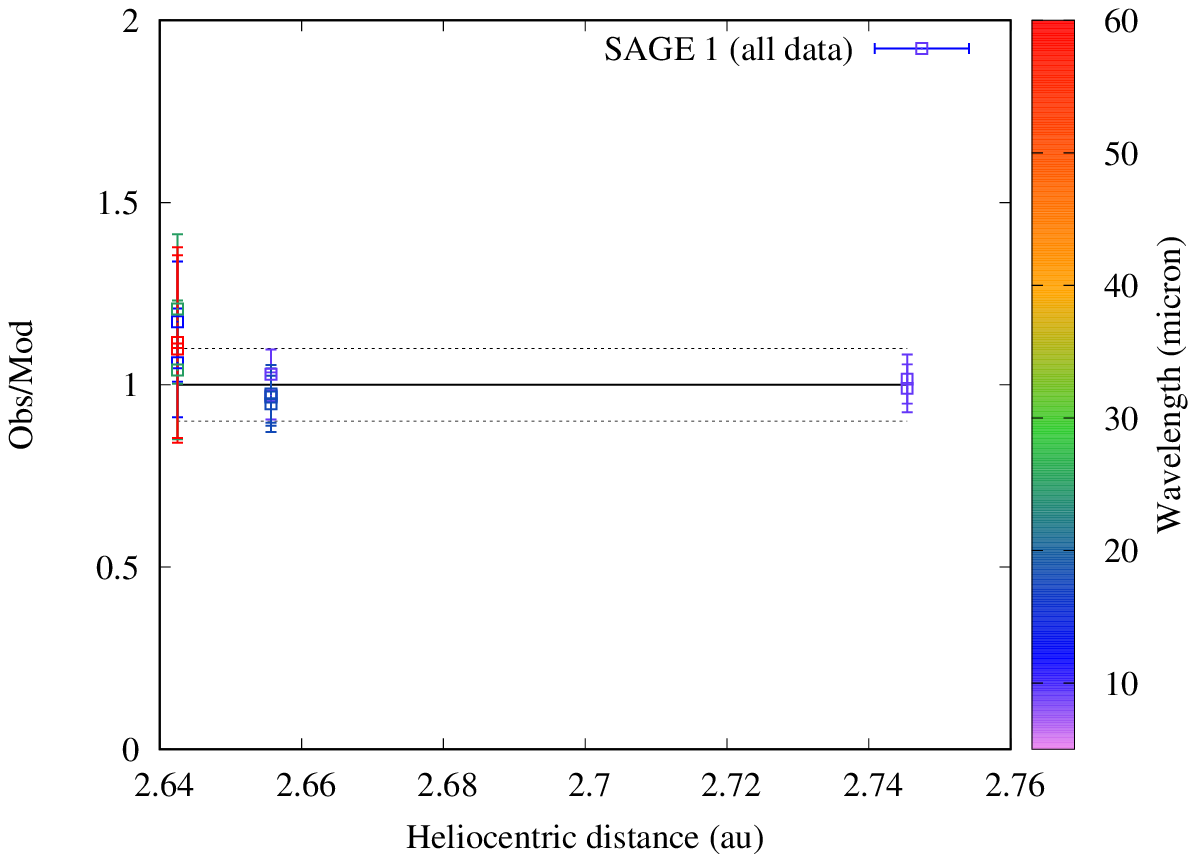}
  
  \includegraphics[width=0.8\linewidth]{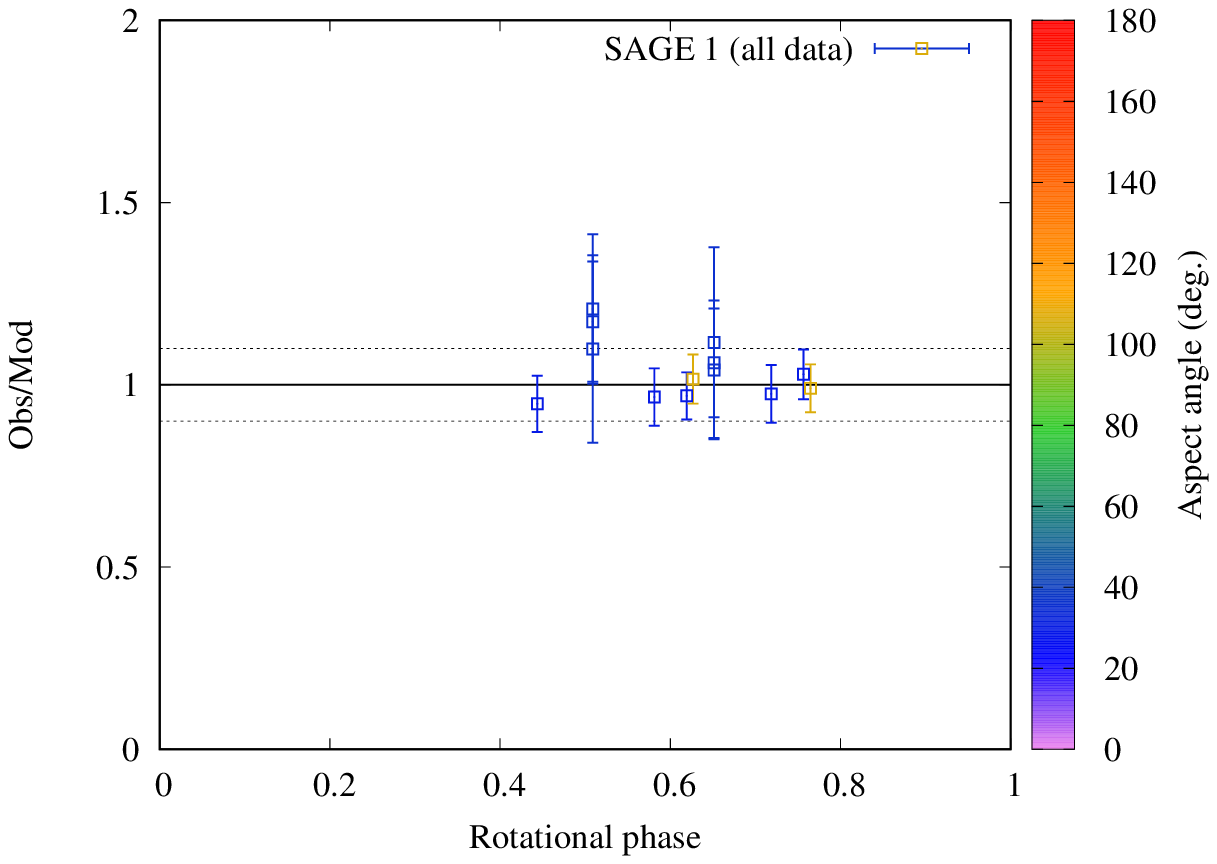}
  
  \includegraphics[width=0.8\linewidth]{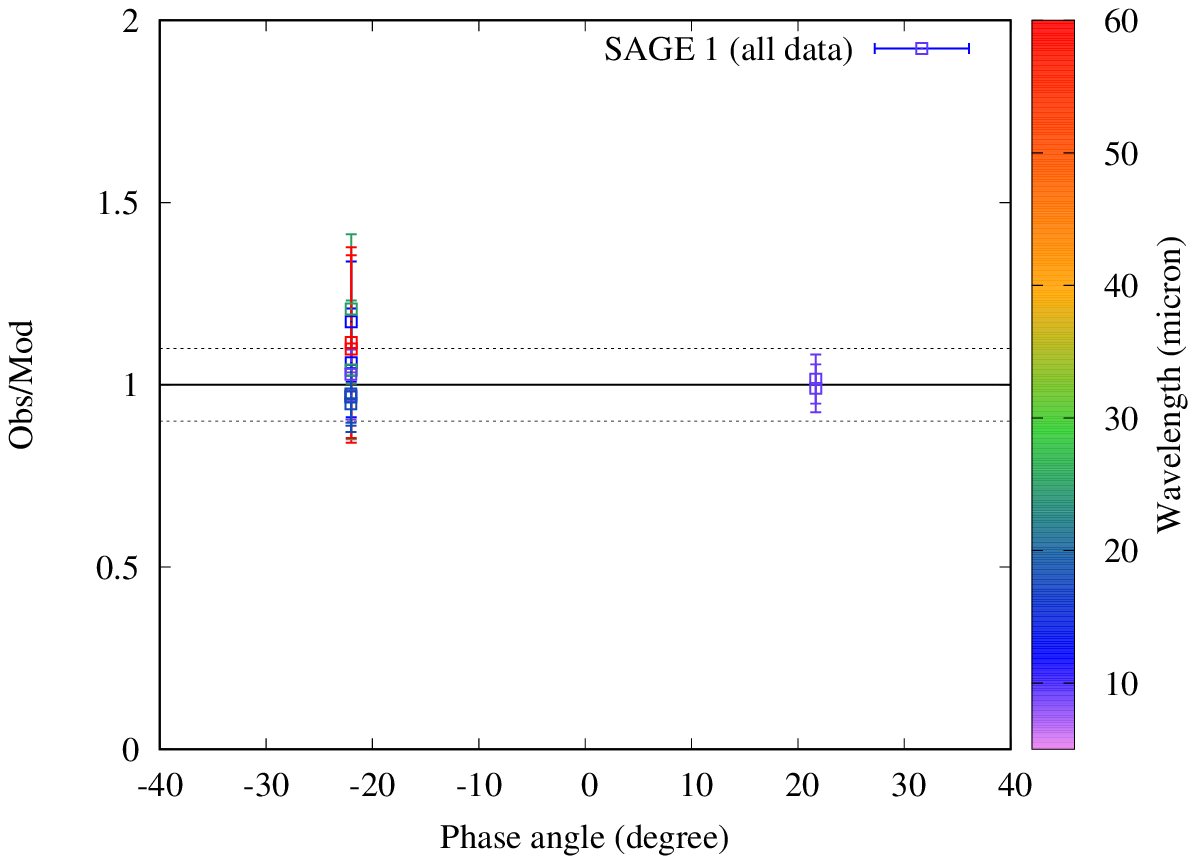}

  \caption{(308) Polyxo. 
  }\label{fig:00308_OMR}
\end{figure}

\begin{figure}
  \centering
  
  \includegraphics[width=0.8\linewidth]{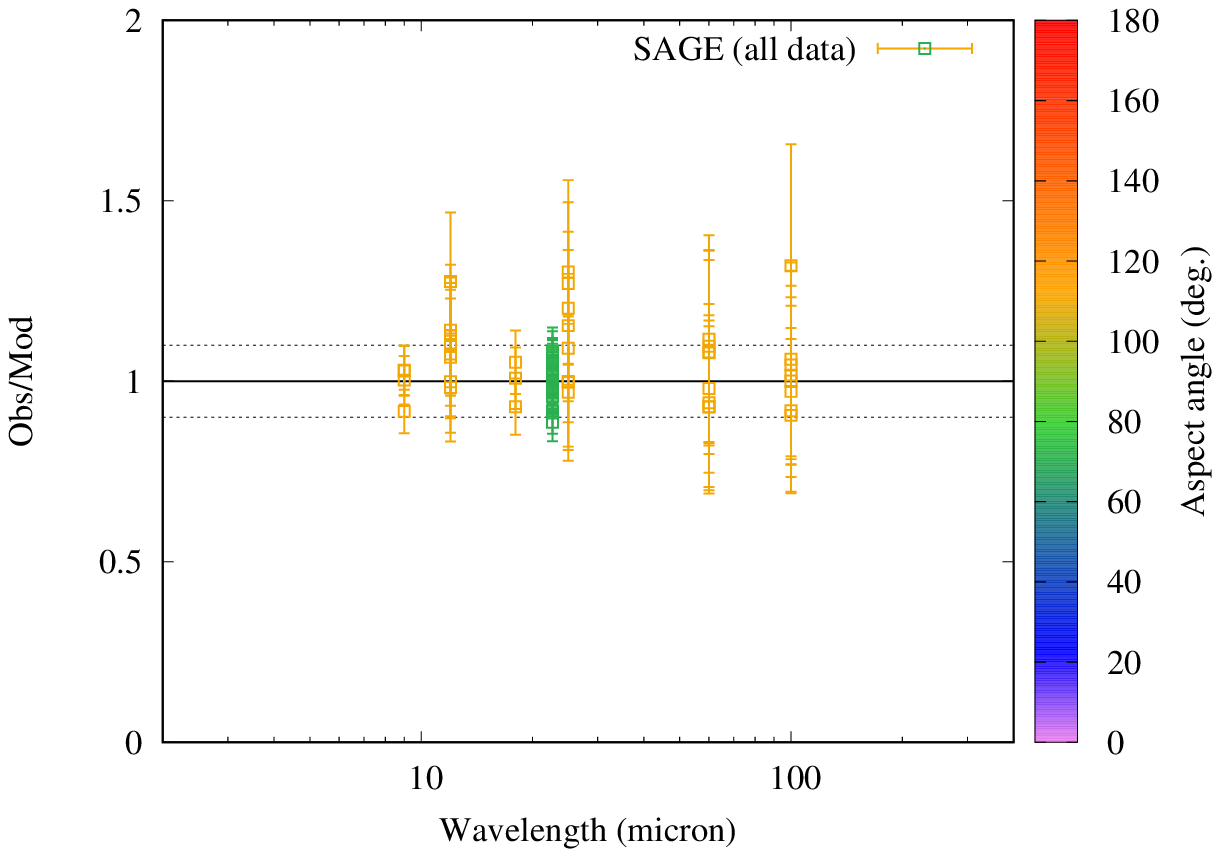}
  
  \includegraphics[width=0.8\linewidth]{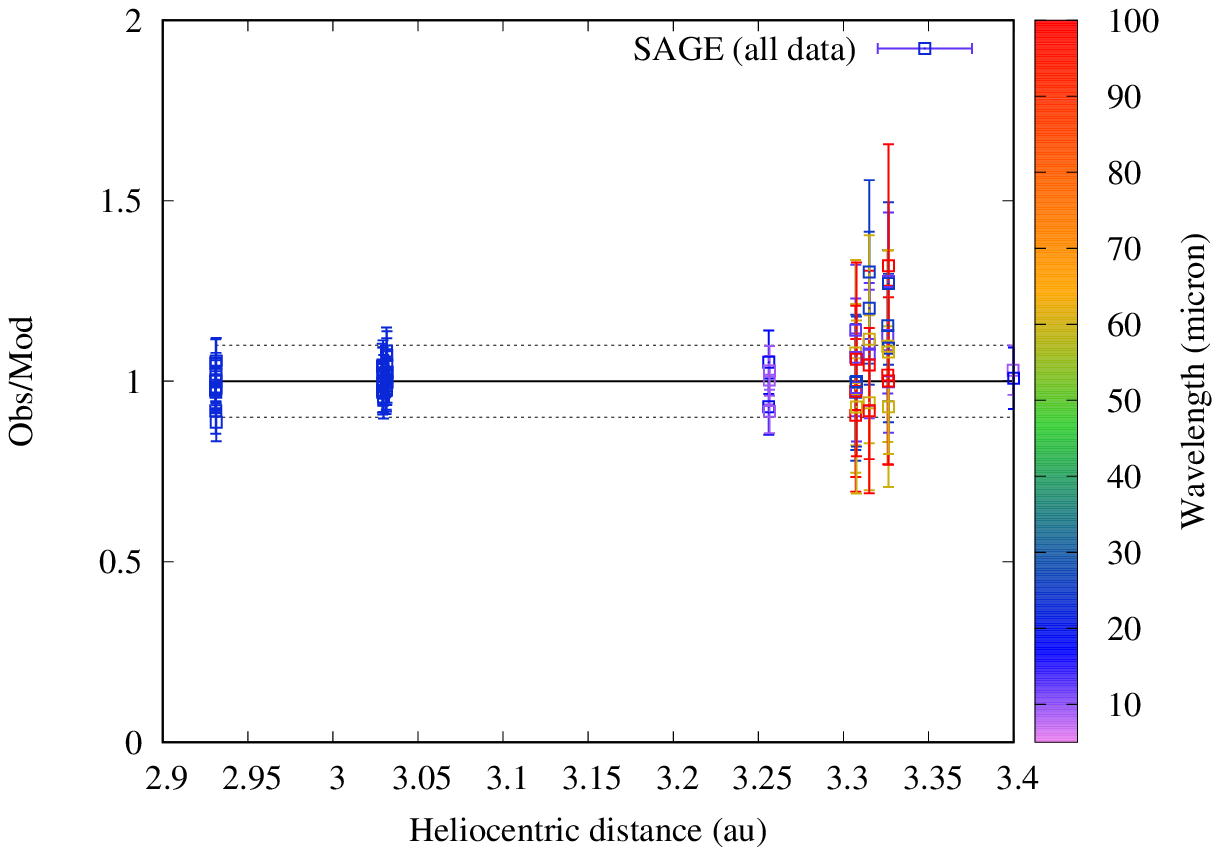}
  
  \includegraphics[width=0.8\linewidth]{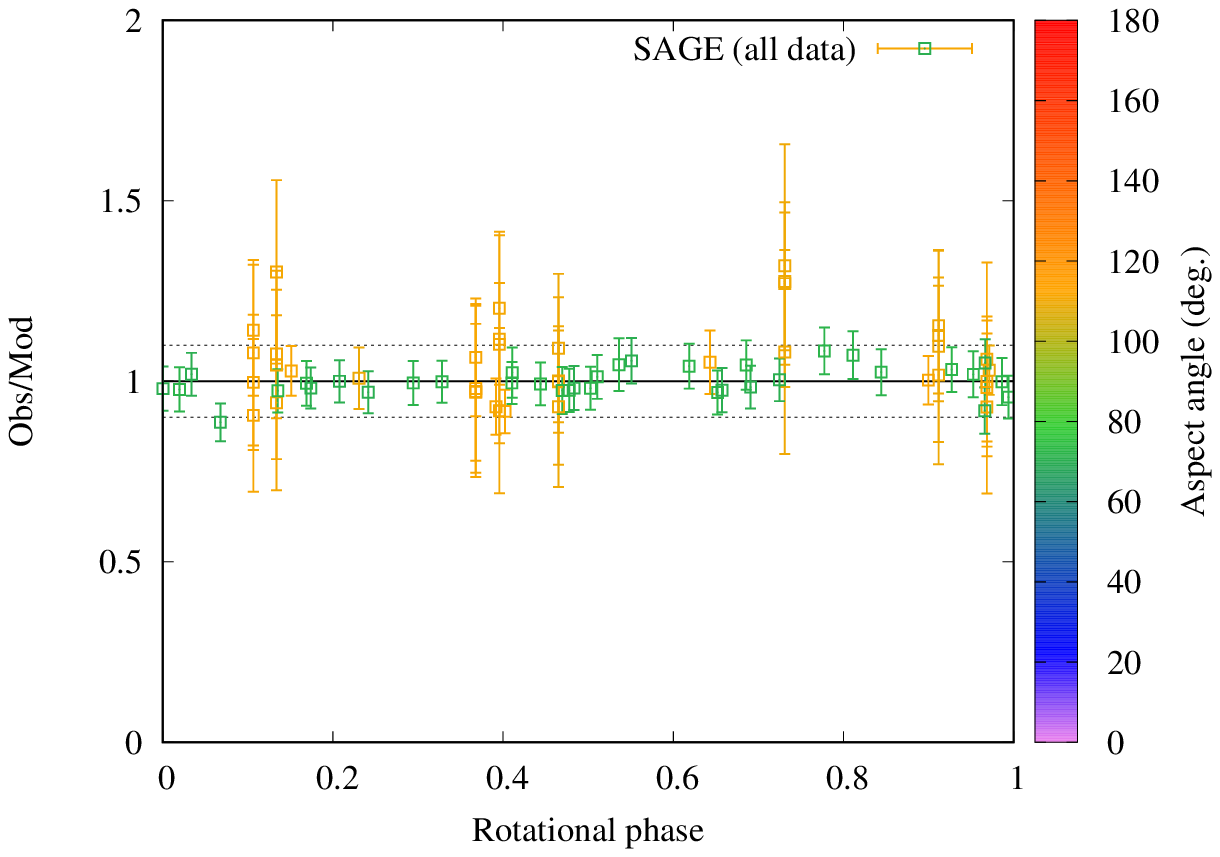}
  
  \includegraphics[width=0.8\linewidth]{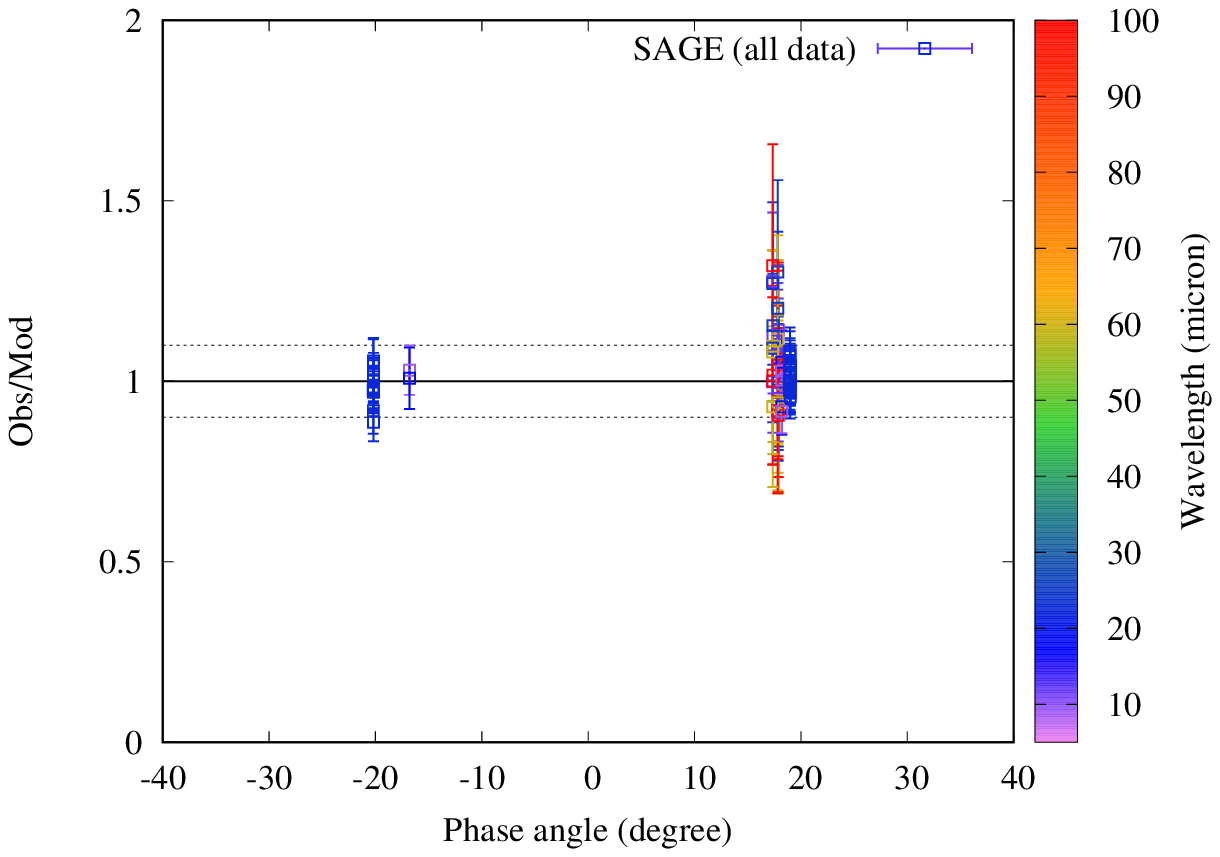}

  \caption{(381) Myrrha.
    There are some waves in the rotational phase plot that suggest small shape
    issues (see also Fig.~\ref{fig:00114_ThLC}), but overall, the fit has a
    low $\chi^2$ and is much better than the sphere with the same spin axis. 
  }\label{fig:00381_OMR}
\end{figure}

\begin{figure}
  \centering
  
  \includegraphics[width=0.8\linewidth]{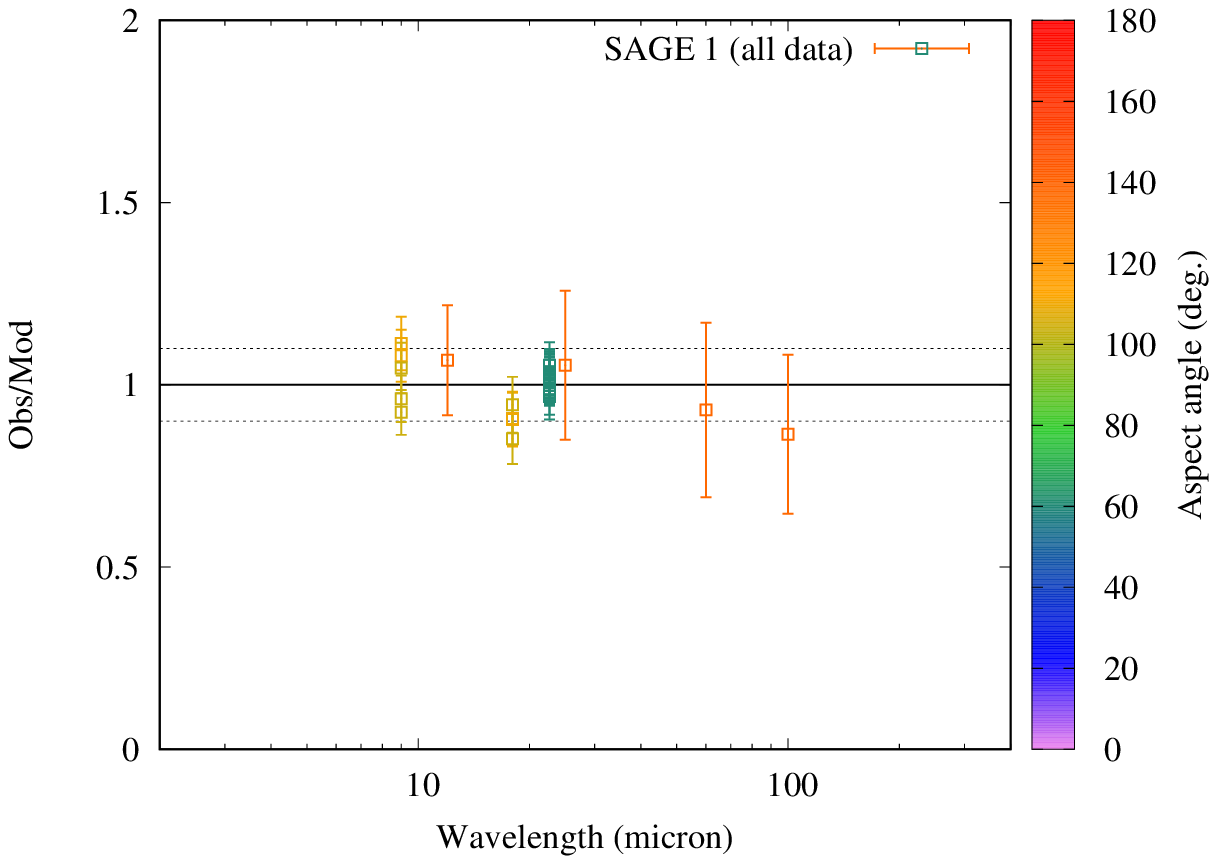}
  
  \includegraphics[width=0.8\linewidth]{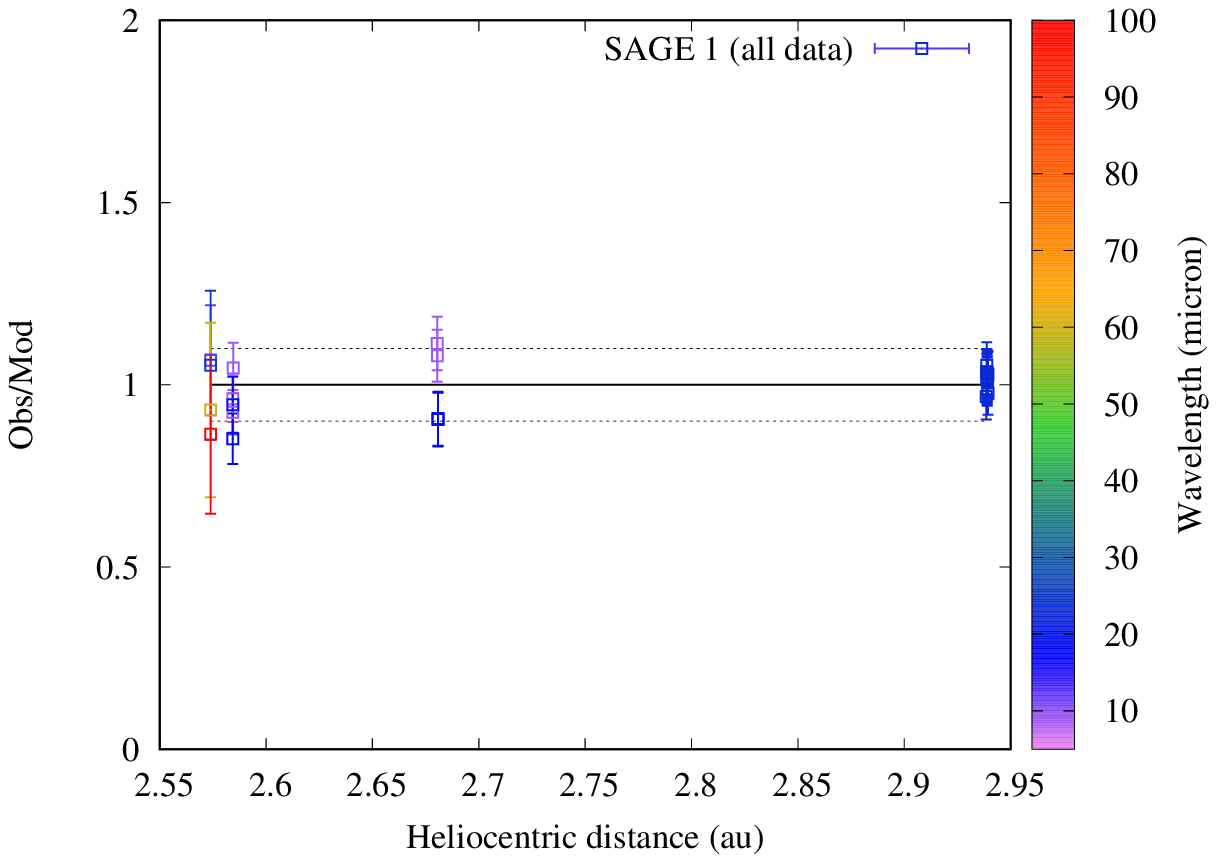}
  
  \includegraphics[width=0.8\linewidth]{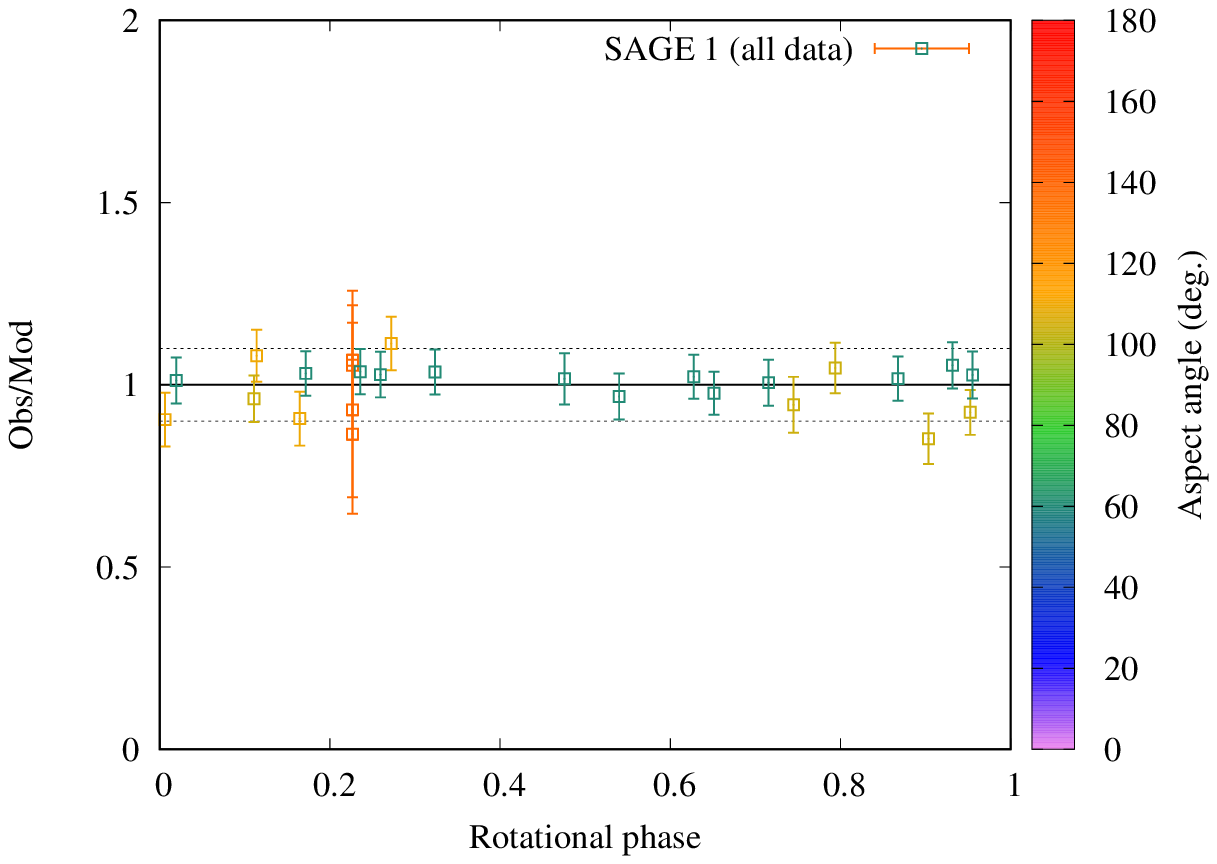}
  
  \includegraphics[width=0.8\linewidth]{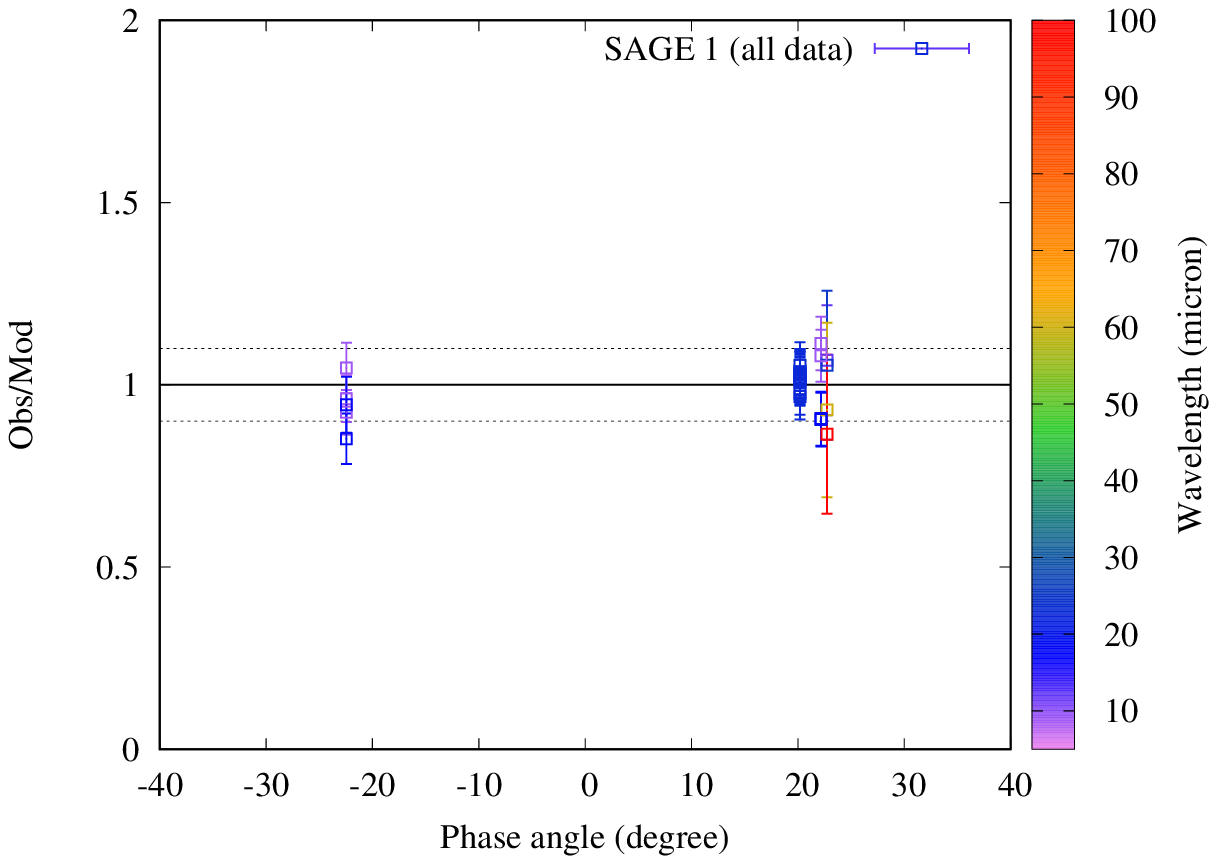}

  \caption{(441) Bathilde. 
  }\label{fig:00441_OMR}
\end{figure}

\begin{figure}
  \centering
  
  \includegraphics[width=0.8\linewidth]{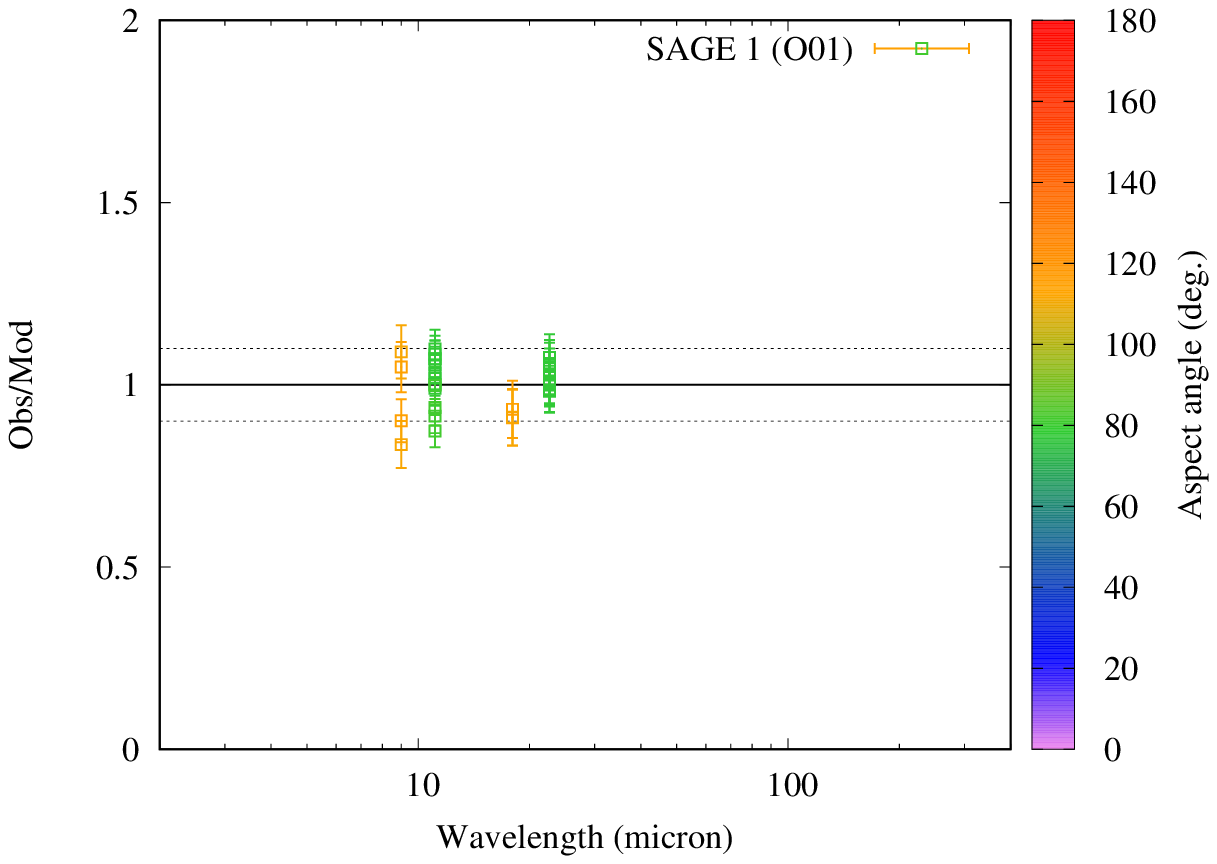}
  
  \includegraphics[width=0.8\linewidth]{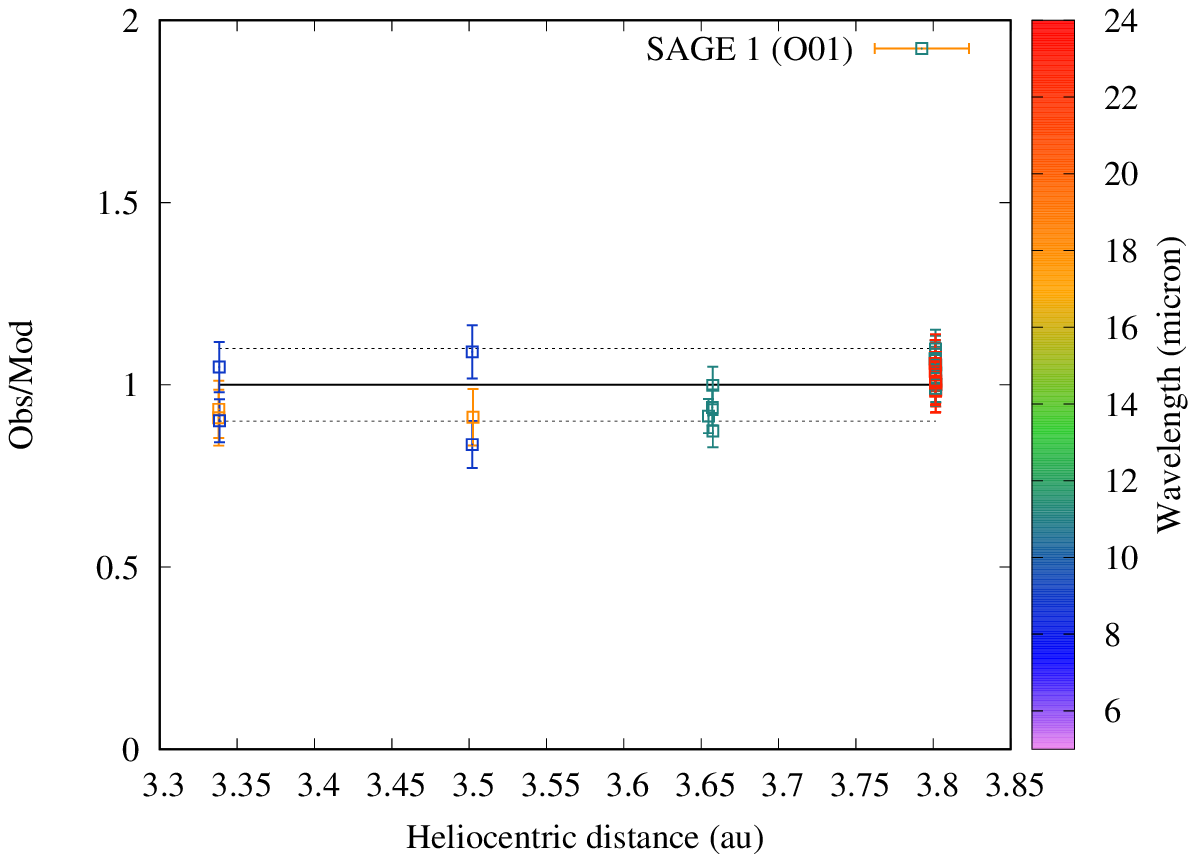}
  
  \includegraphics[width=0.8\linewidth]{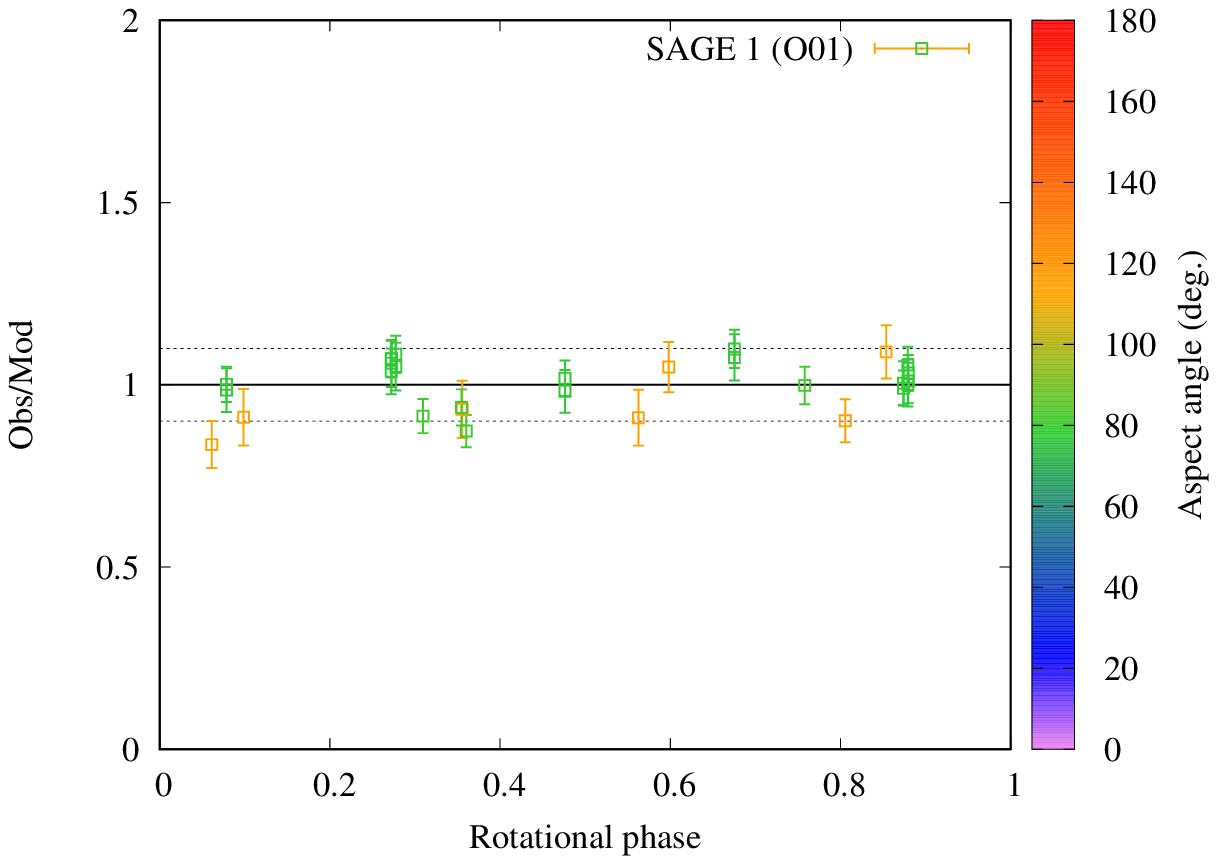}
  
  \includegraphics[width=0.8\linewidth]{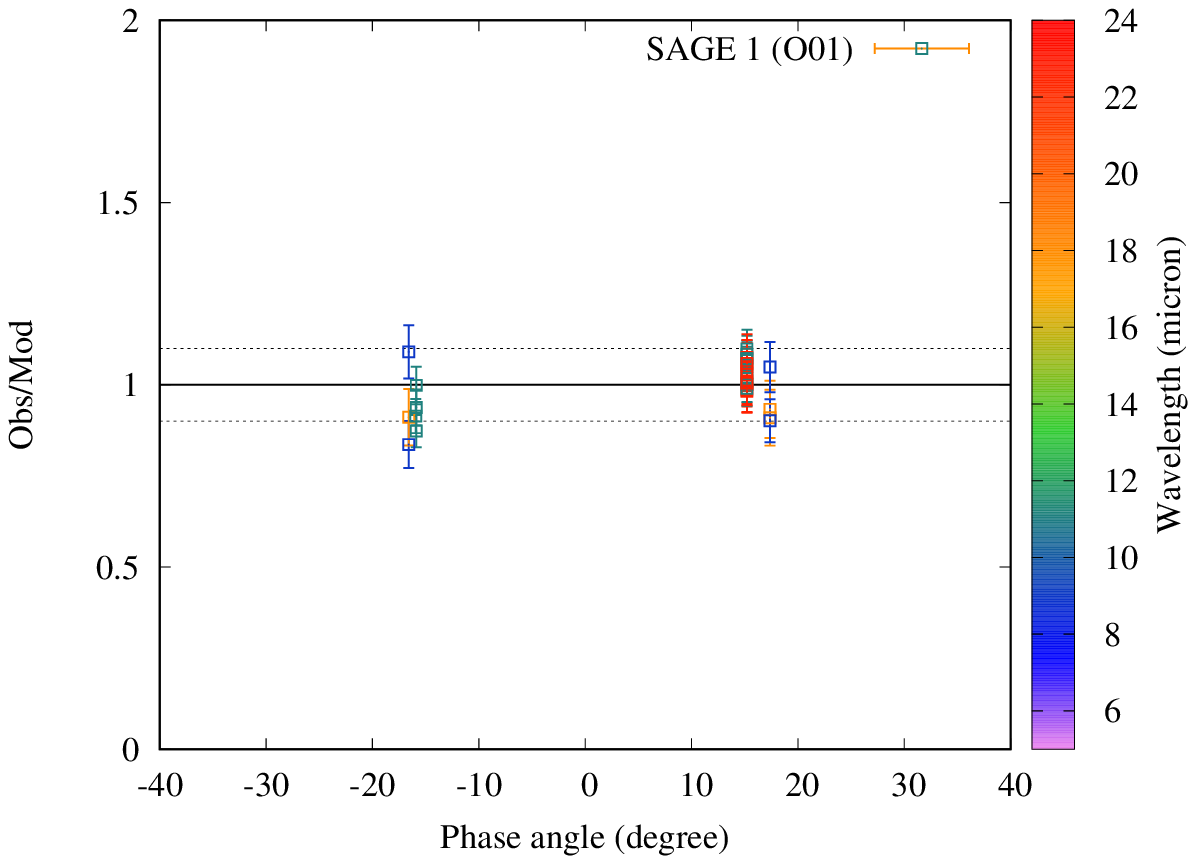}

  \caption{(721) Tabora. 
  }\label{fig:00721_OMR}
\end{figure}

\clearpage

\section{Stellar occultation records fitting}
In this Section we present the model fit to stellar occultation chords.
\begin{figure*}[h]
\centering
\includegraphics[width=0.292\textwidth]{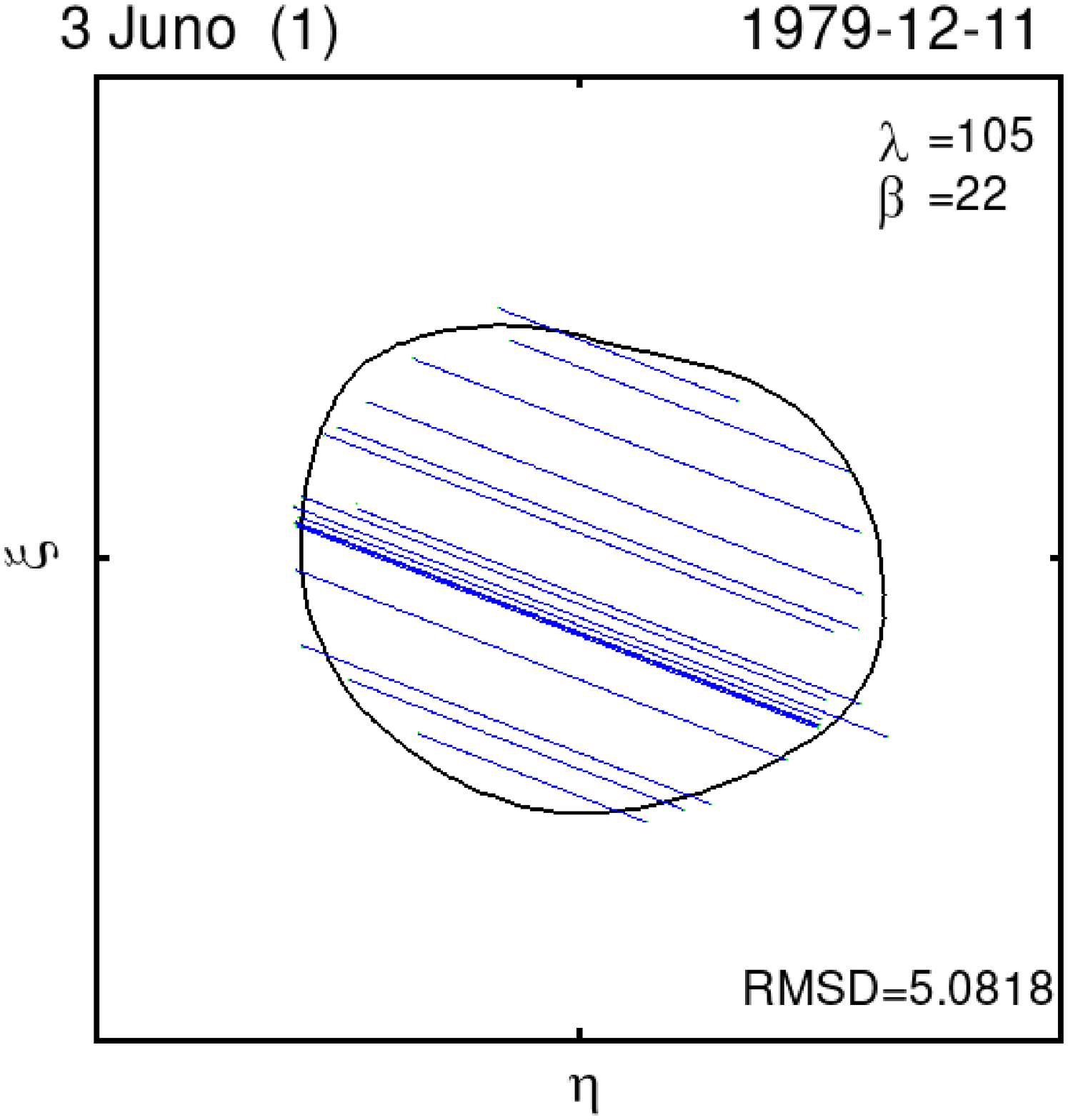}
\includegraphics[width=0.3\textwidth]{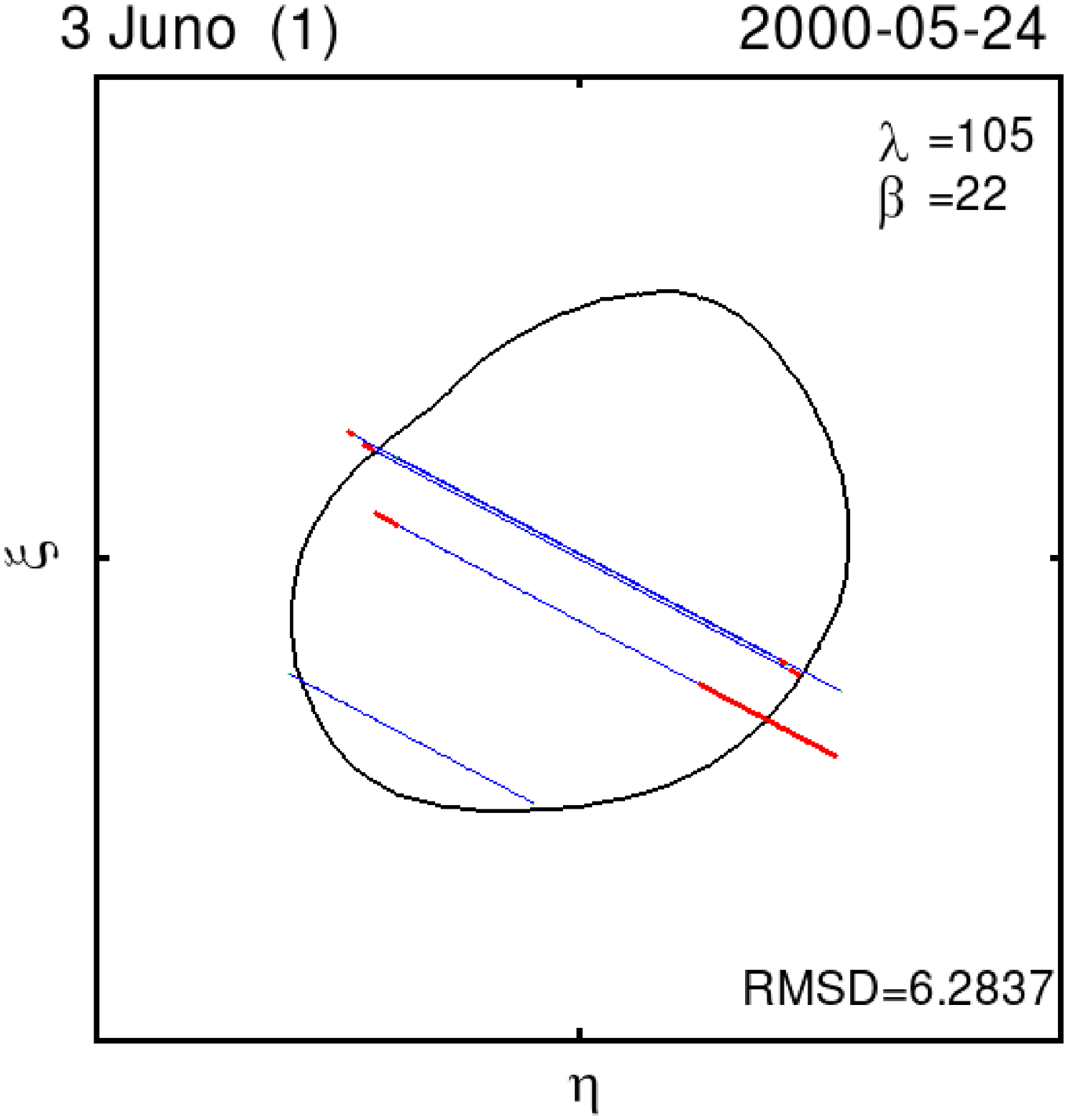}  
\includegraphics[width=0.3\textwidth]{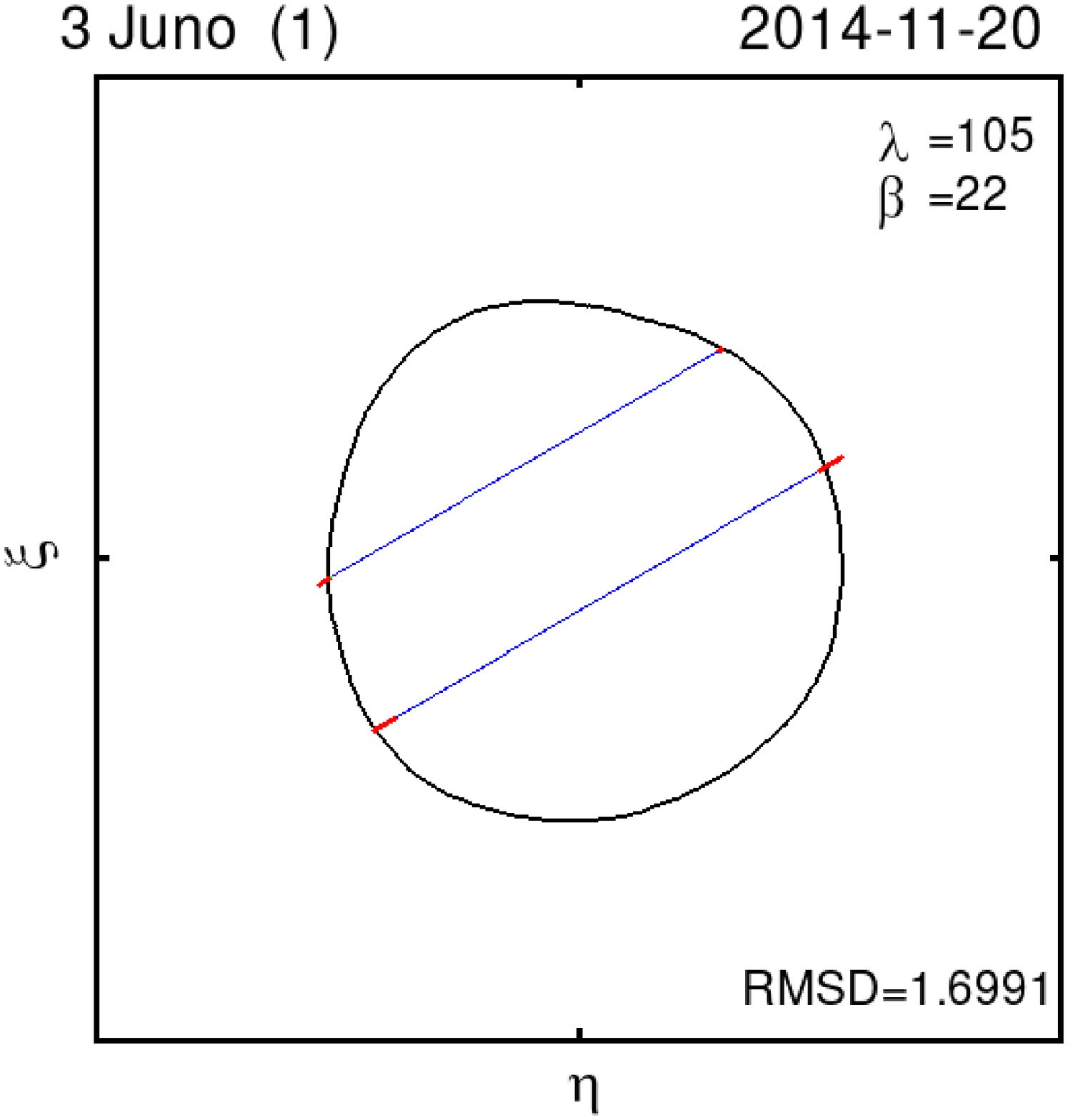}  
\caption{
  Shape model fitting to stellar occultations by 3 Juno.
} 
\label{junoocc}
\end{figure*}  

\begin{figure*}[h]
\centering
\includegraphics[width=0.292\textwidth]{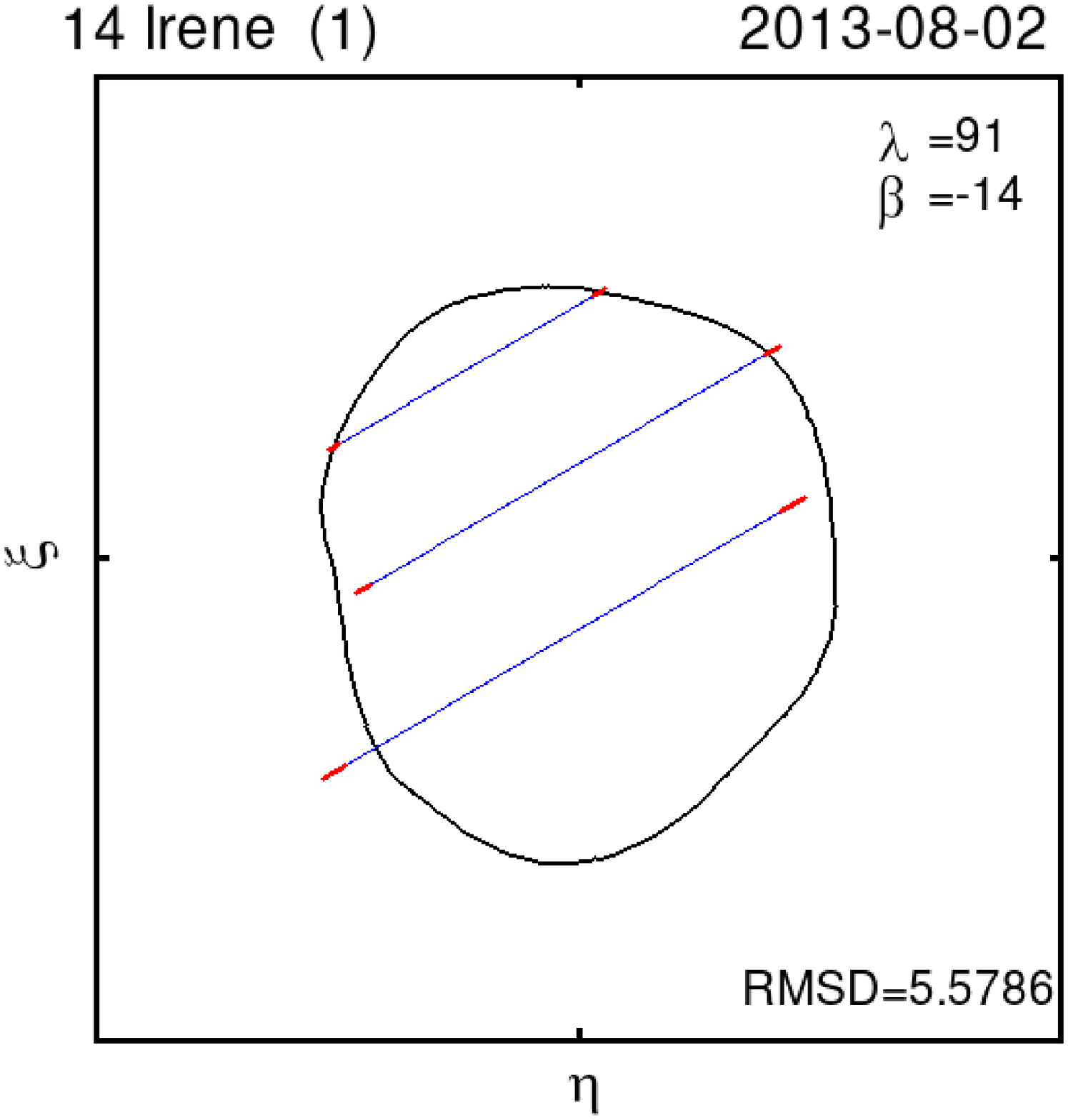}
\includegraphics[width=0.3\textwidth]{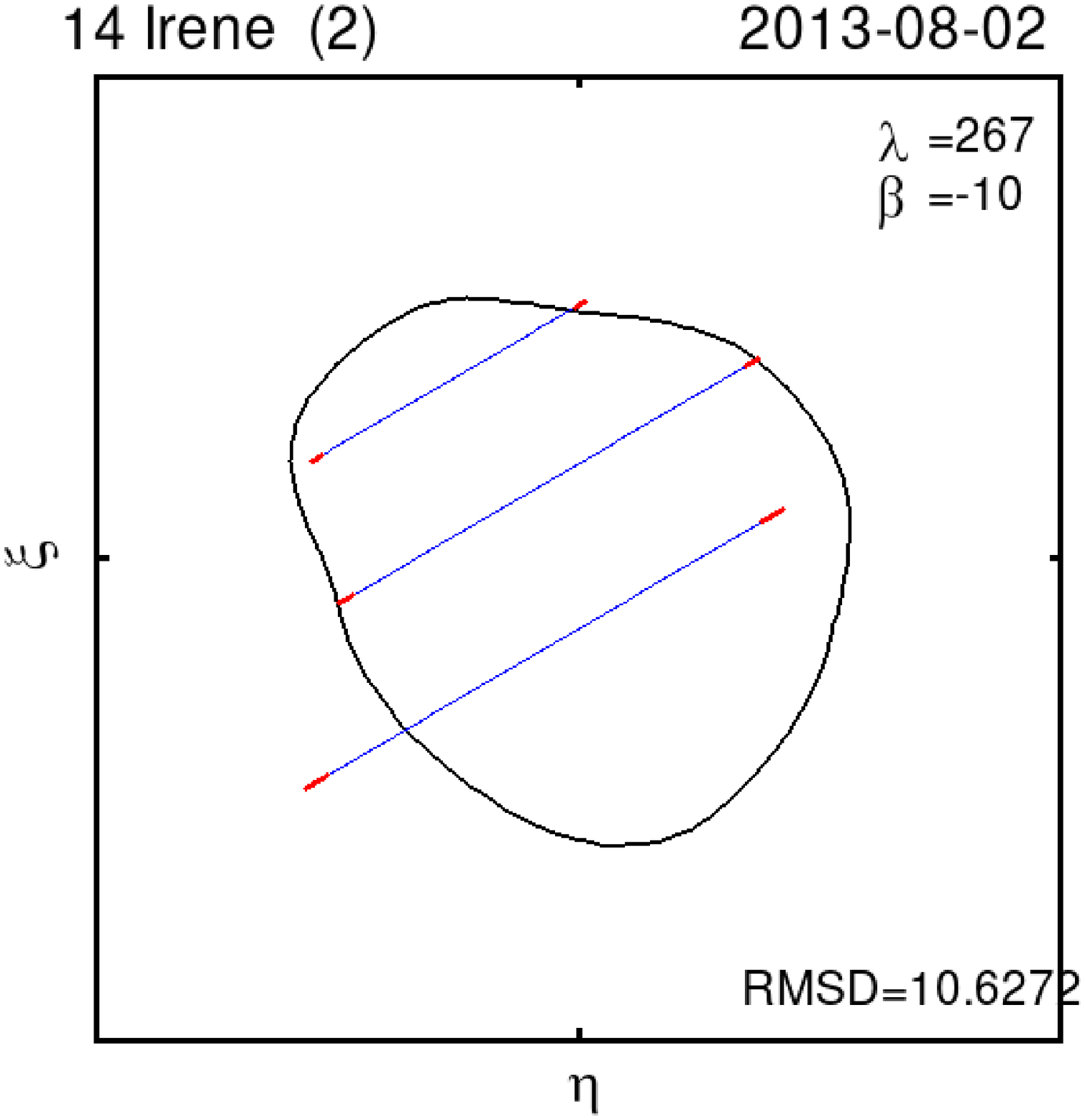}
\caption{
  Shape model fitting to stellar occultations by 14 Irene.
} 
\label{ireneocc}
\end{figure*}

\begin{figure*}[h]
\centering
\includegraphics[width=0.292\textwidth]{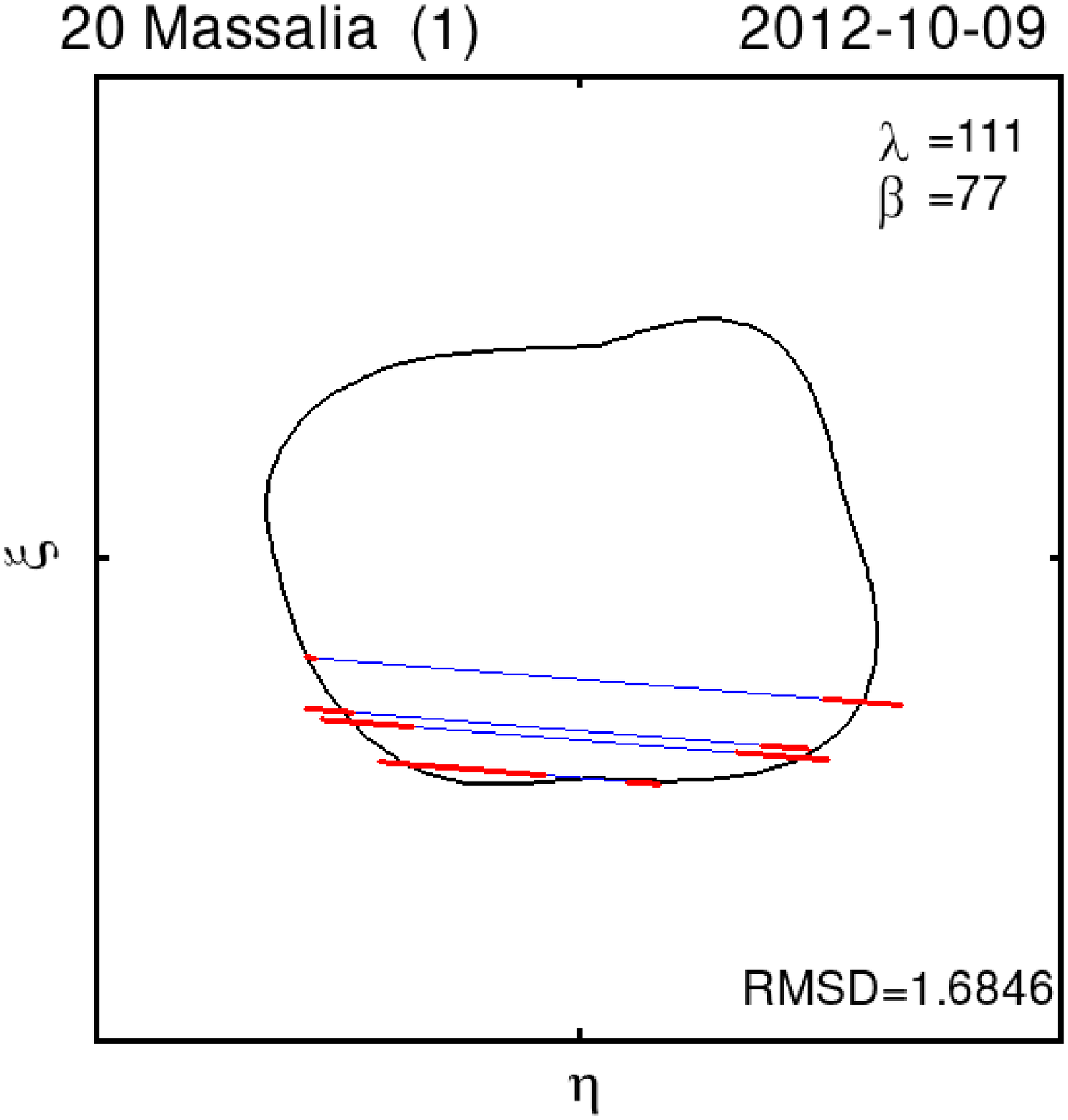}
\includegraphics[width=0.3\textwidth]{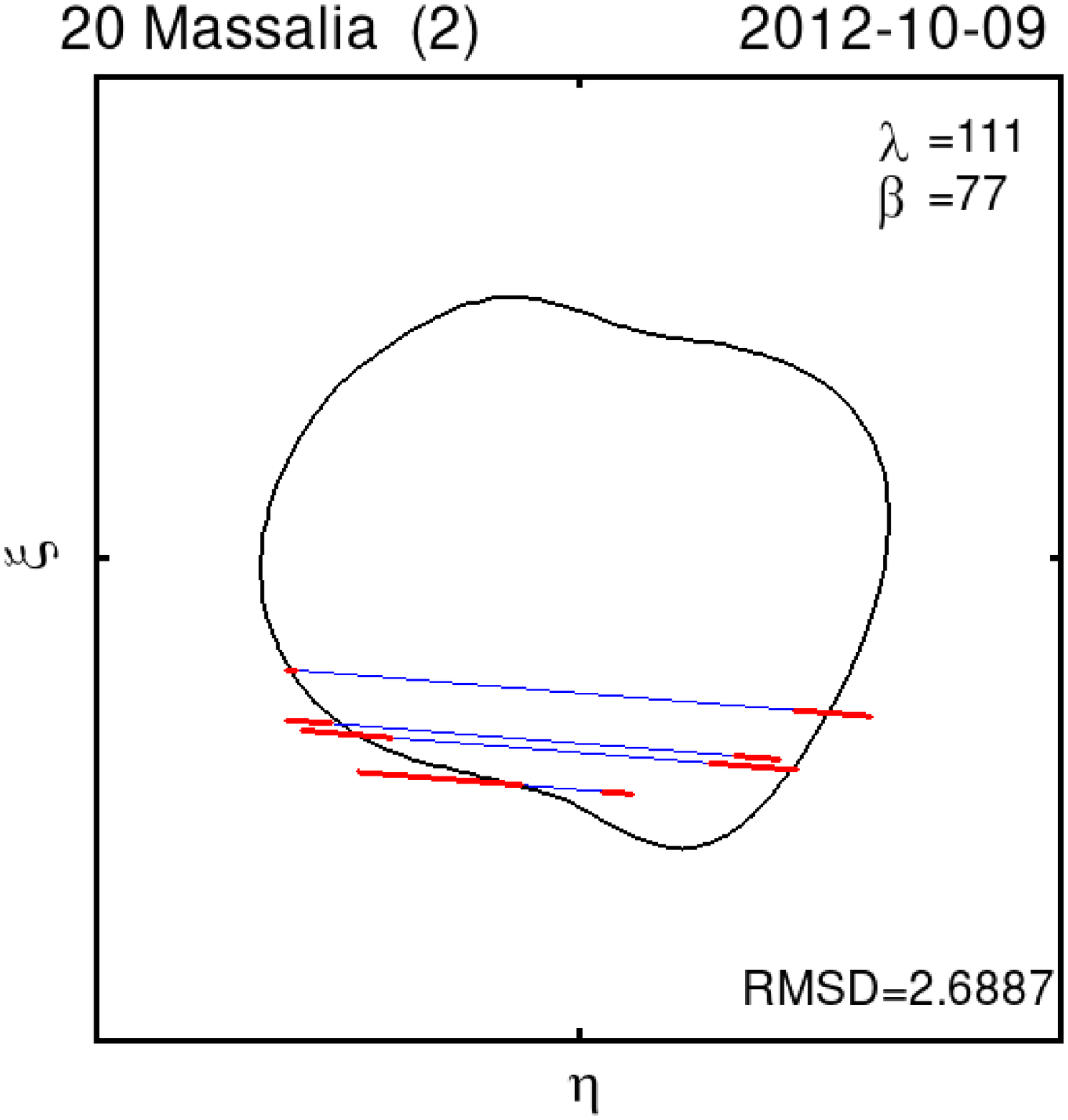}
\caption{
  Shape model fitting to stellar occultations by 20 Massalia.                  
}
\label{massaliaocc}
\end{figure*}

\clearpage

\begin{figure*}[h]
\centering
\includegraphics[width=0.292\textwidth]{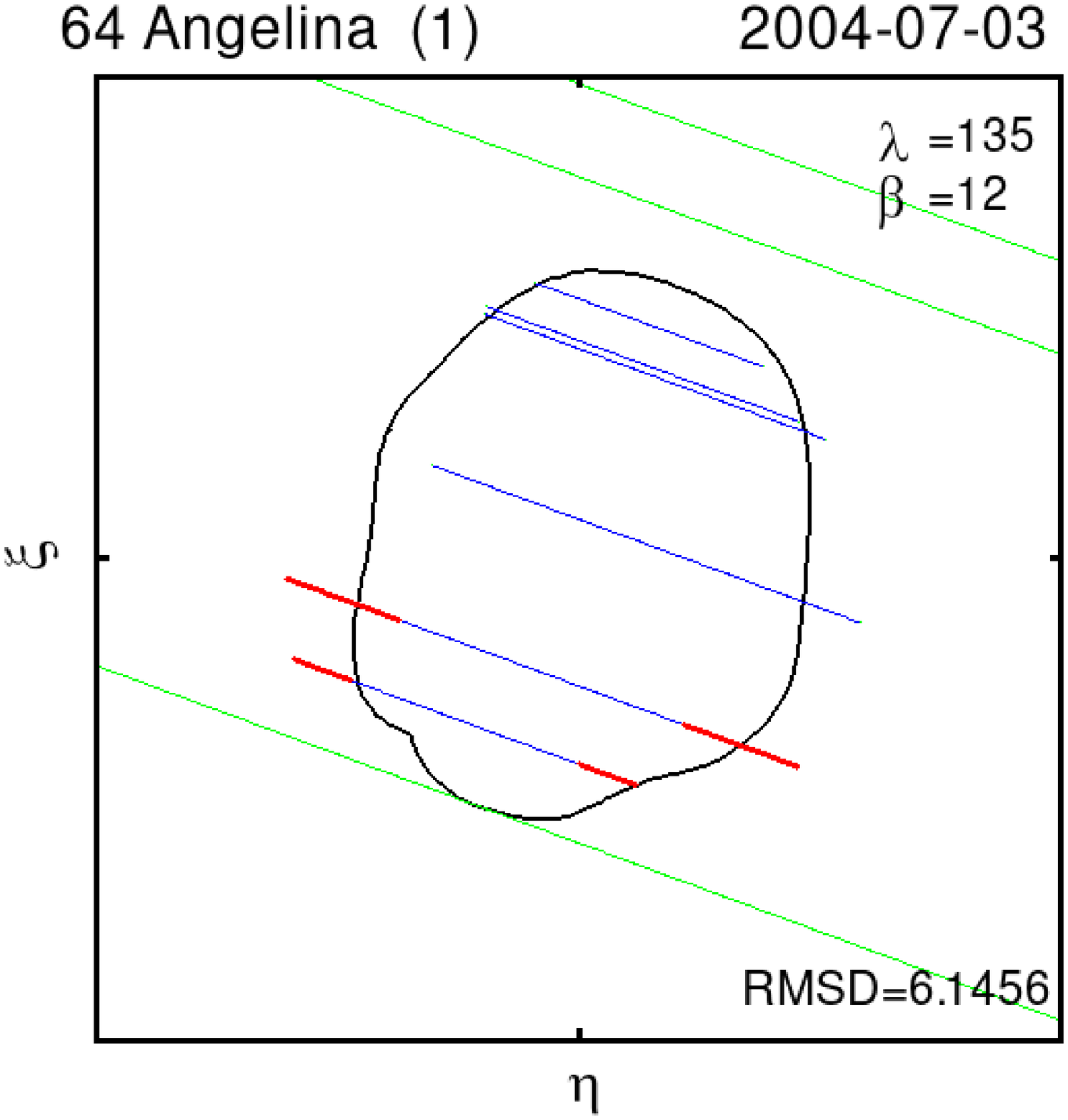}
\includegraphics[width=0.3\textwidth]{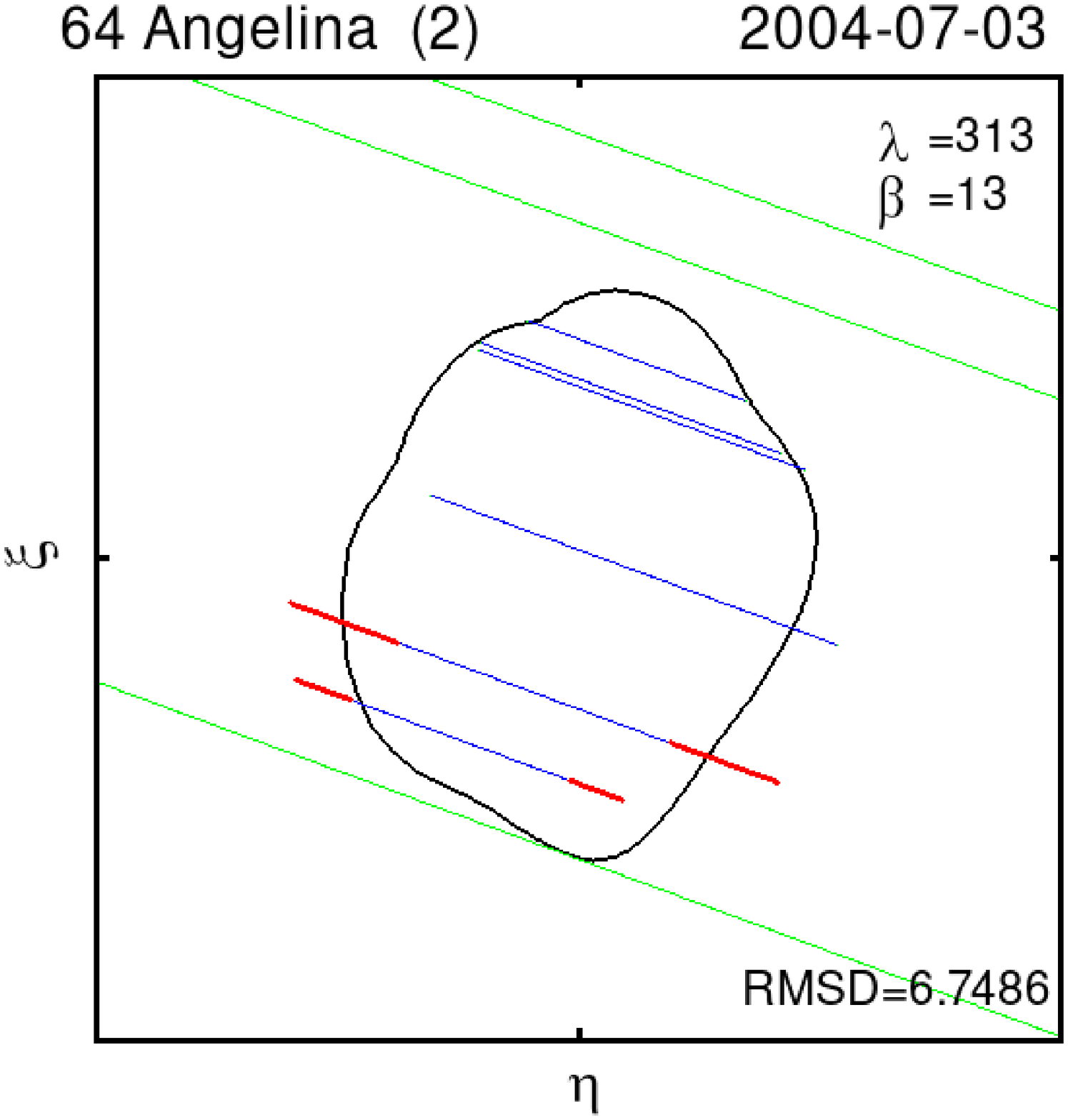}
\caption{
  Shape model fitting to stellar occultations by 64 Angelina.
}
\label{angelinaocc}
\end{figure*}

\begin{figure*}[h]
\centering
\includegraphics[width=0.292\textwidth]{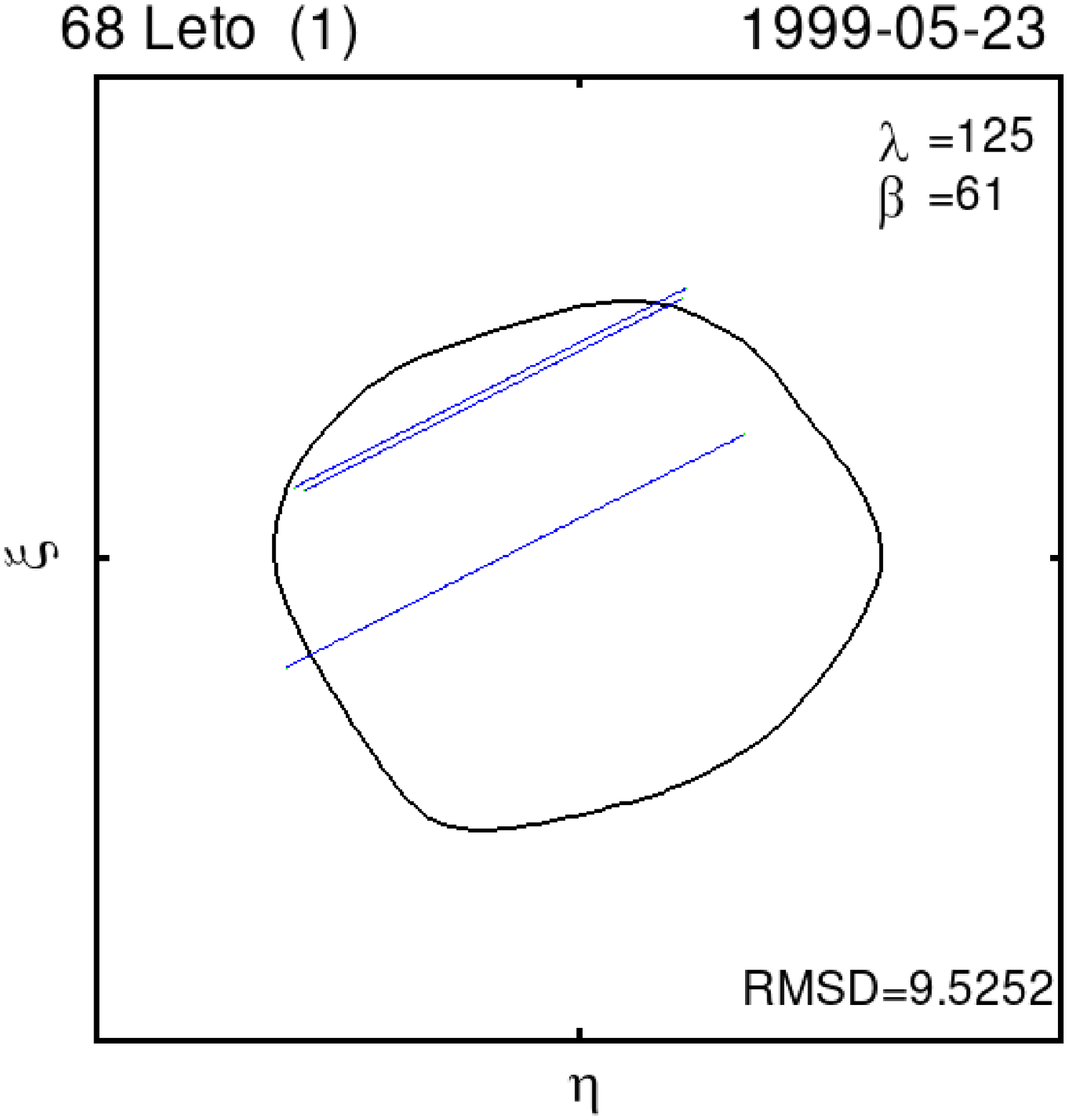}
\includegraphics[width=0.3\textwidth]{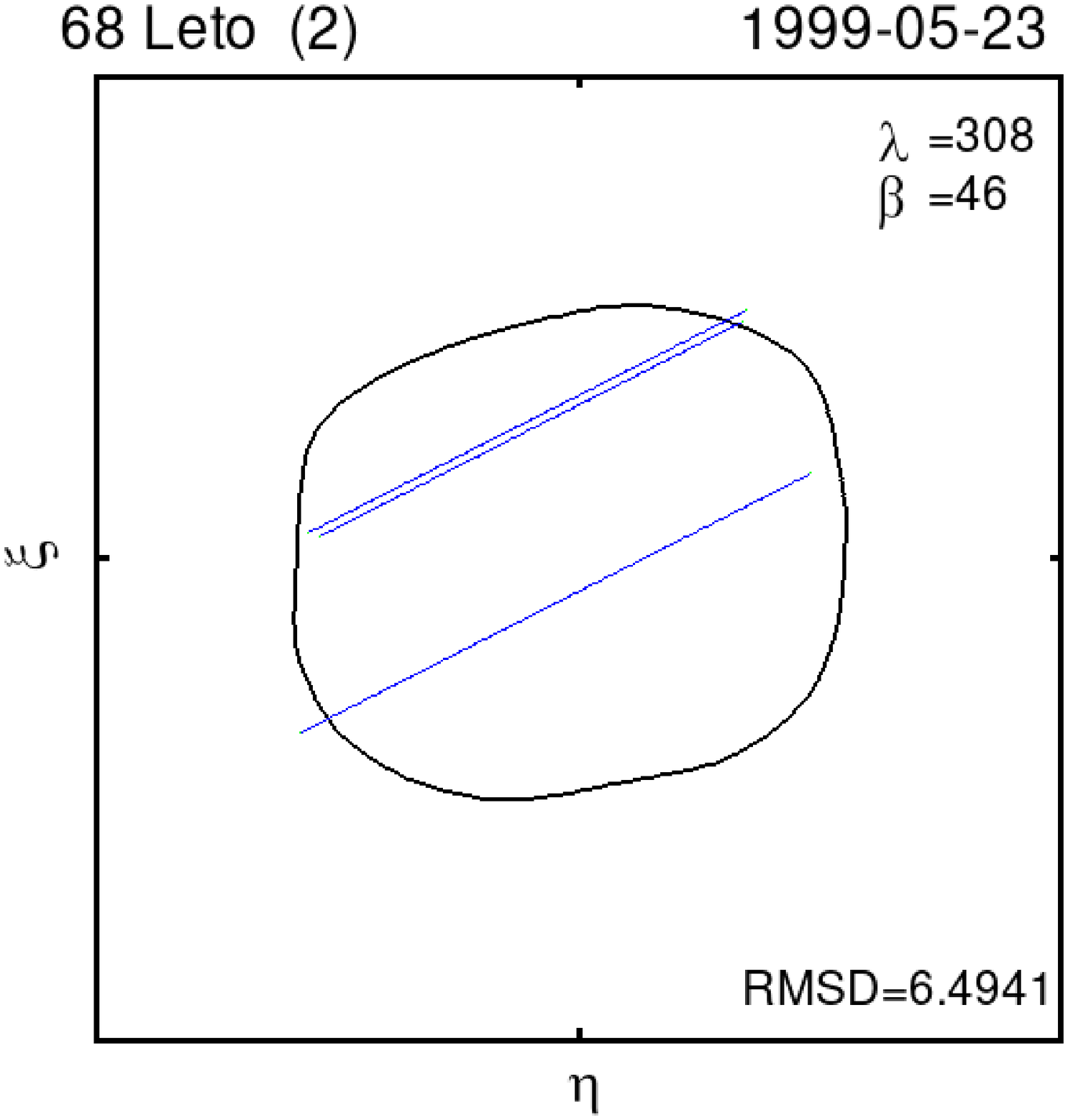}  
\caption{
  Shape model fitting to stellar occultations by 68 Leto.
}
\label{letoocc}
\end{figure*}

\begin{figure*}[h]
\centering
\includegraphics[width=0.292\textwidth]{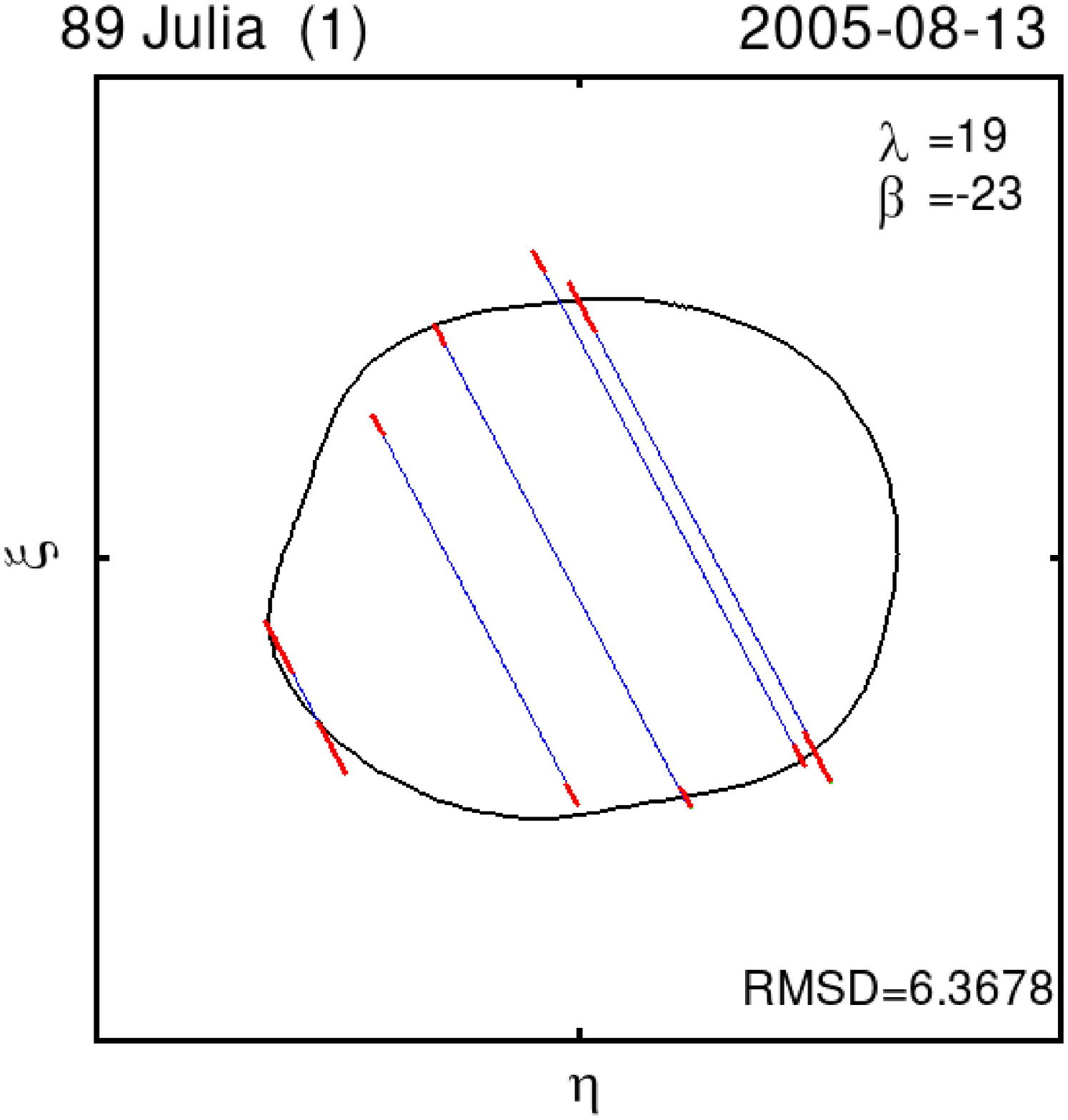}
\includegraphics[width=0.3\textwidth]{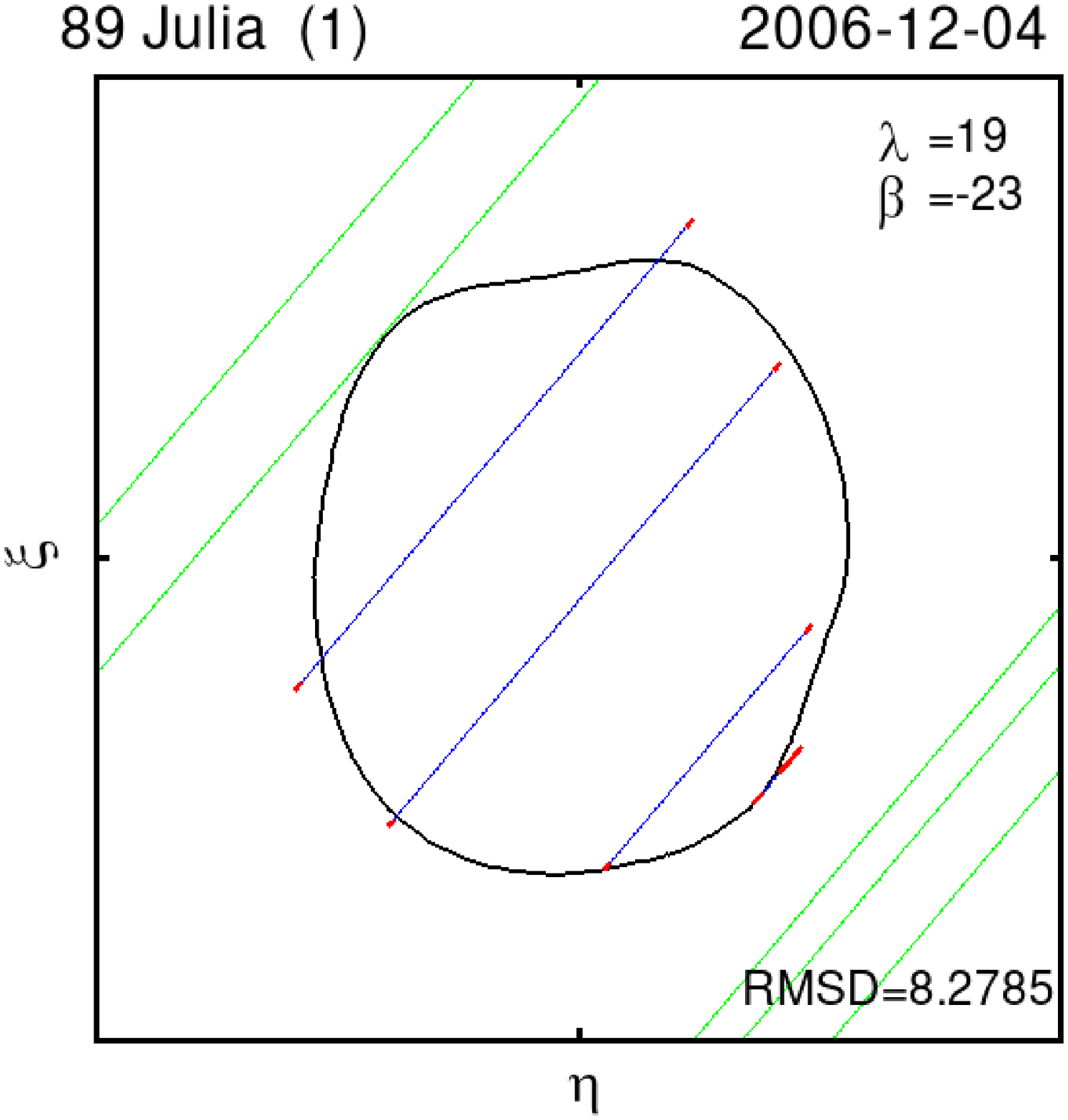}  
\caption{
  Shape model fitting to stellar occultations by 89 Julia.
}
\label{juliaocc}
\end{figure*} 

\clearpage

\begin{figure*}[h]
\centering
\includegraphics[width=0.292\textwidth]{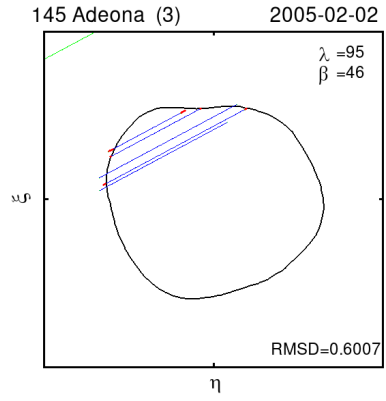}
\caption{
  Shape model fitting to stellar occultations by 145 Adeona.
}
\label{adeonaocc}
\end{figure*} 

\begin{figure*}[h]
\centering
\includegraphics[width=0.292\textwidth]{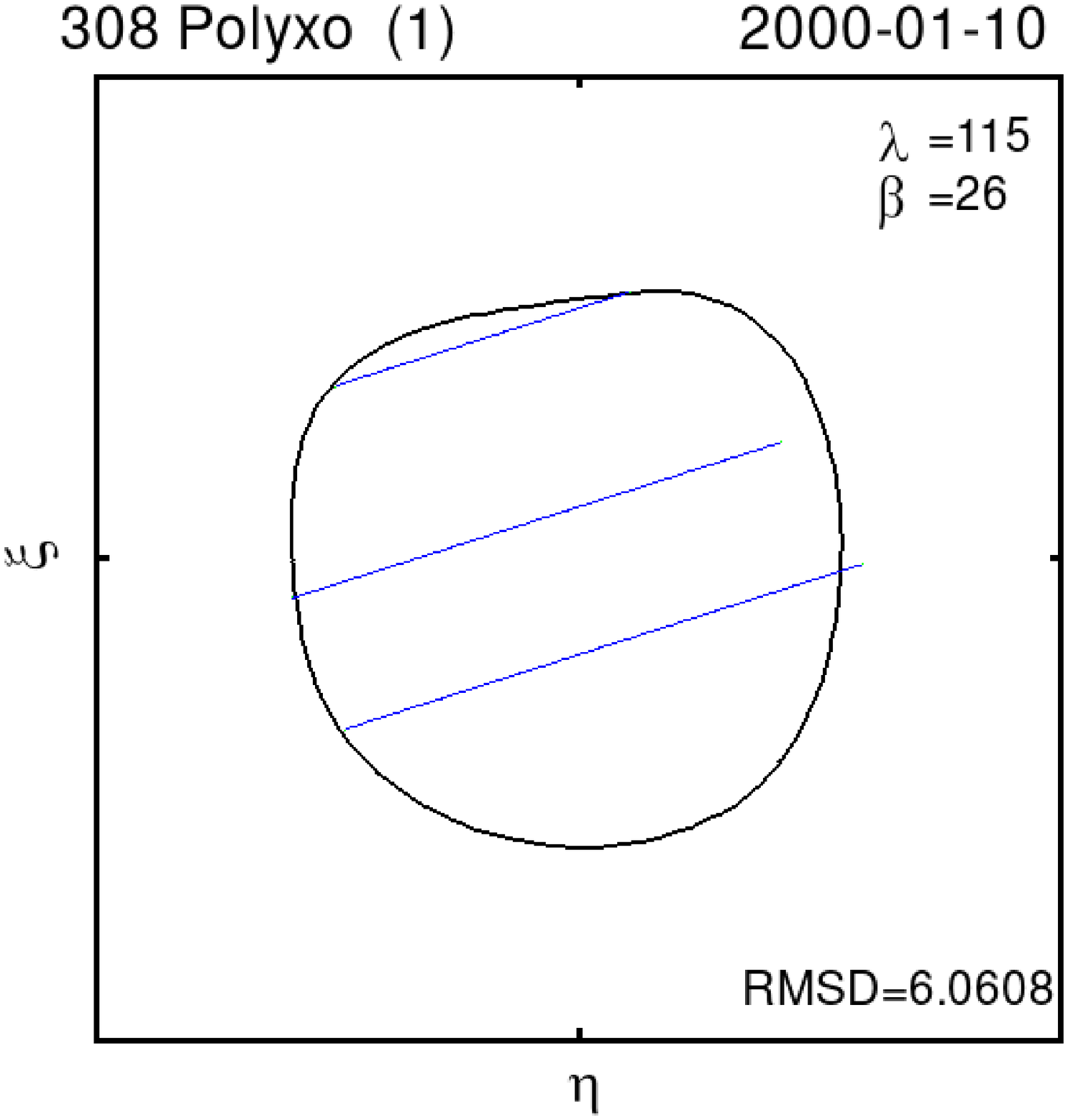}  
\includegraphics[width=0.3\textwidth]{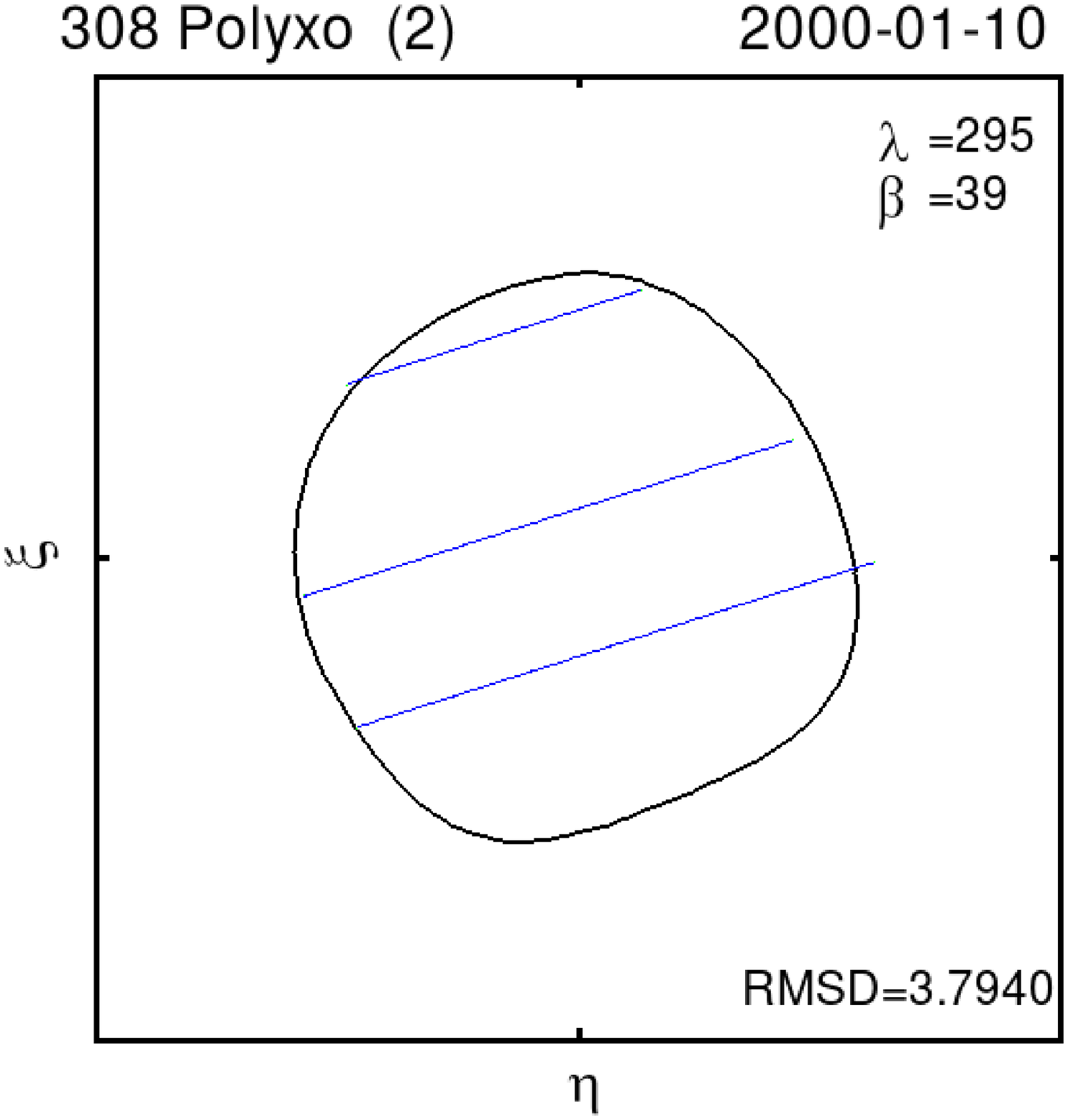}
\includegraphics[width=0.3\textwidth]{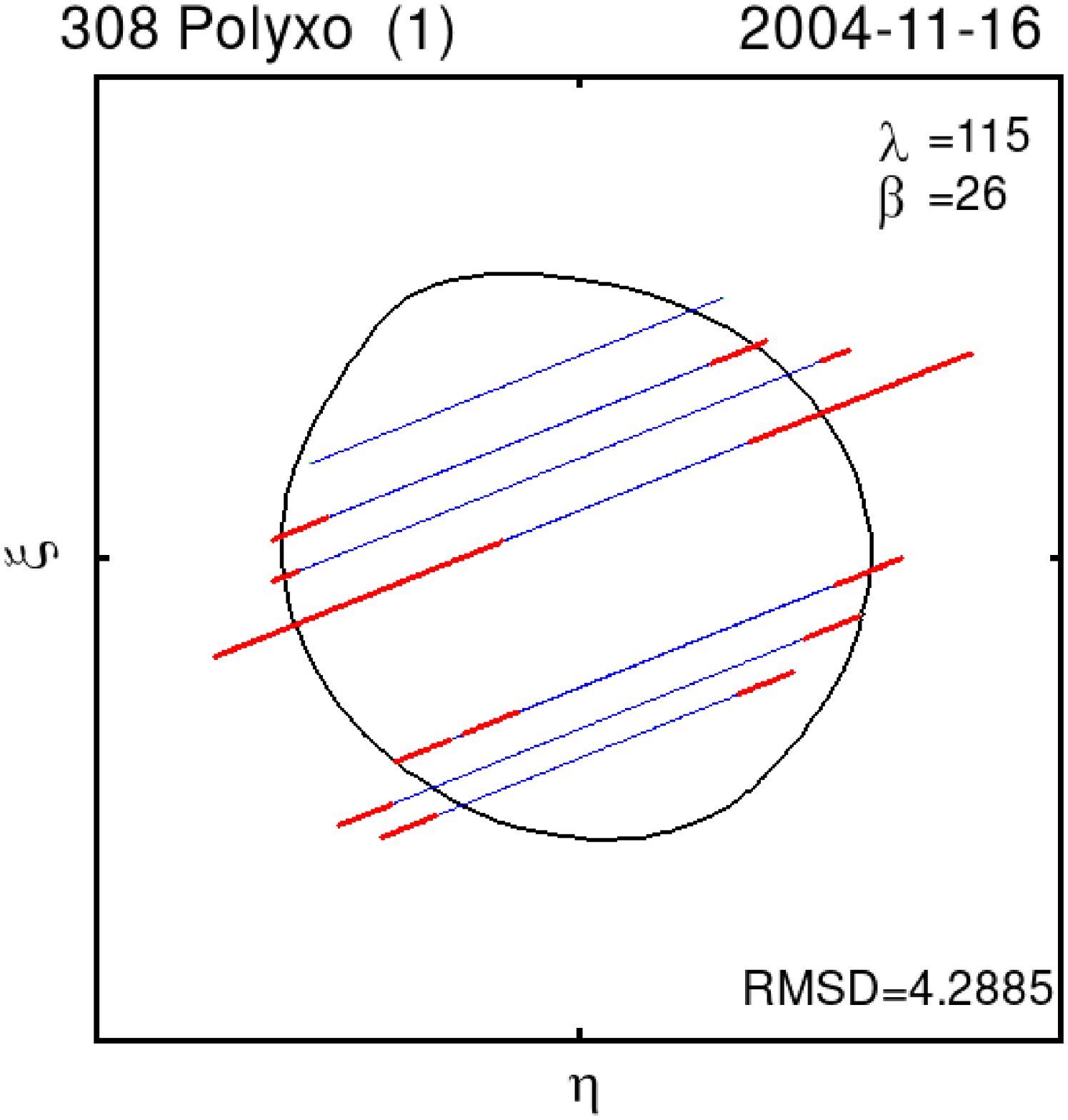}  
\includegraphics[width=0.3\textwidth]{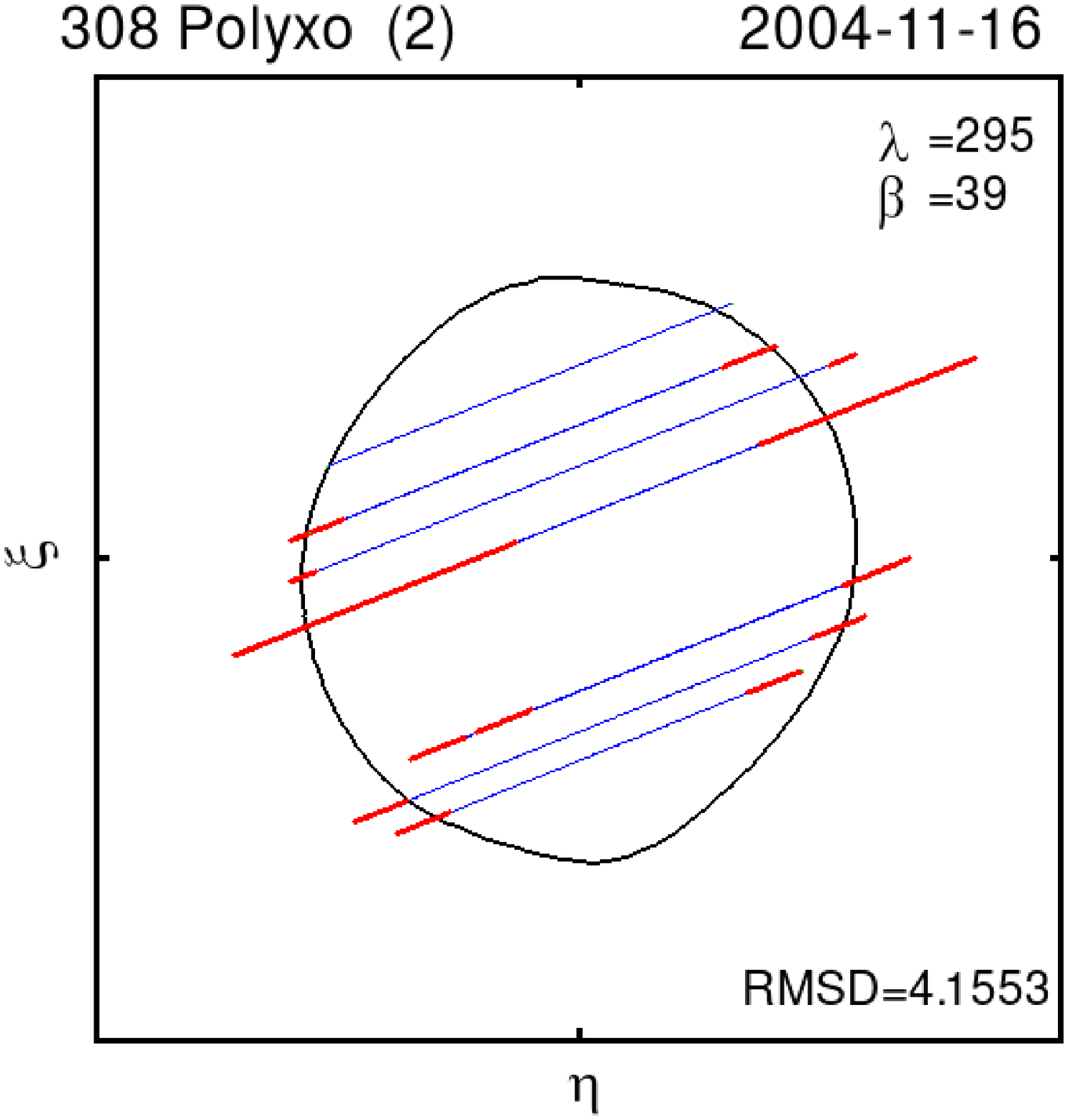}
\includegraphics[width=0.3\textwidth]{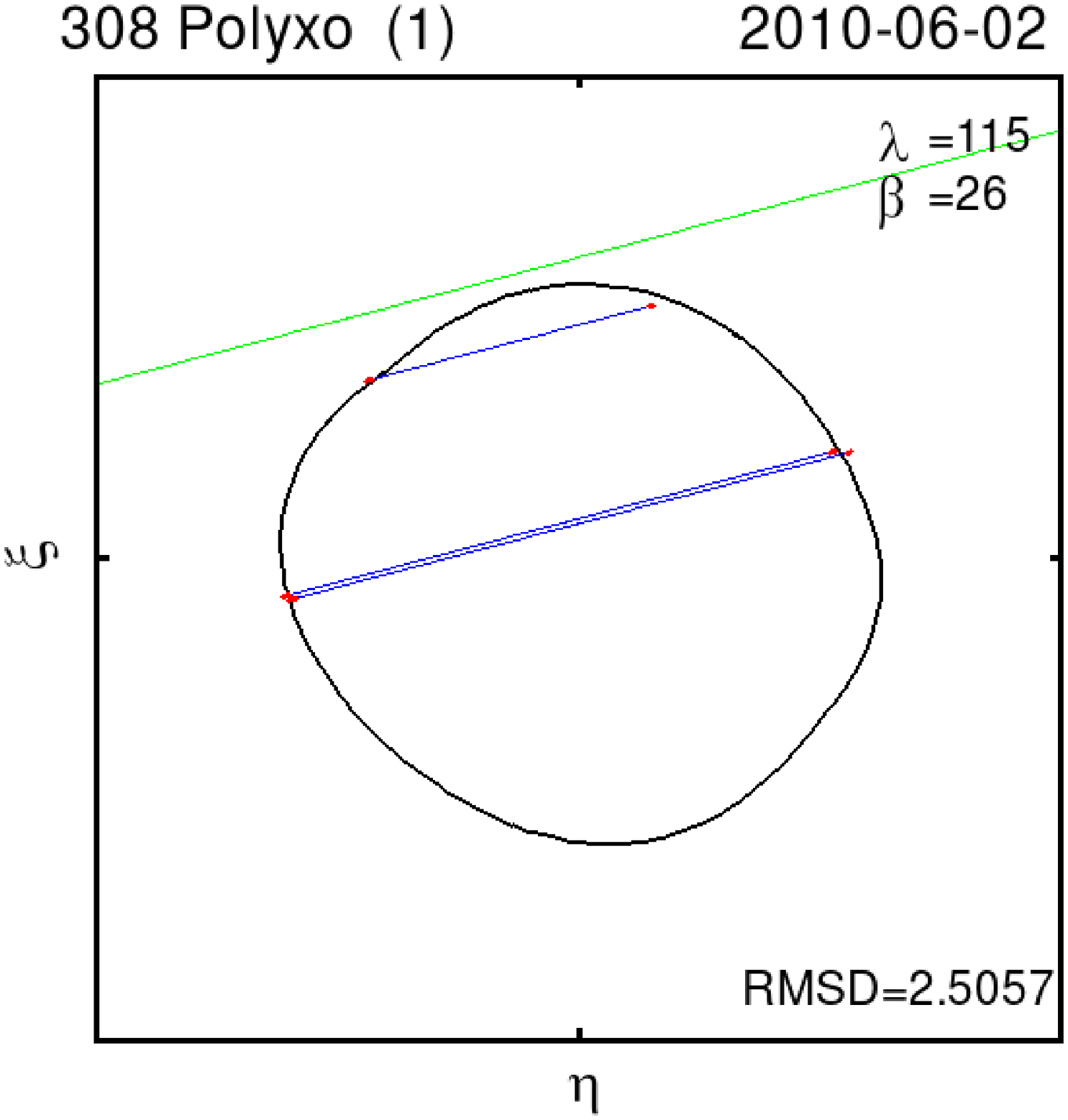}
\includegraphics[width=0.3\textwidth]{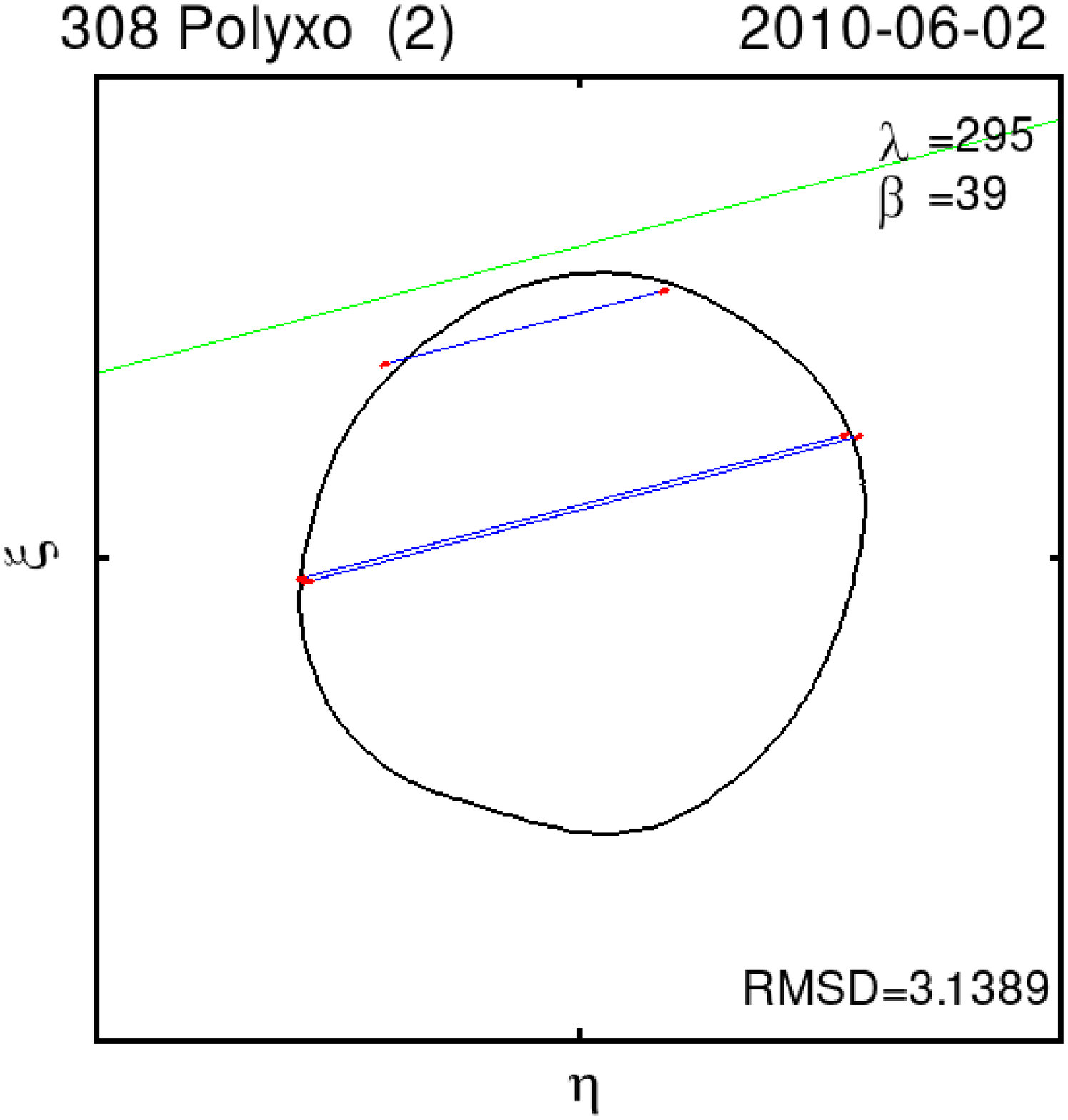}
\caption{
  Shape model fitting to stellar occultations by 308 Polyxo.
}
\label{polyxoocc}
\end{figure*} 

\clearpage

\begin{figure*}[h]
\centering
\includegraphics[width=0.292\textwidth]{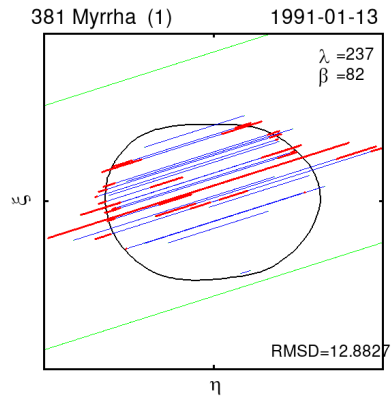}
\caption{
  Shape model fitting to stellar occultations by 381 Myrrha.
}
\label{myrrhaocc}
\end{figure*} 

\begin{figure*}[h]
\centering
\includegraphics[width=0.292\textwidth]{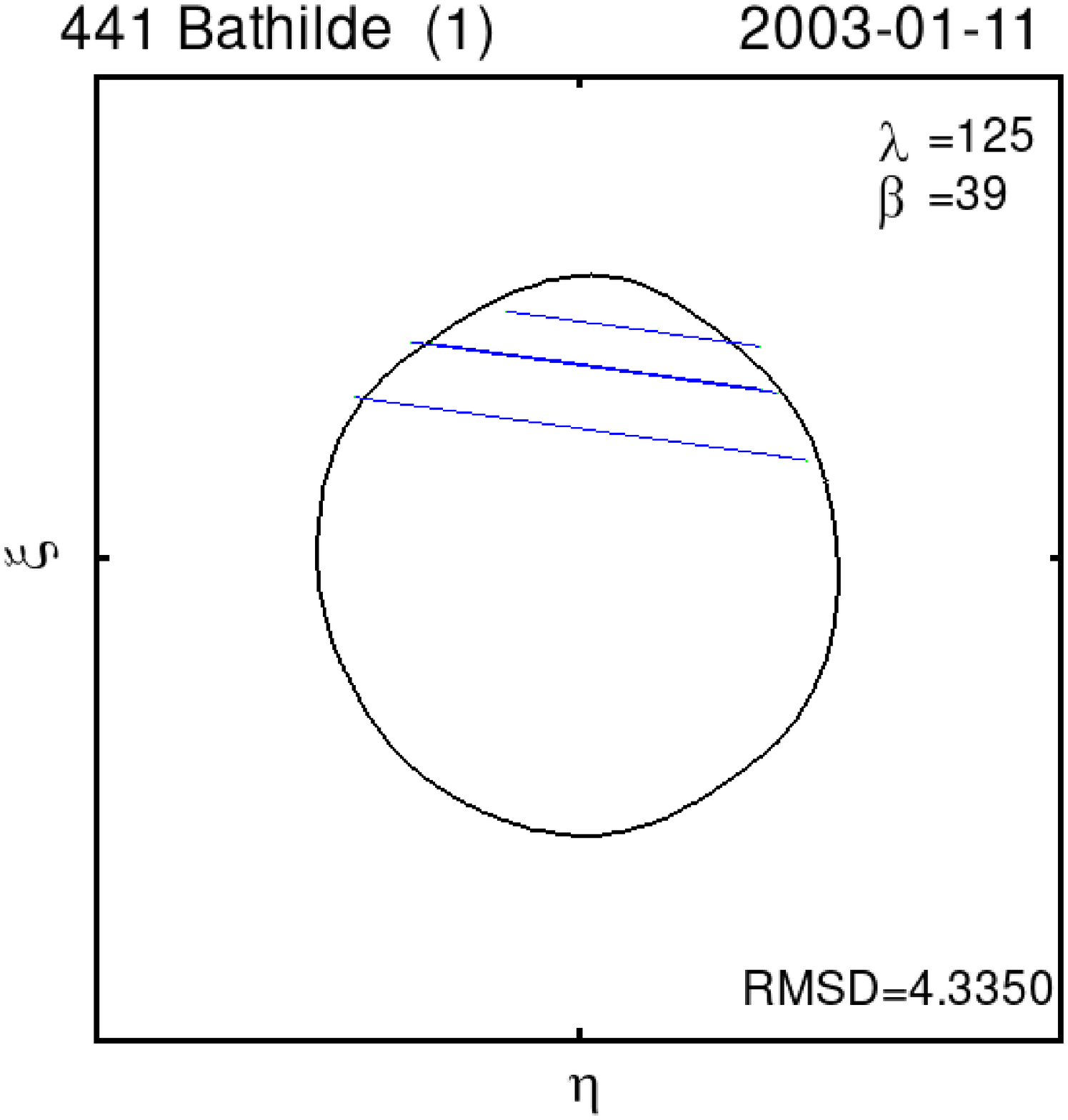}
\includegraphics[width=0.3\textwidth]{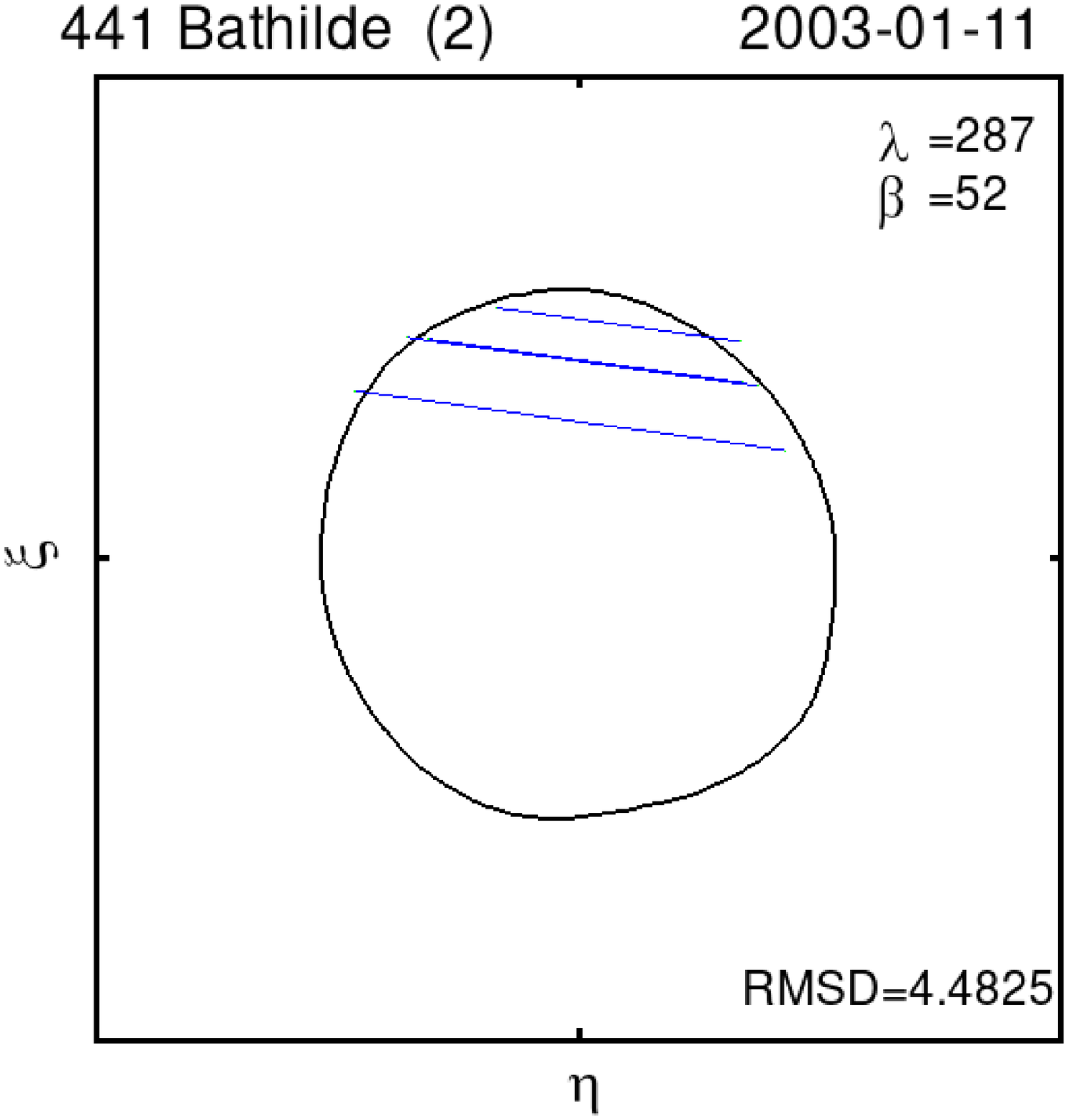}  
\caption{
  Shape model fitting to stellar occultations by 441 Bathilde.
}
\label{bathildeocc}
\end{figure*}

\end{appendix}

\end{document}